\documentclass[a4paper,11pt]{article}
\usepackage{graphicx}
\usepackage{epsf,amsmath,bbold,amsfonts,stmaryrd}

\usepackage[utf8]{inputenc}
\usepackage{mathrsfs}
\usepackage{appendix}
\usepackage{amssymb}
\usepackage{float}
\usepackage{color}
\usepackage{cite}
\usepackage{hyperref}
\hypersetup{pageanchor=false}
\usepackage{indentfirst}
\usepackage{url}
\usepackage{float}
\usepackage{caption}
\usepackage[numbers,square,comma,sort&compress,merge]{natbib}
\usepackage{esint}
\usepackage{overpic}
\usepackage{textcomp}
\usepackage{ulem}
\DeclareMathOperator\arctanh{arctanh}
\def\a{\alpha}
\def\b{\beta}

\def\d{\delta}
\def\D{\Delta}
\def\e{\varepsilon} %
\def\f{\frac}
\def\g{\gamma}

\def\lm{\lambda} %
\def\l{\left}

\def\nn{\nonumber}
\def\om{\omega}

\def\pl{\partial}
\def\td{\tilde} %
\def\r{\right}

\def\t{\theta}
\def\T{\Theta}

\def\vp{\varphi}

\def\be{\begin{equation}}
\def\ee{\end{equation}}
\def\bag{\begin{aligned}}
\def\eag{\end{aligned}}
\def\bea{\begin{eqnarray}}
\def\eea{\end{eqnarray}}
\def\ba{\begin{array}}
\def\ea{\end{array}}

\def\bc{\begin{center}}
\def\ec{\end{center}}
\def\bl{\begin{flushleft}}
\def\el{\end{flushleft}}
\def\br{\begin{flushright}}
\def\er{\end{flushright}}
\def\bi{\begin{itemize}}
\def\ei{\end{itemize}}

\def\bt{\begin{tabular}}
\def\et{\end{tabular}}
\newcommand{\mC}{{\mathcal C}}
\newcommand{\si}{\text{sign}}
\newcommand{\In}{\text{in}}
\newcommand{\Out}{\text{out}}

\newcommand{\M}[1]{\mathcal{#1}}
\newcommand{\ab}[1]{\left|#1\right|}

\numberwithin{equation}{section}

\numberwithin{equation}{section}

\begin{document}
\title{\textbf{Multi-level images around Kerr-Newman black holes}}

\author{
Yehui Hou$^{1}$, Peng Liu$^{2}$, Minyong Guo$^{3}$, Haopeng Yan$^{4*}$,
Bin Chen$^{1,4,5}$}
\date{}

\maketitle

\vspace{-10mm}

\begin{center}
{\it

$^1$Department of Physics, Peking University, No.5 Yiheyuan Rd, Beijing
100871, P.R. China\\\vspace{1mm}

$^2$Department of Physics and Siyuan Laboratory,
Jinan University, Guangzhou 510632, China\\\vspace{1mm}

$^3$Department of Physics, Beijing Normal University, Beijing 100875, P. R. China\\\vspace{1mm}

$^4$Center for High Energy Physics, Peking University,
No.5 Yiheyuan Rd, Beijing 100871, P. R. China\\\vspace{1mm}

$^5$ Collaborative Innovation Center of Quantum Matter,
No.5 Yiheyuan Rd, Beijing 100871, P. R. China\\\vspace{1mm}

}
\end{center}

\vspace{8mm}

\begin{abstract}
We study the lensing effects of Kerr-Newman (KN) black holes and the multi-level images of luminous sources around the black holes. We obtain the analytic forms of the lens equations in KN spacetime and  derive approximately their limiting forms for photon emissions near the bounded photon orbits. After deriving the expressions of three key parameters $\gamma$, $\delta$ and $\tau$, we investigate the primary, the secondary and the higher-order images of spherical sources and equatorial disk-like sources. In particular, we find that the charge of KN black hole has a significant effect on the higher-order images.
\end{abstract}

\vfill{\footnotesize $^\ast$ Corresponding author: haopeng.yan@pku.edu.cn}

\maketitle

\newpage

\section{Introduction}\label{sec:introduction}
Recently, the Event Horizon Telescope (EHT) Collaboration has released the horizon-scale images of the supermassive black holes M87* \cite{EventHorizonTelescope:2019dse} and Sgr A* \cite{EventHorizonTelescope:2022xnr}.
These images have opened up a new window to study the strong gravity effects through gravitational lensing by a black hole (see \cite{Cunha:2018acu,Perlick:2021aok} for recent reviews).
As we know, the EHT image consists of a central dark area (``black hole shadow") and a surrounding bright ring (``photon ring"). Since the light rays in the vicinity of a black hole are highly bent \cite{Teo:2003spo,Beckwith:2004ae,Edery2006SecondOK}, the bright ring is actually composed of an infinite sequence of lensed photon emissions from a nearby accretion disk \cite{Luminet:1979nyg,Gralla:2019xty,Gralla:2019drh,
Himwich:2020msm,Johnson:2019ljv,Peng:2020wun,Bisnovatyi-Kogan:2022ujt,
Peng:2021osd,Zeng:2020dco, Guerrero:2021ues, Guo:2021wid, Li:2021ypw, Gan:2021xdl, Cunha:2019hzj, Zhang:2021hit}. While the observed profile depends on the less-understood plasma physics in the accretion disk as well, the light bending effect is mainly determined by the spacetime geometry. Thus, the characteristics of the multi-level lensed images may offer us a promising way to test the strong gravity \cite{Bronzwaer:2021lzo,
Gralla:2020srx,Lara:2021zth,Wielgus:2021peu,EventHorizonTelescope:2021dqv,Li:2021mzq,
Broderick:2021ohx,Younsi:2021dxe,Ozel:2021ayr,Kocherlakota:2022jnz,
Ayzenberg:2022twz,Bambi:2019tjh,Vagnozzi:2019apd}.

In General Relativity, the most general asymptotically flat and stationary black hole solution to the Einstein-Maxwell equations is the Kerr-Newman (KN) metric, which describes charged rotating black holes with three parameters: mass $M$, electric charge $Q$ and angular momentum $J=Ma$. The KN metric reduces to the Kerr metric for $Q=0$. The Kerr metric is the most relevance solution in astrophysical application, because astrophysical black holes are thought to be electrically neutral and rotating. The Kerr metric further reduces to the Schwarzschild metric for $a=0$. A Schwarzschild black hole is static and only has one parameter, thus it is the simplest model to study astrophysical problems for black holes.
Although we would not expect an astrophysical black hole to be considerably charged, it is still reasonable to assume a black hole is slightly charged, and it is thus interesting to study the effects of the charge on the observational feature of a black hole \cite{Adamo:2014baa,Bozzola:2020mjx,EventHorizonTelescope:2021dqv,Li:2021zct}.

The lensing effects by a Schwarzschild black hole have been revisited in \cite{Gralla:2019xty}, where the authors  have argued that the results in the Schwarzschild spacetime are qualitatively unchanged in the Kerr spacetime. In order to study the images of an astrophysical source near a black hole, it is convenient to introduce a fractional number $n$ for the photon orbits from the source to an observer. The photons having $n\leq\f{1}{2}$ produce a ``direct" primary image of the source, and  photons having $\f{1}{2}<n\leq1$ produce a diminished secondary image of the source surrounding the edge of the black hole shadow. We call this edge the critical curve since it corresponds to the critical photon emissions from the unstable bounded photon orbits.
The higher-order photons having $n>1$  contribute to exponentially diminished successive images which approach the critical curve rapidly. That is, the higher-order images produce the observed feature of photon ring and they actually corresponds to the near-critical photon emissions.
The demagnification between two successive images is determined by the Lyapunov exponent $\g$, which characterizes the instability of the bounded photon orbits \cite{Darwin,Bardeen:1972fi,Luminet:1979nyg,Teo:2003spo}.

Later in \cite{Gralla:2019drh}, the lensing by Kerr black holes has also been restudied. In addition to the Lyapunov exponent $\g$, two more parameters, $\d$ and $\tau$, are introduced to describe the successive images. These two parameters characterize the rotation and time delay of the critical photons over a polar half-libration, respectively. The primary image depicts the overall apparent shape of a given source and thus describes the source-dependent features. On the other hand, the secondary image starts to show the effects of spacetime, and the higher-order images are almost independent of the source and describe the universal feature of a black hole spacetime. The higher-order images can be characterized by the above three key parameters, $\g$, $\d$ and $\tau$, of the bounded photon orbits, which are functions of the black hole spin $a$. In this paper, we will study the multi-level lensed images around KN black holes, generalizing the discussions in the Kerr case, and especially study the influence of black hole charge $Q$ on the images.

The remaining part of this paper is organized as follows.
In Sec.~\ref{sec:nullgeodesics}, we review the KN null geodesics and introduce the KN lens equations by integrating the null geodesics along the photon trajectories from a source to a distant observer. We study the critical light rays and define three key parameters $\g$, $\d$ and $\tau$  in Sec.~\ref{sec:criticalrays}. We then derive the near-critical lens equations in Sec.~\ref{sec:nearcriticalrays}, which are expressed in terms of these key parameters. In Sec.~\ref{sec:ObserverScreen}, we introduce two kind of screen coordinates for the observers at distinct inclinations. In Sec.~\ref{sec:directimage}, we consider the optical appearances of various sources, focusing on the primary and secondary images. In Sec.~\ref{sec:photonring}, we analyze the property of the photon ring (higher-order images) by studying the near-critical lens equations, focusing on the effects of black hole spin and charge. In Sec.~\ref{sec:summary}, we summarize and conclude this work. In Appendix.~\ref{app:radialmotion}, we provide more details in computing the near-critical radial integrals.
We work in the geometric units such that $G=c=1$ throughout the paper and we set the black hole mass $M=1$ in Secs.~\ref{sec:ObserverScreen}, \ref{sec:directimage} and \ref{sec:photonring} for simplicity.

\section{Light bending near a Kerr-Newman black hole}\label{sec:nullgeodesics}
The KN metric in the Boyer-Lindquist coordinates is given by
\be
ds^2=-\f{\D}{\Sigma}(dt-a\sin^2{\t}d\phi)^2+\Sigma\l(\f{dr^2}{\D}+d\phi^2\r)+\f{\sin^2{\t}}{\Sigma}[adt-(r^2+a^2)d\phi]^2,
\ee
where
\bea
\label{delta}
\Sigma(r,\theta)=r^2+a^2\cos^2\theta,\qquad
\Delta(r)=r^2-2Mr+a^2+Q^2,
\eea
with $M,a,Q$ being the mass, spin and charge of the black hole, respectively. Here the spin parameter is defined by $a = J/M$, where $J$ is the angular momentum of the black hole. The coordinate singularities are located at
\bea
r_{\pm}=M\pm\sqrt{M^2-(a^2+Q^2)},
\eea
and $r_+$ is the outer event horizon. The allowed range of the charge parameter is
\be
a^2+Q^2\leq M^2.
\ee
The outer ergosphere locates at
\be
r_e=M + \sqrt{M^2-(a^2\cos^2\theta+Q^2)}.
\ee
Astronomical observations suggest that one should not expect an astrophysical black hole to have a large electric charge, so we may consider $Q$ in a range of $0\leq Q\leq0.5M$ \cite{Bozzola:2020mjx,EventHorizonTelescope:2021dqv}.
A photon trajectory can be described with its conserved quantities (i.e., the impact parameters), which are the energy-rescaled angular momentum $\lm$ and the energy-rescaled Carter integral $\eta$. With these two quantities, the four-momentum of a photon along its trajectory is given by
\bea
\label{pr}
&&\f{\Sigma}{E}p^r=\pm_r\sqrt{\M R(r)}, \\
\label{ptheta}
&&\f{\Sigma}{E}p^{\t}=\pm_{\t}\sqrt{\T(\t)}, \\
\label{pphi}
&&\f{\Sigma}{E}p^{\phi}=\f{a}{\D}(r^2+a^2-a\lm)+\f{\lm}{\sin^2{\t}}-a, \\
\label{pt}
&&\f{\Sigma}{E}p^{t}=\f{r^2+a^2}{\D}(r^2+a^2-a\lm)+a(\lm-a\sin^2{\t}).
\eea
where
\bea
\M R(r)&=&(r^2+a^2-a\lm)^2-\D[\eta+(\lm-a)^2],\\
\T(\t)&=&\eta+a^2\cos^2{\t}-\lm^2\cot^2{\t},
\eea
are the ``potentials'' in radial and angular directions, respectively, and $\pm_r$ and $\pm_\theta$ denote the signs of $p^r$ and $p^\theta$ that switch at the radial and angular turning points, respectively. Obviously the potentials should be non-negative along photon trajectories.

Now we consider a photon emitted from the source $(t_s,r_s,\t_s,\phi_s)$ and reaching a distant observer $(t_o,\infty,\t_o,0)$. Hereafter, the subscripts ``$s$'' and ``$o$'' denote the coordinates of source and observer, respectively.  The integral form of the photon motion equations (\ref{pr}---\ref{pt}) can be obtained by integrating along null geodesics,
\bea
\label{geodesic}
I_r=G_\theta= \M{T},\qquad  
\D\phi
=I_{\phi} + \lm G_{\phi}, \qquad 
\D t
= I_t + a^2 G_t,
\eea
where $\mathcal{T}$ is the so called the ``Mino time", a parameter to characterize the length of the geodesics, $\D\phi=\phi_o-\phi_s$, $\D t=t_o-t_s$ and
\be
\label{defintegrals}
I_i=\fint_{r_s}^{r_o}\frac{\M I_i}{\pm_r \sqrt{\M R(r)}}dr,\qquad
G_j=\fint_{\t_s}^{\t_o}\frac{g_j}{\pm_{\t}\sqrt{\T(\t)}}d\t,\qquad
i\in\{r,\phi,t\},\quad
j\in\{\theta,\phi,t\},
\ee
with
\bea
\label{defintegrals0}
&&\M I_r=1,\qquad
\M I_{\phi} = \frac{a(2Mr-a\lambda)}{ \Delta(r)},\qquad
\M I_t =\frac{r^2\Delta(r)+2Mr(r^2 + a^2 - a\lambda)}{\Delta(r)},\\
&&g_\theta=1,\qquad
g_\phi=\csc^2\t,\qquad
g_t=\cos^2\t.
\eea
The integral symbol ``$\fint$'' represents the integration along the path of a photon trajectory. Let $\om=0,1$ denotes the direct and reflected rays (i.e., for rays with and without radial turning point), respectively, and let $m\geq0$ denotes the numbers of angular turning points. Then the radial and angular integrals can be formally written as
\bea
\label{radialintegrals}
I_i&\sim&\fint_{r_s}^{r_o}dr\dots
\,\,\sim\,\,\int_{r_s}^{r_o}dr\dots+2\om\int_{r_t}^{r_s}dr\dots,\\
\label{angularintegrals}
G_i&\sim&\fint_{\theta_s}^{\theta_o}d\theta\dots
\,\,\sim\,\, 2m\int_{\theta_-}^{\theta_+}d\theta\dots
\pm_s\int_{r_s}^{\pi/2}d\theta\dots \mp_o\int_{\pi/2}^{\theta_o}d\theta\dots,
\eea
where $r_t$ is the largest real root of the radial potential\footnote{The explicit form of $r_t$ is the same as Eq. (A8d) in \cite{Gralla:2019drh} for the Kerr case, while the correction from the charge $Q$ enters the expression through the change of Eq. (A3) in \cite{Gralla:2019drh} by $\M C \rightarrow \M C=-a^2\eta-Q[\eta+(\lambda-a)^2]$.} $\M{R}$, and $\theta_\pm$ are the roots of the angular potential $\Theta$, which are given by
\be
\label{thetapm}
\t_\pm = \arccos{(\mp \sqrt{u_+})},\qquad
u_\pm =\D_\theta  \pm \sqrt{\D_\theta^2+ \frac{\eta}{a^2}},\qquad
\Delta_\theta=\frac{1}{2}\l(1-\frac{\eta + \lambda^2}{a^2}\r).
\ee
In terms of elliptic functions for the angular integrals $G_i$ [see Eq.~\eqref{angularintegrals}], the original geodesic equations \eqref{geodesic} 
can be rewritten as \cite{Gralla:2019drh}
\bea
&&\,\,\,I_r=\frac{1}{a\sqrt{-u_-}}[2mK \pm_s F_s \mp_o F_o ]
\label{G},\\
&&\Delta\phi=I_\phi+\frac{\lambda}{a\sqrt{-u_-}}[2m\Pi \pm_s \Pi_s \mp_o \Pi_o ],
\label{Gphi} \\
&&\Delta t=I_t-\frac{2a u_+}{\sqrt{-u_-}}[2mE' \pm_s E'_s \mp_o E'_o ],
\label{Gt}
\eea
where $\pm_{s/o}$ denote the sign of $p^\theta$ at the source and at the observer, respectively, and
\bea
&&F_{s/o}=F\l(\Psi_{s/o};\frac{u_+}{u_-}\r),\qquad\quad\,\,\,\,
K=K\l(\frac{u_+}{u_-}\r)=F\l(\frac{\pi}{2};\frac{u_+}{u_-}\r),\\
&&\Pi_{s/o}=\Pi\l(u_+,\Psi_{s/o};\frac{u_+}{u_-}\r),\qquad
\Pi=\Pi\l(u_+;\frac{u_+}{u_-}\r)=\Pi\l(u_+,\frac{\pi}{2};\frac{u_+}{u_-}\r)\\
&&E^\prime_{s/o}=E^\prime_{s/o}\l(\Psi_{s/o};\frac{u_+}{u_-}\r),\qquad\,\,\,
E^\prime=E^\prime\l(\frac{u_+}{u_-}\r)=E^\prime\l(\frac{\pi}{2};\frac{u_+}{u_-}\r),
\eea
with $\Psi=\arcsin\l(\frac{\cos\theta}{\sqrt{u_+}}\r)$ and $E'=\pl_n E(\theta;n)$, and $F_{s/o}$, $\Pi_{s/o}$, $E_{s/o}$ being the incomplete elliptic integrals of the first, second and third kinds, respectively, while $K$, $\Pi$, $E$ being the complete elliptic integral of the first, second and third kinds, respectively.
Note that, until now the differences between the geodesics equations for the KN and the Kerr \cite{Gralla:2019drh} cases are only implicitly contained in the function $\D(r)$.

Photons may orbiting around a black hole for several orbits along their pathes, so we introduce the so-called fractional number of orbits \cite{Gralla:2019drh},
\be
\label{defn}
n=\frac{G_\theta}{2\int^{\theta_+}_{\theta_-}d\theta\l[\Theta(\theta)\r]^{-1/2}}
=\frac{a\sqrt{-u_-}}{4K}I_r.
\ee
For later reference, we derive the relation between the fractional number $n$ and the number of angular turning points $m$ by using Eqs.~\eqref{G} and \eqref{defn}, which is give by
\bea
\label{nm}
n = \f{1}{4K}[2mK\pm_s F_s \mp_o F_o]=\frac{1}{2}m\pm_o\frac{1}{4}\l[(-1)^m\frac{F_s}{K}-\f{F_o}{K}\r].
\eea

We call the equations \eqref{G}, \eqref{Gphi} and \eqref{Gt} the KN lens equations, which are the main equations that we will deal with throughout the rest of the paper.

\subsection{Critical rays and key parameters}\label{sec:criticalrays}
Critical rays are defined by the extremely bent bounded photon orbits, which correspond to double roots of the radial potential, $\M R(\td r) = \M R^{\prime}(\td r) = 0$. If $\td r>r_+$, there are the solutions \cite{Gralla:2019ceu,Gralla:2019drh} 
\bea
\label{criticalL}
\td{\lm}(\td r) &=& 
a+\frac{\td r}{a}\left[\td r-\frac{2\td \Delta}{\td r-M}\right]
\ ,\\
\label{criticalEt}
\td{\eta}(\td r) &=&
\frac{\td r^2}{a^2}\left[\frac{4\td\Delta(M \td r-Q^2)}{(\td r-M)^2}-\td r^2\right]
\ .
\eea
These orbits are parameterized by the location of the double root $\td{r}$. Hereafter, the quantities with tildes are for those corresponding to the critical photon emissions.
It follows from the nonnegativity of the angular potential ($\eta\geq0$) that $\td r$ is in the range of $\td{r}_- \leq \td{r} \leq \td{r}_+$ with $\td r_\pm$ being the solutions for $\eta=0$. These spherical photon orbits $[\td\lm(\td r),\td\eta(\td r)]$ form a region called the photon shell.
The pole-crossing rays lie at the intermediate radius $\td r=\td r_0$ for which $\td \lm=0$ and $[\t_-,\t_+]=[0,\pi]$.

Now we consider the photons moving along a precisely critical ray with $(\td\lambda,\td \eta)$ at the radii $r_i=\td r(1+\d r_i)$ with $\ab{\d r_i}\ll1$, executing $n_1$ and $n_2$ fractional orbits, respectively. For the radial motion, we have\footnote{See Appendix \ref{app:expansions} for the case of $\d\lambda=\d q=0$, or see Eq.~\eqref{nearintegrals} for $\d r_0=0$. When expanding in $\d r_i$, we note that a factor $\f{Q}{\td\D}$ appears at the leading order in the $I_\phi$ equation, which is absent in the Kerr results \cite{Gralla:2019drh}.}
\be
\label{criticalIr}
I_r=\int_{r_1}^{r_2}\frac{dr}{\sqrt{\M R (r)}}\approx
\frac{1}{2\td{r}\sqrt{\td\chi}}\log\l(\frac{\d r_2}{\d r_1}\r).
\ee
\be
\label{criticalIphit}
I_\phi\approx a\l(\f{\td{r}+M}{\td{r}-M}+\f{Q}{\td{\D}}\r)I_r, \qquad
I_t\approx\td{r}^2\l(\f{\td{r}+3M}{\td{r}-M}\r)I_r.
\ee
For the angular motion, we have $G_i=\td G_i(\td\lambda,\td \eta)$.
Then by applying Eqs.~\eqref{G} and \eqref{defn} we can see that the radial integral has a logarithmic divergence along the path between $r_1$ and $r_2$,
\be
\frac{r_2-\td r}{r_1-\td r}\approx \exp[2\g (n_2-n_1)],
\ee
where the Lyaponent exponent $\g$ is defined by \cite{Gralla:2019drh}
\be
\label{gamma}
\g
=\frac{4\td r\sqrt{\td \chi}}{a\sqrt{-\td u_-}}\td K,\qquad
\td \chi=1-\frac{M \td \Delta}{\td r(\td r-M)^2}.
\ee
By applying the Eqs.~\eqref{Gphi} and \eqref{Gt} respectively
for a complete orbit with $\theta_s=\theta_o$ and ${m}=2$, we can define the change in azimuthal angle over a half polar libration, $\d$, and the change in period over a half polar libration, $\tau$, respectively, which are given by\footnote{We have introduced a Heaviside function $H(\td r-\td r_0)$, since $\D\phi$ is discontinuous at the pole-crossing orbit ($r=\td r_0$) for which we have \cite{Gralla:2019drh}
\be
\label{polecrossing}
\lim\limits_{\lambda\rightarrow0^\pm} \f{2\lambda\Pi}{a\sqrt{-u_-}}=\pm\pi.
\ee}
\be
\d\equiv\frac{1}{2}\D \phi|_{\theta_s=\theta_o,m=2}+2\pi H(\td r-\td r_0),\qquad
\tau\equiv\frac{1}{2}\D t|_{\theta_s=\theta_o,m=2}.
\ee
Writing in terms of elliptical functions, we obtain
\bea
\label{deltaexp}
&&\d=\f{2}{a\sqrt{-\td{u}_-}} \l[ a\l(\f{\td{r}+M}{\td{r}-M}+\f{Q}{\td{\D}}\r)\td{K}+\td{\lm} \td{\Pi} \r]
+2\pi H(\td r-\td r_0),       \\
\label{tauexp}
&&\tau= \f{2}{a\sqrt{-\td{u}_-}} \l[ \td{r}^2\l( \f{\td{r}+3M}{\td{r}-M}\r)\td{K} - 2\td{u}_+a^2\td{E}'  \r].
\eea

The three quantities $\g$, $\d$ and $\tau$ are the key parameters for critical photon emissions. The influence of charge $Q$ on these parameters can be seen from the expression of $\td\D$ [Eq.~\eqref{delta}] and from the elliptical integrals implicitly through $\td \lambda(\td r)$ and $\td q(\td r)$ [see Eqs.~\eqref{criticalL} and \eqref{criticalEt}]. These are the main analytical results for this paper.
We will discuss these in more detail in Sec.~\ref{sec:photonring} to see how these parameters control the behavior of higher-order images.

\subsection{Near-critical lens equations}\label{sec:nearcriticalrays}
Now we consider near-critical rays whose impact parameters can be written as
\be
\lambda=\td\lambda(1+\d \lambda),\qquad
q=\sqrt{\eta}=\td q(1+\d q),
\ee
with $\ab{\d\lambda}\sim\ab{\d q}\sim\epsilon \ll1$.
We will discuss the lens equations \eqref{G}, \eqref{Gphi} and \eqref{Gt} in this near-critical limit.
\begin{figure}[H]
\centering
\includegraphics[scale=0.4]{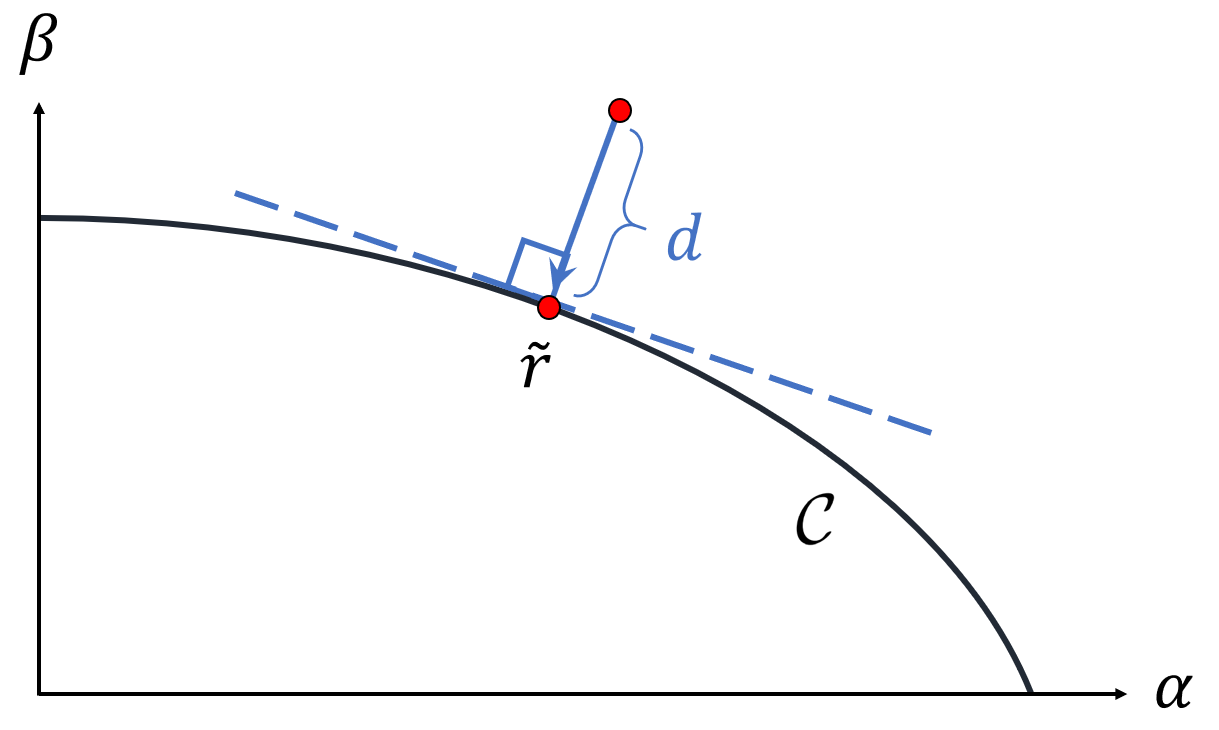}
\caption{Perpendicular distance $d$ of a nearby point (around $\td r$) from the critical curve $\M C$ [Eq.~\eqref{criticalcurve}] on a distant observer's screen [Eq.~\eqref{offcoordinates}].}
\label{d}
\end{figure}

By expanding the angular integrals $G_i$ in $\e$, we have $G_i\approx\td G_i(\td \lambda,\td q)$. On the other hand, we may compute the leading order radial integrals approximately (see Appendix \ref{app:radialmotion}).
It is convenient to introduce a dimensionless quantity $\d r_0$ to describe the near-criticality [see Eq.~\eqref{dr0}],
\be
\d r_0^2=\frac{\td\D}{2\td r^2\td \chi}\left[\frac{\td q^2}{\td r^2}\d q-\frac{\td \lambda}{a}\left(\frac{\td r-3M}{\td r-M}\right)\d\lambda\right].\nn
\ee
This quantity can be connected with the perpendicular distance $d$ from the critical curve for each choice of a $\td r$ through (see Fig.~\ref{d} and Sec.~\ref{sec:ObserverScreen} for details) \cite{Gralla:2019drh}
\bea
\label{distance}
d=\f{\d r_0^2}{ \sqrt{(\partial_{\a}\d r_0^2)^2+(\partial_{\b}\d r_0^2)^2}}
=\d r_0^2 \l[\frac{\td{\D}}{2\td{r}^4\td\chi}\sqrt{\td{\b}^2+\l[ \td{\a}-\l( \f{\td{r}+M}{\td{r}-M}\r)a\sin{\t_o} \r]^2}\r]^{-1}.
\eea
If $\d r_0=0$, then it is for the precisely critical rays and the radial integrals are given in Eqs.~\eqref{criticalIr} and \eqref{criticalIphit}. If $\d r_0\neq0$, the radial integrals becomes (see Appendix \ref{app:radialmotion}) \cite{Gralla:2019drh}
\bea
\label{Ir}
&&I_r\approx-\frac{1}{2\td r\sqrt{\td\chi}}\log[C_\pm(r_s,\td r)d],\\
\label{Iphi}
&&I_\phi\approx a\l(\f{\td{r}+M}{\td{r}-M}+\f{Q}{\td{\D}}\r)I_r+D_\pm(r_s,\td r), \\
\label{It}
&&I_t\approx\td{r}^2\l(\f{\td{r}+3M}{\td{r}-M}\r)I_r+H_\pm(r_s,\td r),
\eea
where $C_\pm(r_s,\td r)$, $D_\pm(r_s,\td r)$ and $H_\pm(r_s,\td r)$ are independent of $d$ and can be obtained from the results in Appendix \ref{app:radialmotion}. We also give an example for the expressions of $C_\pm$, $D_\pm$ and $H_\pm$ in Appendix \ref{app:results}.

By using the relations \eqref{defn} and \eqref{nm}, together with \eqref{Ir}, \eqref{Iphi} and \eqref{It}, and expressing $\td G_i$ in terms of elliptical integrals, the lens equations \eqref{G}, \eqref{Gphi} and \eqref{Gt} can be expanded as
\be
\label{ncequations}
d\approx\frac{1}{C_\pm}e^{-2 n\g},\qquad
\D \phi\approx 2n[\d-2\pi H(\td r-\td r_0)]-J^\phi_\pm, \qquad
\D t\approx 2n\tau-J^t_\pm
\ee
where 
\bea
&&J^{\phi}_\pm = \pm_o \f{\td{\lm} \td \Pi}{a\sqrt{-\td{u}_-}}
\l[(-1)^m [\f{\td{F}_s}{\td K}-\f{\td{\Pi}_s}{\td \Pi}]-[\f{\td{F}_o}{\td K}-\f{\td{\Pi}_o}{\td \Pi}]\r]-D_\pm(r_s,\td r),\\
&&J^{t}_\pm = \mp_o \f{2a\td{u}_+}{\sqrt{-\td{u}_-}}\td E'
\l[(-1)^m [\f{\td{F}_s}{\td K}-\f{\td{E}'_s}{\td E'}]-[\f{\td{F}_o}{\td K}-\f{\td{E}'_o}{\td E'}]\r]-H_\pm(r_s,\td r).
\eea
These near-critical lens equations \eqref{ncequations} 
determine the higher-order images which will be discussed in detail in Sec.~\ref{sec:photonring}.

\section{Observer's screen and critical curve}\label{sec:ObserverScreen}
We now introduce screen coordinates to describe the image positions of light sources relative to a distant static observer.
The screen coordinates can be defined by using the reception angles of photons reaching the observer. Regarding to the observed inclination $\theta_o$ relative to the spin axis of the black hole, we will consider the off-axis and on-axis cases separately.

\begin{figure}[h]
\centering
\includegraphics[scale=0.31]{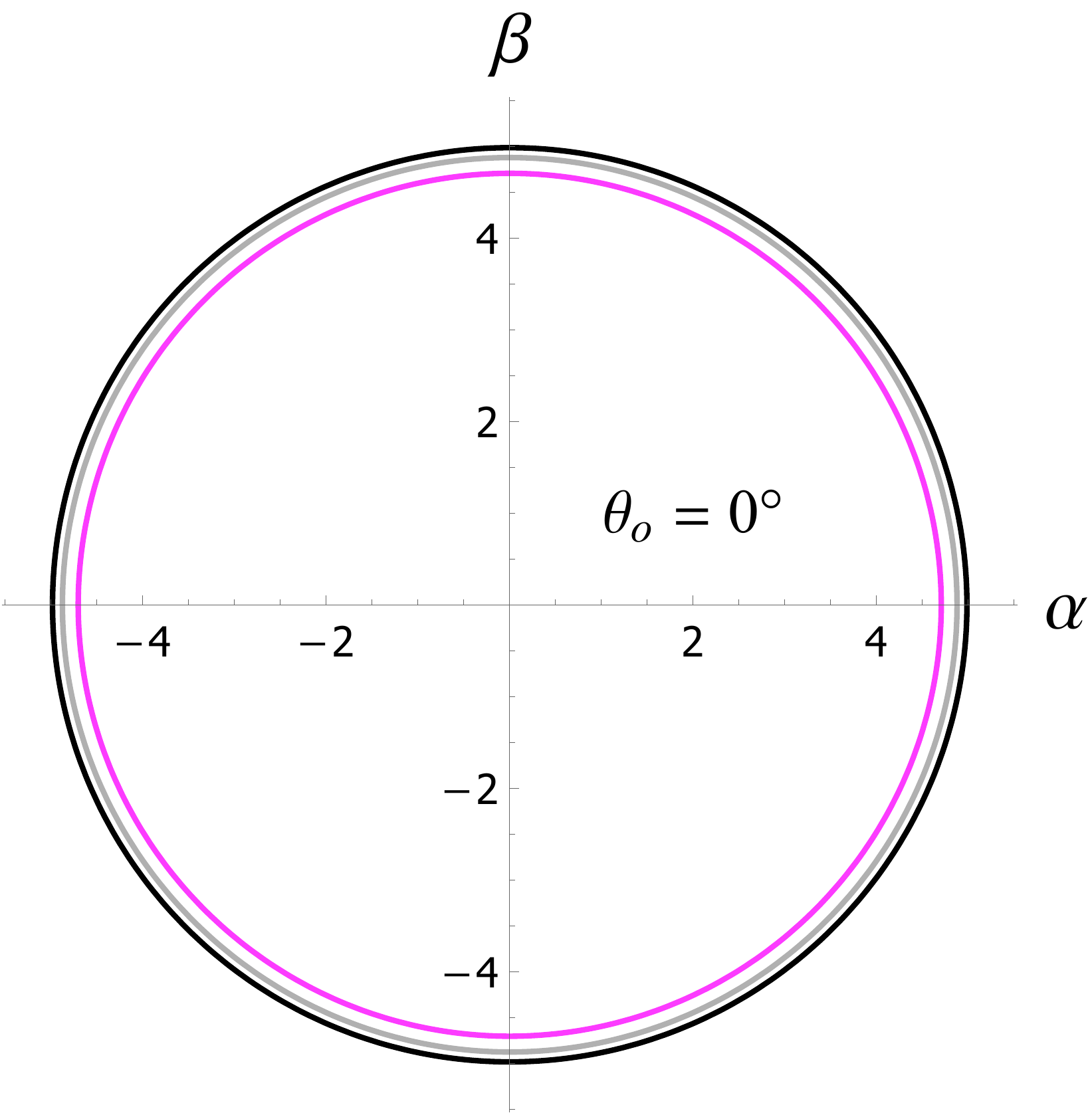}  \,
\includegraphics[scale=0.31]{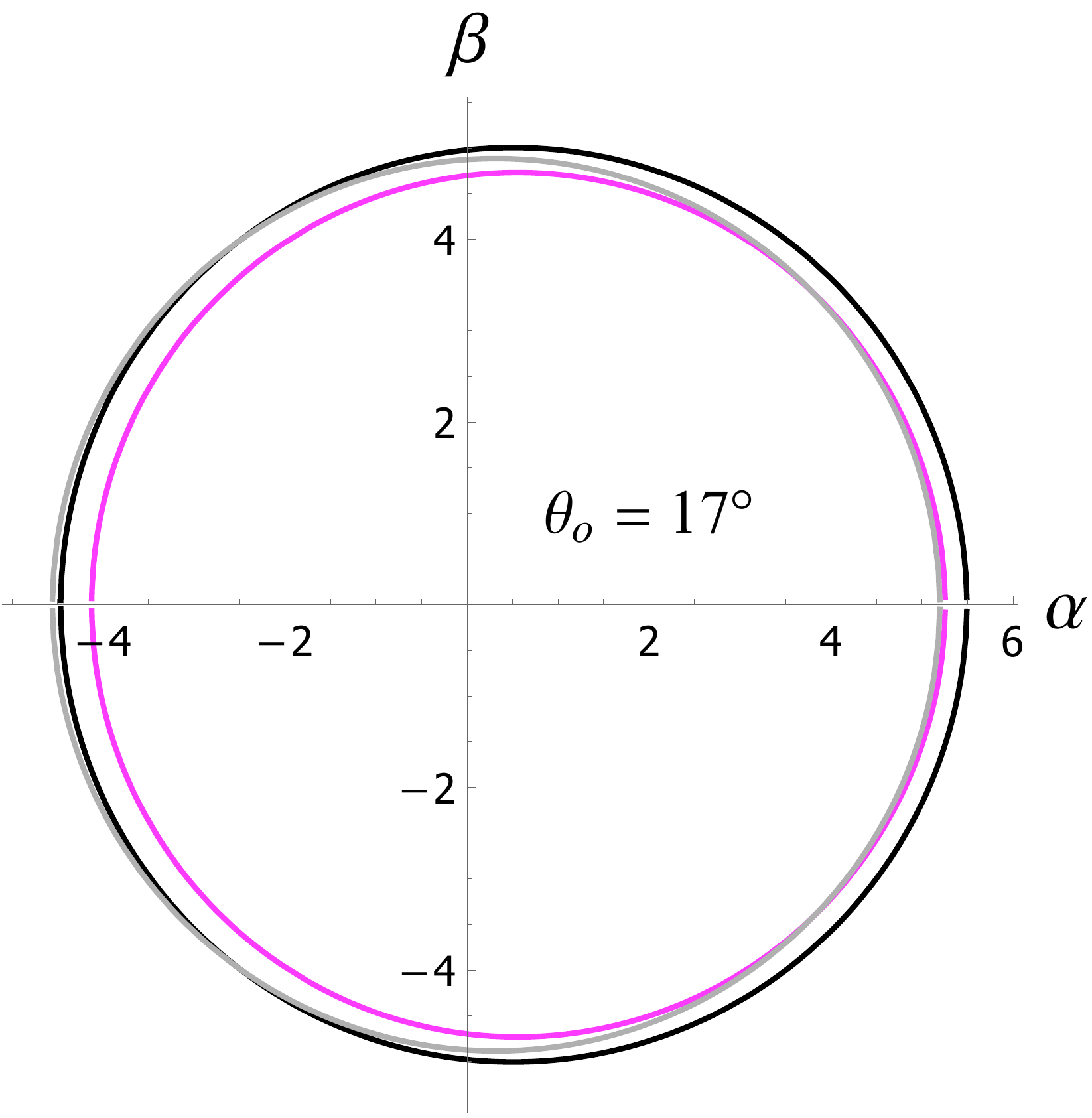} \,
\includegraphics[scale=0.31]{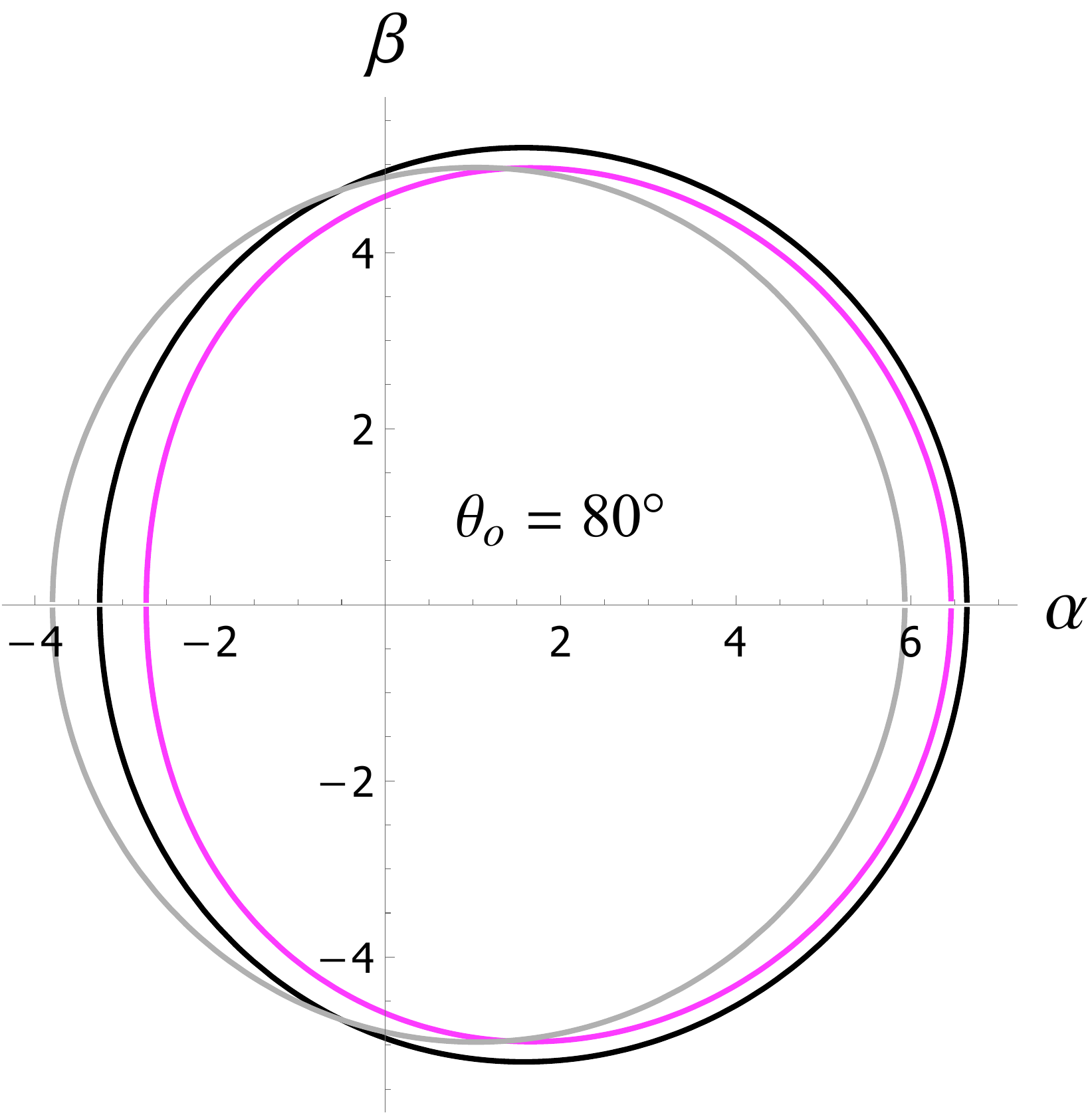} \\
\includegraphics[scale=0.32]{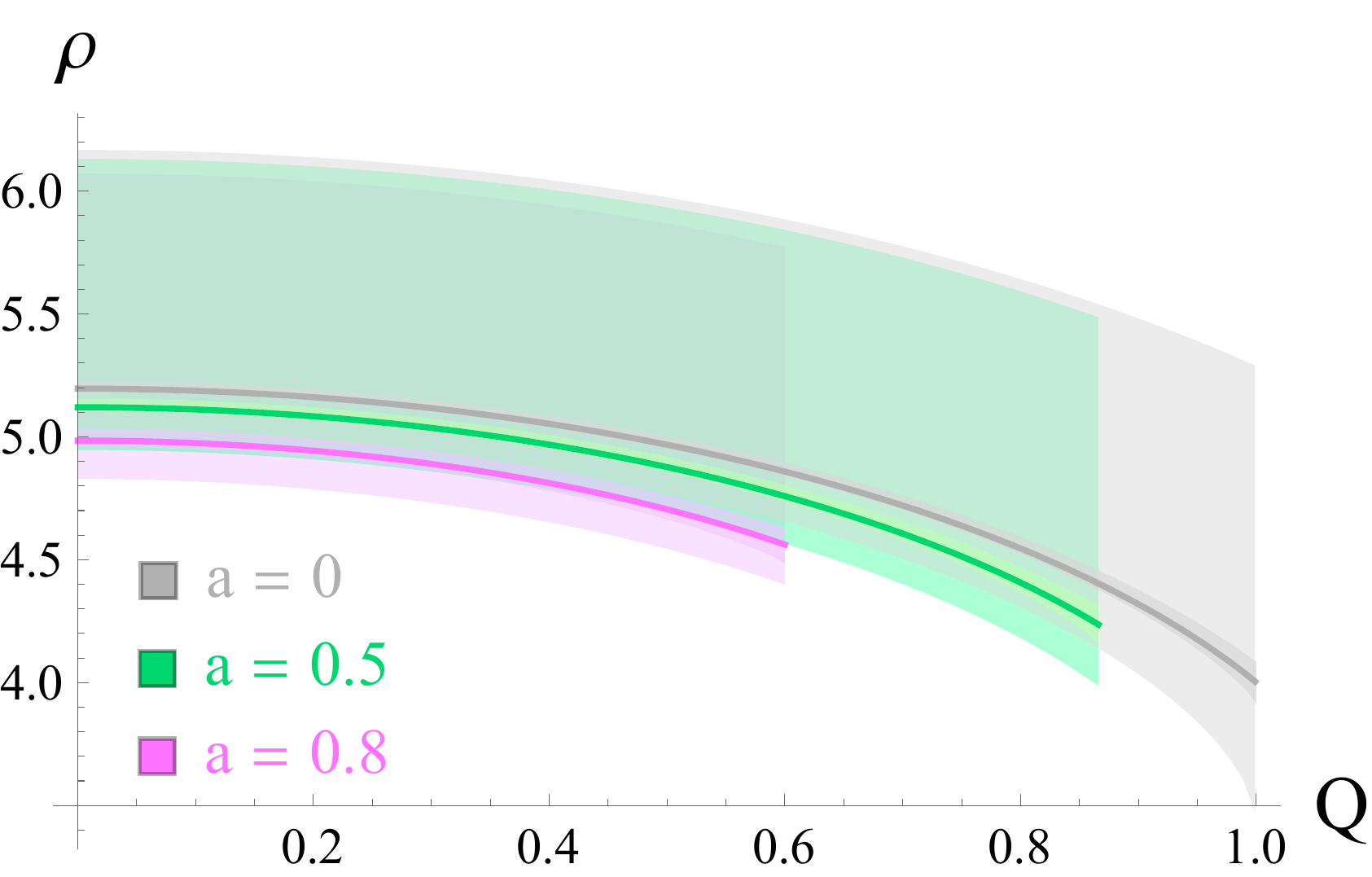} \,
\includegraphics[scale=0.31]{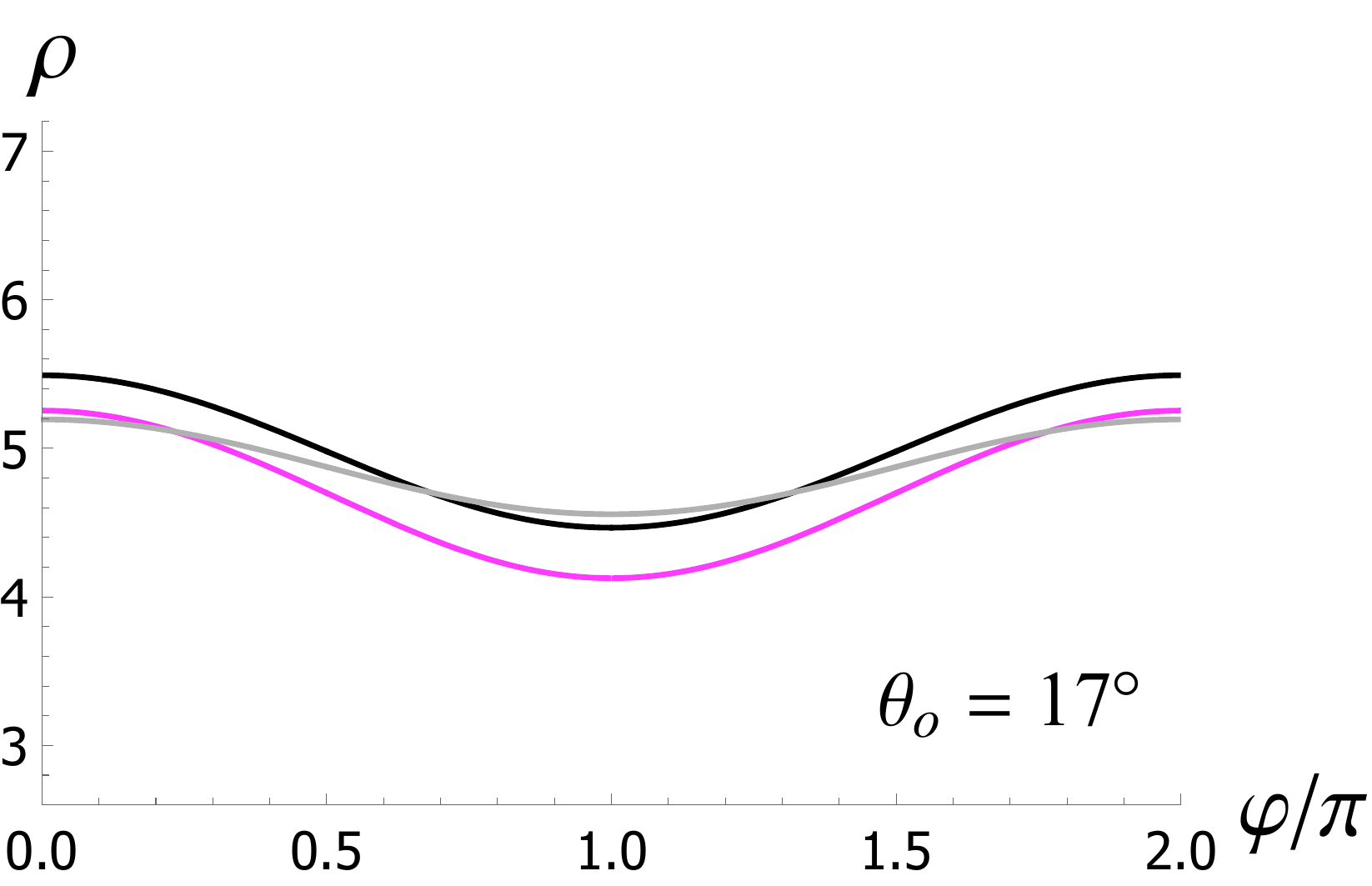} \,
\includegraphics[scale=0.31]{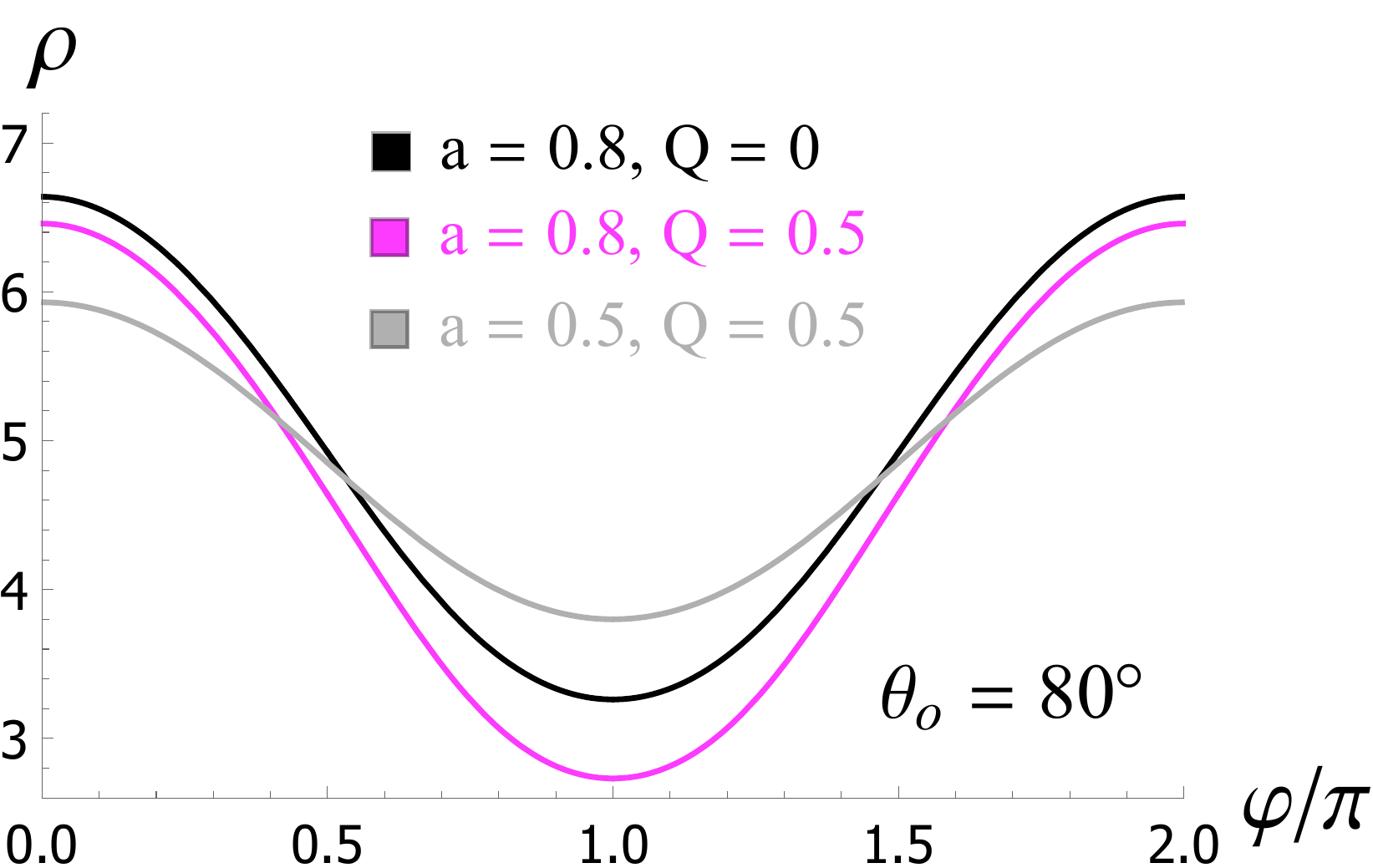} \\
\caption{Critical curves $\M C$ for the selected black hole parameters. Left column: above we show $\M C$ in the Cartesian coordinates \eqref{onCarcoord}, and below we show the dependence of $\td\rho$ (central lines in the colored bands), $\D\rho^{(1)}$ (lighter colored bands) and $\D\rho^{(2)}$ (darker colored bands) on black hole spin $a$ and charge $Q$ in the polar coordinates \eqref{oncoordinates}. Here $\D\rho^{(m)}$ [Eq.~\eqref{defDrho}] represents the width of the $m$-th image ring for an equatorial source ring with $r_s\in(r_+,\infty)$. Middle and right columns: we show $\M C$ in the Cartesian coordinates \eqref{offcoordinates} (above) and polar coordinates \eqref{offpolar} (below).}
\label{fig:criticalcurves}
\end{figure}

First we consider an off-axis observer at $r_o\rightarrow\infty$ and $0<\t_o<\pi/2$. For such an observer, we can set $\phi_o=0$ due to the axisymmetry of the spacetime. It is convenient to use the well-established Cartesian coordinates $(\a,\b)$ to cover its screen, which are expressed in terms of the impact parameters $(\lm,\eta)$ as \cite{Bardeen:1973tla}
\be
\label{offcoordinates}
\a=-\f{\lm}{\sin{\t_o}} \ , \qquad \b=\pm_o\sqrt{\eta+a^2\cos^2{\t_o}-\lm^2\cot^2{\t_o}} \ .
\ee
Note that we have $\pm_o=\si(\b)$ in this case. We may also transfer $(\a,\b)$ to a pair of polar coordinates $(\rho,\vp)$ through
\be
\label{offpolar}
\rho=\sqrt{\a^2 + \b^2} =\sqrt{\eta+a^2\cos^2\theta_o+\lm^2},\qquad
\tan\vp=\f{\b}{\a}.
\ee
Photons arrivaling with critical impact parameters $[\td\lambda(\td r),\td\eta(\td r)]$ [see Eqs.~\eqref{criticalL} and \eqref{criticalEt}] provide a critical curve on the screen
\be
\label{criticalcurve}
\M C=\l\{ [\td \a(\td r),\td\b(\td r)] \,\Big|\,\td r_-<\td r<\td r_+  \r\},
\ee
which is parameterized by $\td r$. Therefore, a point on the critical curve $\M C$ can also be depicted with $[\td r, \si(\td\b)]$. We show several examples for the critical curve observed off the pole on the middle and right columns of Fig.~\ref{fig:criticalcurves}. In the following discussions,  we select two inclination angles for the off-pole observers. One is at $\t_o=17^\circ$, the same as in the M87* observations in EHT, and the other is at $\t_o=80^\circ$, nearly edge-on scenario which is quite different from the M87* case.

Next we consider an on-axis observer at $r_o\rightarrow\infty$ and $\t_o=0$.
The photons reaching such an observer would always have zero azimuthal angular momentum and negative polar momentum, that is $\lambda=0$ and $\pm_o=\si(p^\theta_o)=-1$.
Then there is only $\eta$ left for a free parameter to use, and
both of the choices \eqref{offcoordinates} and \eqref{offpolar} are no longer good in this case.
Nevertheless, since the spacetime is axisymmetric relative to the on-axis observer, the angle of arrival photons $\phi_o$ can also serve as an independent parameter to describe the image. Therefore, we may introduce another pair of polar coordinates $(\rho, \varphi)$ to cover its screen, which varied from \eqref{offpolar} to \cite{Gralla:2019drh}
\bea
\label{oncoordinates}
\rho 
=\sqrt{\eta+a^2},\qquad
\varphi=\phi_o.
\eea
We may also define the Cartesian coordinates
\be
\label{onCarcoord}
\a=\rho\cos\vp,\qquad
\b=\rho\sin\vp,
\ee
for an on-axis observer.
When observed form the pole, the critical curve $\M C$ is a circle at radius $\td \rho=\sqrt{\td\eta+a^2}$ and this curve is instead parameterized by the angle $\vp$. In this case, the perpendicular distance from the critical curve to a point $(\rho,\vp)$ is then given by $d=\rho-\td\rho$. We show several examples for the critical curve observed on the pole on the left column of Fig.~\ref{fig:criticalcurves}.

\section{Observational appearance}\label{sec:directimage}
In this section, we consider the optical appearances of a spherical source and a thin equatorial disk. We assume axissymmetric for the sources, then the shape of each source can be described by a given function of $r_s$ and $\t_s$. Note that we should consider both the direct ($\om=0)$ and reflected ($\om=1)$ emissions for radial integrals \eqref{radialintegrals} and consider $m\geq0$ for angular integrals \eqref{angularintegrals}.
Thus we have a series of imaging maps between each source and an observer's screen, which gives multi-level images: the first (primary), second (secondary) and third images are respectively for $m=0$, $1$, $2$ when viewed form the pole (or respectively for $\bar{m}=0$, $1$, $2$ in Eq.~\eqref{mbar} when viewed from inclined observers), and the higher-order images are for $m\geq2$ or $\bar{m}\geq2$.
Here we focus on the primary images and a few lower-order images and
discuss the overall feature of these images by applying the lens equations \eqref{G}, \eqref{Gphi} and \eqref{Gt}, while the higher-order images will be discussed in detail in the next section by using the near-critical lens equations \eqref{ncequations}.

\subsection{Spherical source viewed by an on-axis observer}\label{Spherical source}
\begin{figure}[t]
\centering
\includegraphics[scale=0.4]{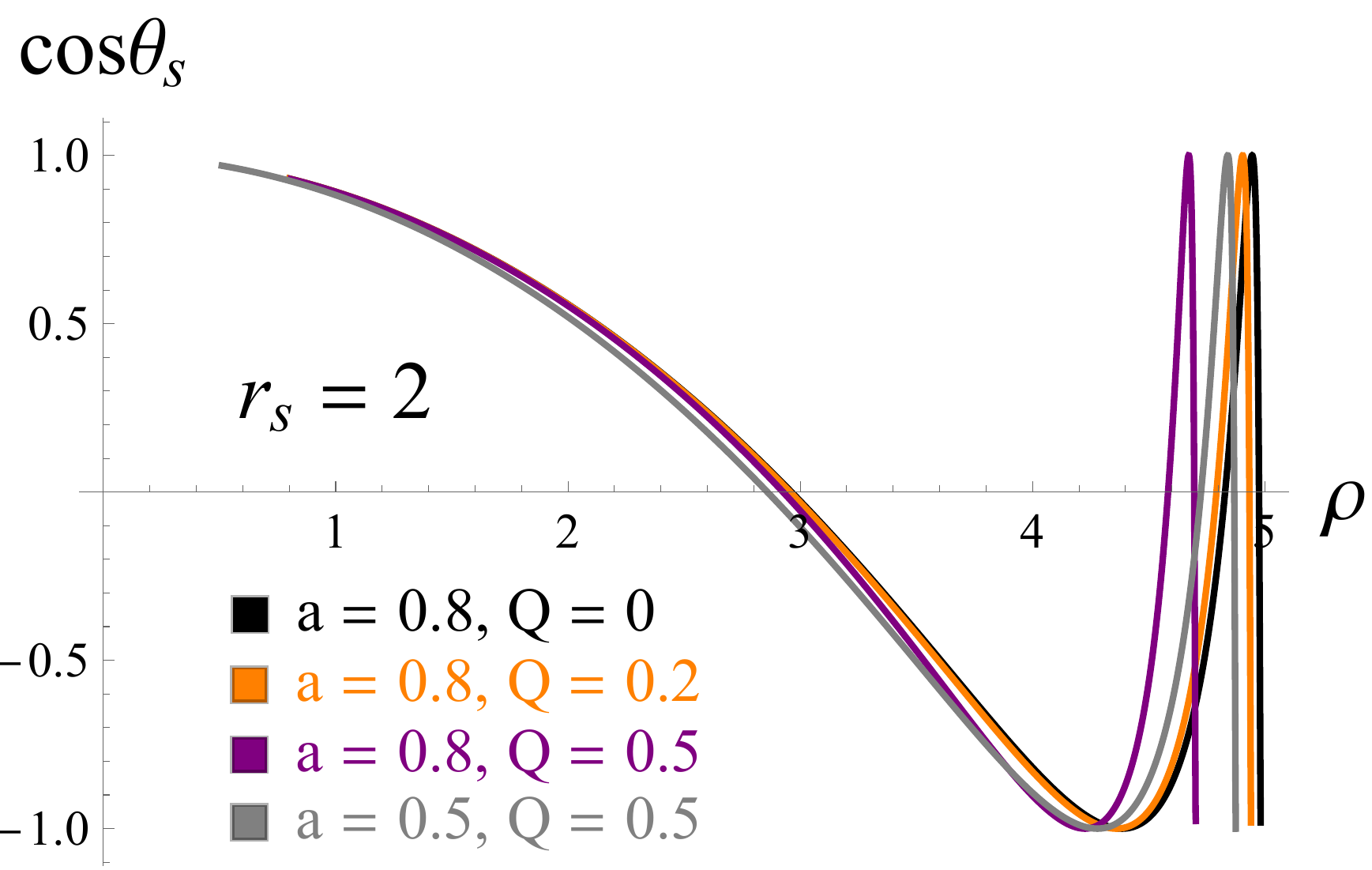}\,\,\,\,\,\quad
\includegraphics[scale=0.4]{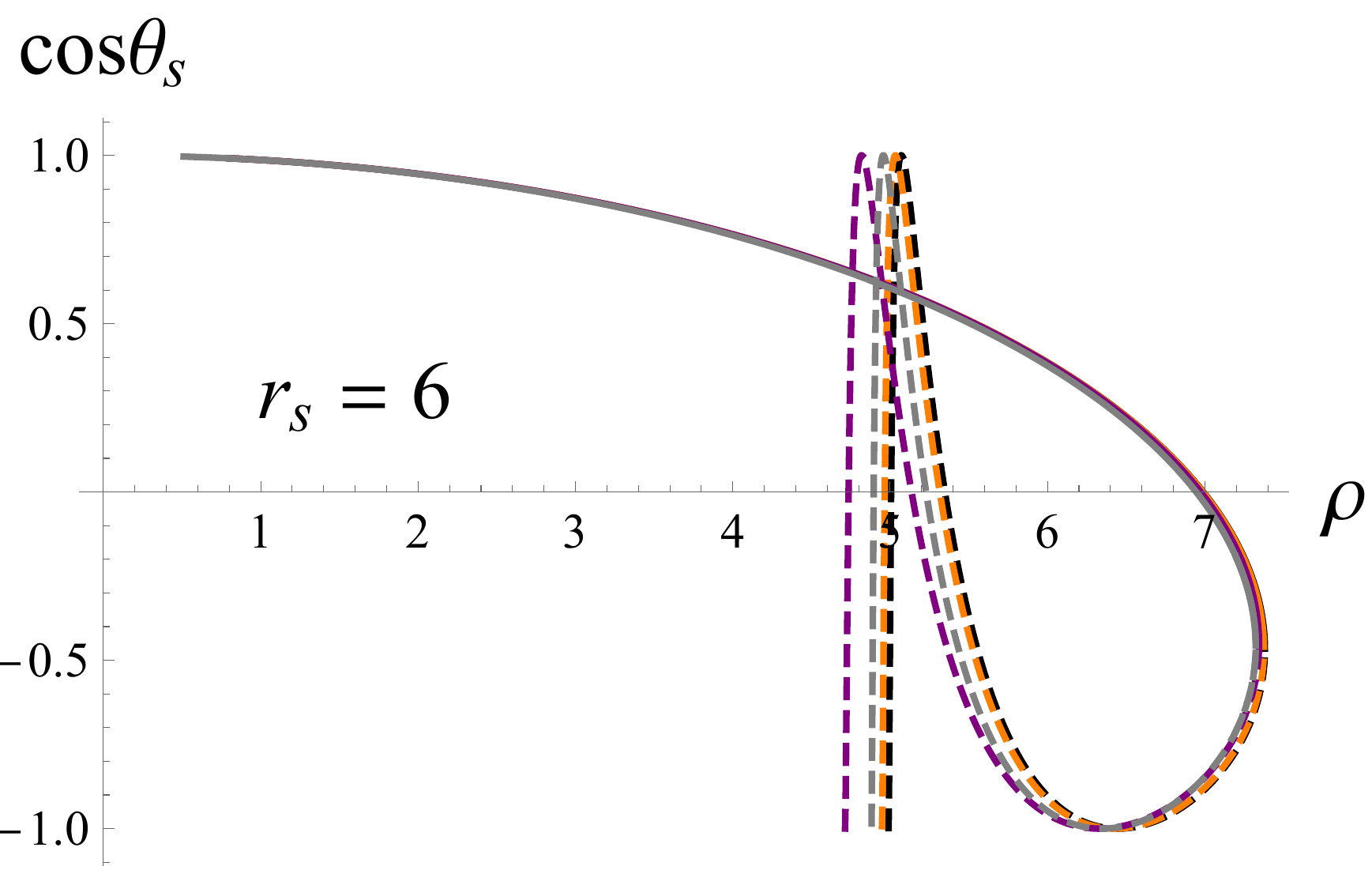}\\~\\
\quad\begin{overpic}[scale=0.4]{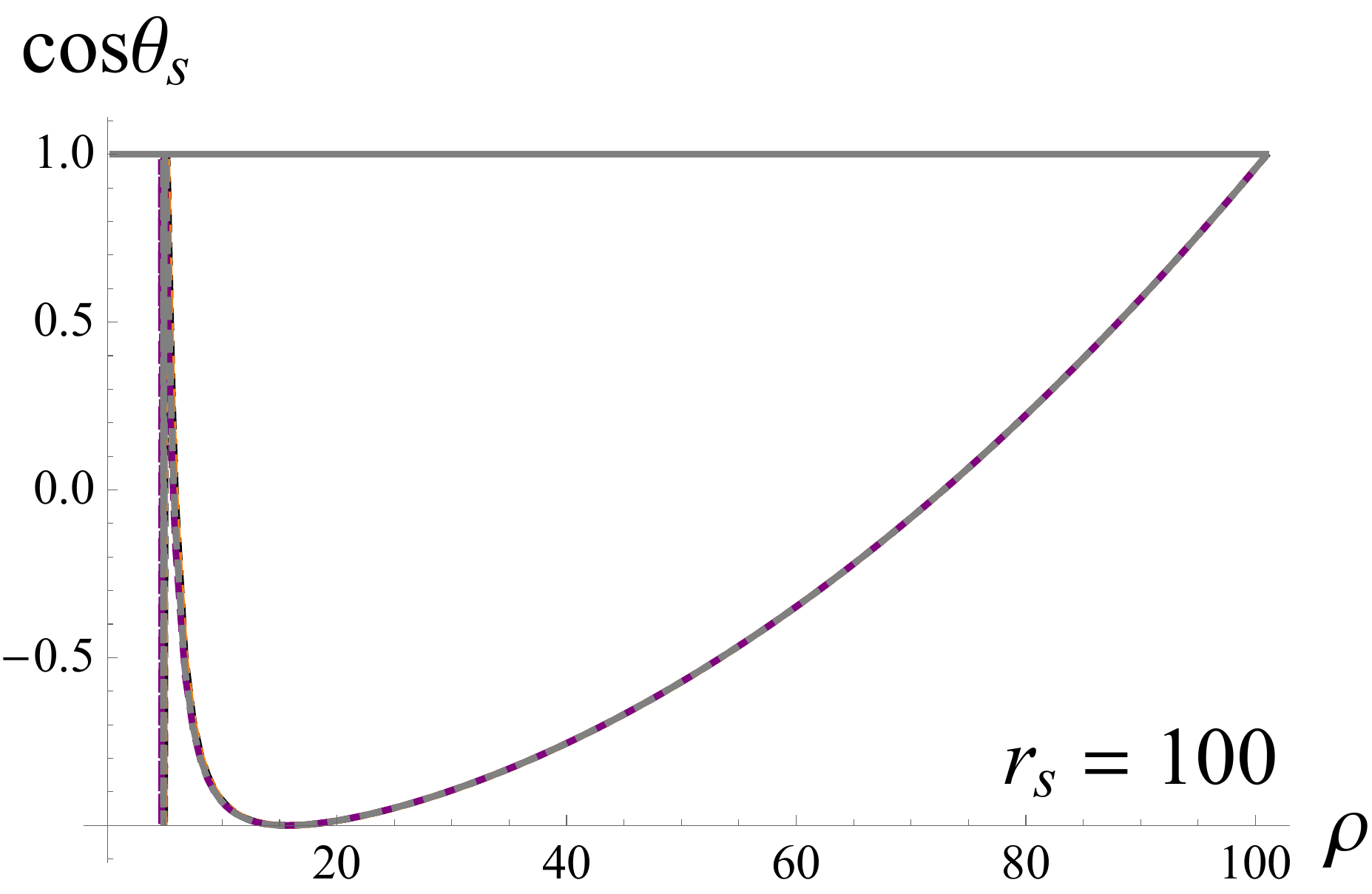}
	\put(15,22){\includegraphics[scale=0.3]%
		{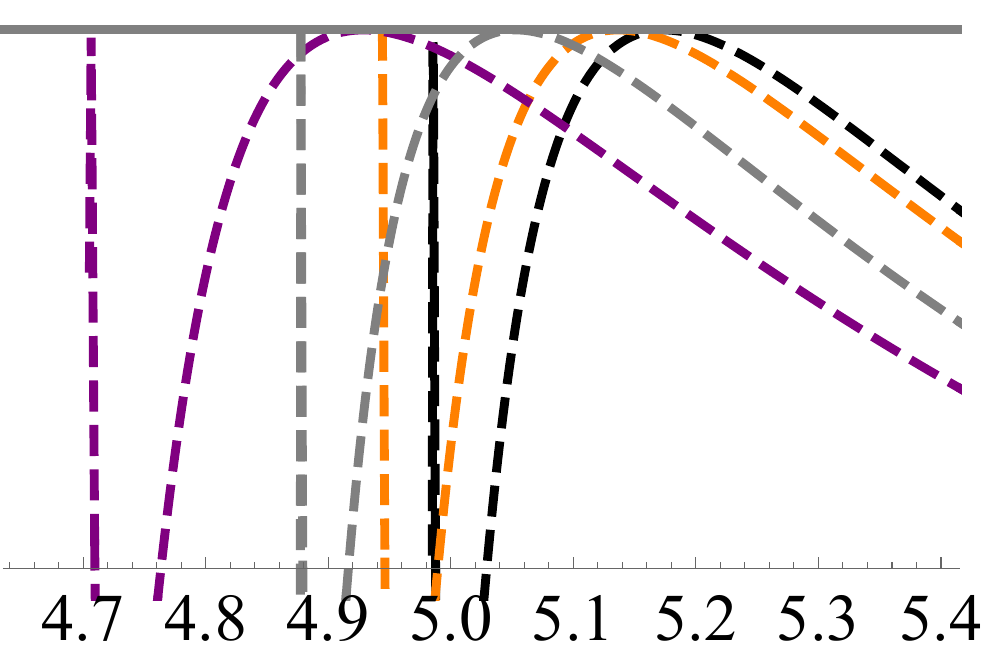}}
\end{overpic}\,\,\,\,\,
\begin{overpic}[scale=0.4]{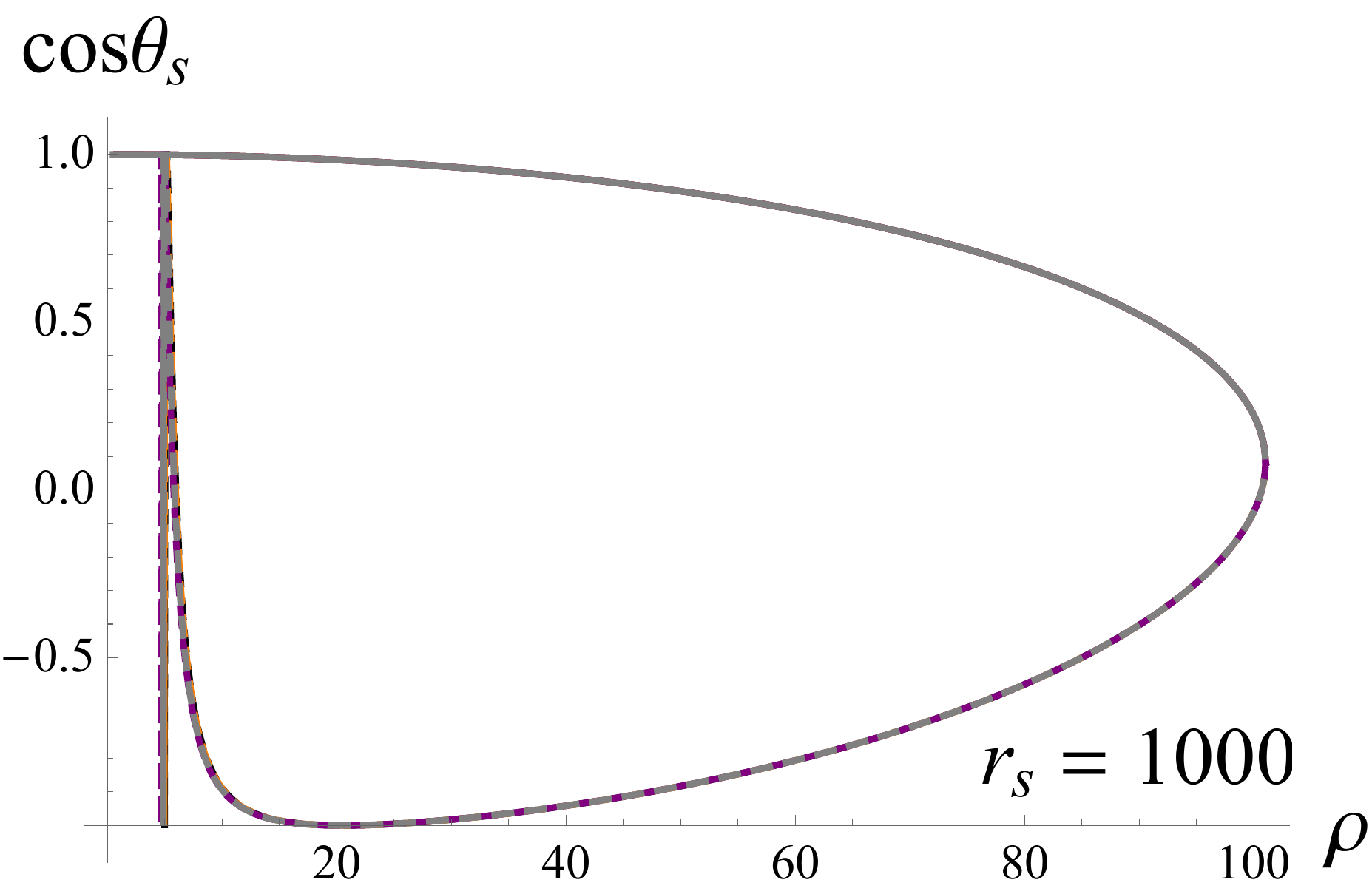}
	\put(15,22){\includegraphics[scale=0.3]%
		{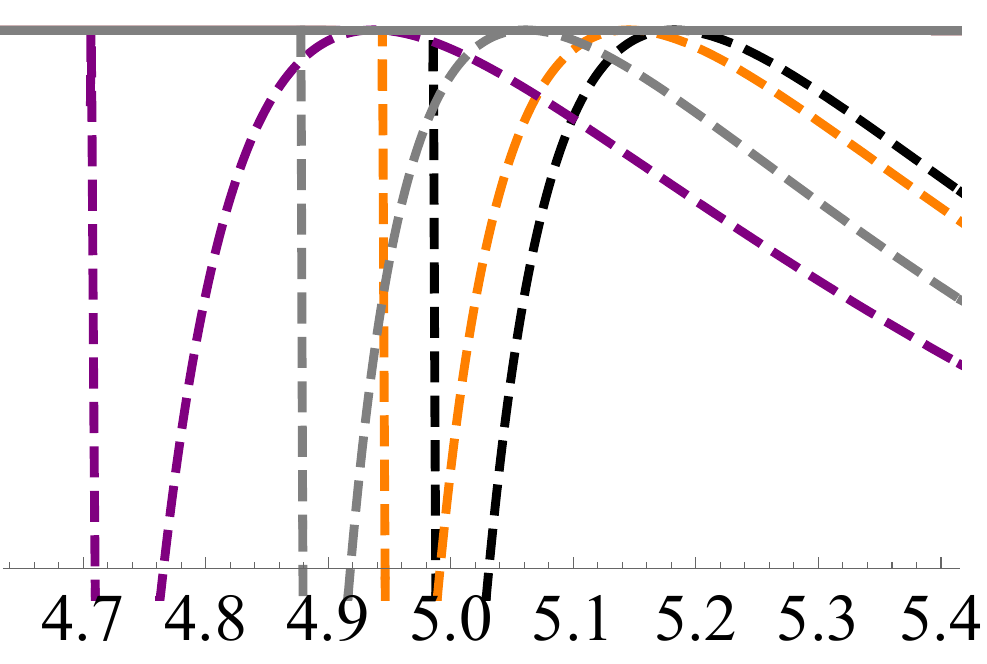}}
\end{overpic}
\caption{Transfer functions $\cos\theta_s(\rho)$ for spherical sources with radii $r_s=2,\,6,\,100$ and $1000$ viewed from an on-axis observer at $r_o=100$. Solid and dashed lines have $\omega=0$ and $1$, respectively.}
\label{sphere}
\end{figure}

We now discuss the image positions of a spherical source viewed by an on-axis observer from above ($\t_o=0$ and $\pm_o = -1$).
For a spherical source we have
\bea
r_s = \text{const},  \qquad
\t_s \in (0, \pi),
\eea
then Eq. \eqref{G} becomes
\bea
I^{(\omega)}_r(r_s,\rho) = \frac{1}{\sqrt{\rho^2-a^2}}[(2m+1)K(\rho) \pm_s F(\t_s, \rho)].
\label{ieg}
\eea
Inserting this for a sphere with radius $r_s$, we get the transfer function $\cos\theta_s(\rho)|_{r_s}$\cite{Gralla:2019ceu},
\bea
&&\cos{\t_s}=\text{cd}\l( \sqrt{\rho^2-a^2}I(\rho) \bigg| \f{a^2}{a^2-\rho^2} \r),
\label{inverse1}
\eea
where ``cd" is the Jacobi cd function. Notice that, due to the periodicity of the elliptic function $K$, the first term on the right side of Eq.~(\ref{ieg}) does not contribute. This transfer function shows directly the connection between the emission latitude $\t_s$ on the emission sphere and the observed radius $\rho$ on the observer's screen.
In Fig.~\ref{sphere}, we show several transfer functions $\cos\theta_s(\rho)|_{r_s}$ for selected values of black hole parameters $a$ and $Q$. Meanwhile, we consider distinct cases for source radii such that $r_s<\td r_0$, $\td r_0<r_s\ll r_o$, $r_s\sim r_o$ and $\td r_0\ll r_s\ll r_o$, respectively.
For $\rho<\td\rho$, the transfer function \eqref{inverse1} is single-valued, corresponding to only the direct emission. For $\rho>\td\rho$, the transfer function \eqref{inverse1} becomes double-valued, corresponding to both the direct and reflected emissions. Each oscillation of $\theta_s$ from $0$ to $\pi$ corresponds to an image of the source sphere. Near the critical radius $\rho=\td\rho$, the oscillations become extremely dense\footnote{However, the extremely dense oscillations are not visible in the plots due to a limited numerical precision.}, which corresponds to infinitely many images labeled with $m$.
We can see that, the primary image in each case mainly reflects the property of the source, while the higher-order images reflect the property of the spacetime which are almost independent of the source directly\footnote{The positions of emitting spheres (relative to the photon sphere and observer) also make differences, showing different behavior of photons (direct or reflected) from a source to an observer. For example, for $r_s=2$ ($r_s<\td r_0$), no photon is reflected and the primary image can already reflect properties of the spacetimes.}. A remarkable feature of the the higher-order images is that these images in each plot approach the critical curve at $\rho=\td \rho$ in a narrow region. As the charge $Q$ or spin $a$ varies, the position of $\td \rho$ is changed correspondingly (see Fig.~\ref{fig:criticalcurves}).

\subsection{Thin equatorial disk viewed by an on-axis observer}\label{disk on axis}
\begin{figure}[H]
\centering
\includegraphics[scale=0.4]{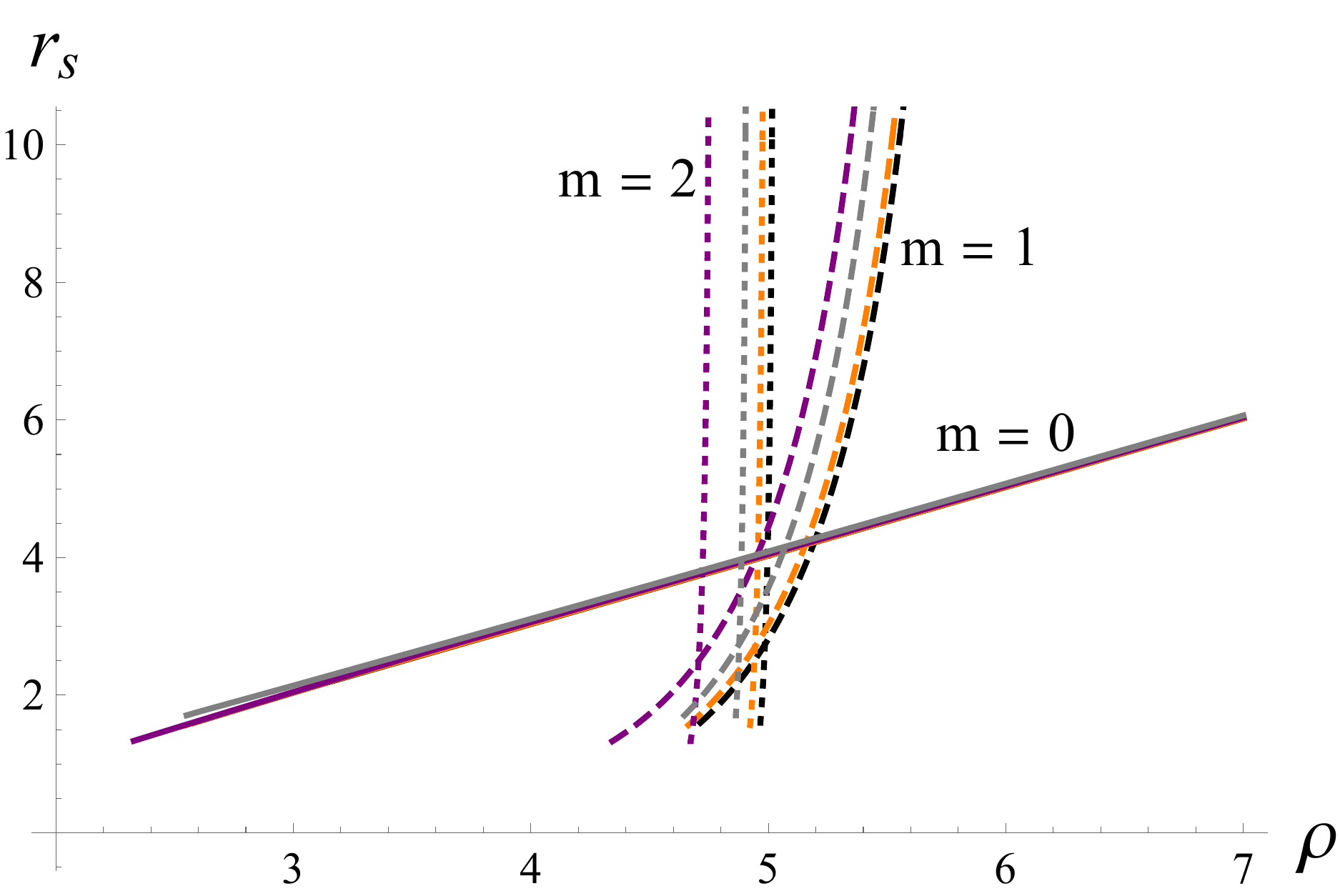}\,\,\,\,\,\,
\includegraphics[scale=0.4]{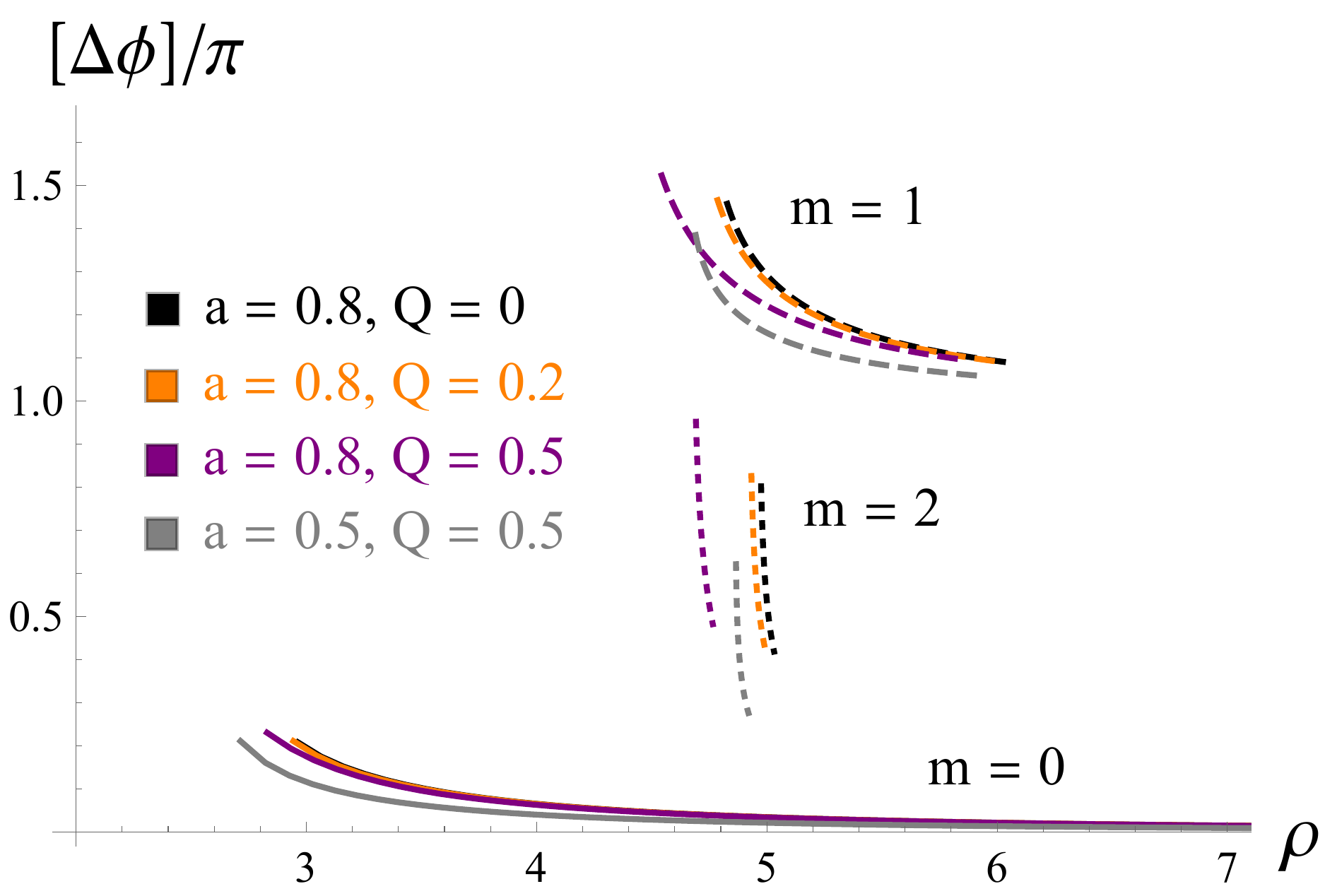}
\caption{Left: positions of the images of an equatorial disk with $r_s\in(r_+,\infty)$ viewed by an on-axis observer at $r_o=1000$. Right: changes of azimuthal angles of the images of an equatorial disk with $r_s\in(r_e,\infty)$ viewed by an on-axis observer at $r_o=1000$.}
\label{equatorial2}
\end{figure}

We now discuss the images of an equatorial thin disk consisting of emitting rings with $r_s>r_+$ and $\theta_s=\pi/2$, viewed by an on-axis observer from above ($\theta_o=0$ and $\pm_o = -1$).
We first consider the apparent positions of the source rings. From Eq.~\eqref{G} we have
\bea
\label{diskmap}
I_r^{(\omega)}(r_s, \rho) = \frac{1}{\sqrt{\rho^2-a^2}}(2m+1)K\l(\f{a^2}{a^2-\rho^2}\r).
\eea
This gives a series of transfer functions ${r_s}^{(m)}(\rho)$ for $m=0,1,2,\dots$, respectively. Different from the spherical case [Eq.~\eqref{ieg}], for each certain $m$,  Eq.~\eqref{diskmap} is solved either by the direct ray with $\omega=0$ or by the reflected ray with $\omega=1$, and in no case can both of rays solve this equation.
We will consider a disk model that extends to the event horizon $r_+$.
In Fig.[\ref{equatorial2}], we show the first three transfer functions, $r_s^{(0)}(\rho)$, $r_s^{(1)}(\rho)$ and $r_s^{(2)}(\rho)$, under different black hole parameters. We can see that the properties of the disk images at different orders are similar as those of the spherical case.
For a black hole with certain parameters, the first transfer function $r_s^{(0)}(\rho)$ is approximately linear,
$r_s \approx \rho -1 $, representing that the primary image is a direct reflection of the source. Note that the lower bound of $\rho$ on the screen corresponds to $r_s=r_+$.
The slope of the transfer function $d r_s/d\rho$ becomes larger as the image level goes higher, which corresponds to narrower and narrower bands covering the critical radius $\td\rho$ on the screen. Like in the spherical case, the black hole charge $Q$ and spin $a$ also affect the position of $\td\rho$. Besides, we see from Fig.~\ref{fig:criticalcurves} that the charge and spin affect the widths of the first and second narrow bands ($\D\rho^{(1)}$ and $\D\rho^{(2)}$) as well. Thus, the higher-order images again show the dependence on the spacetime parameters clearly.

Next, we consider the change of azimuthal angles between the angles at arrival $\phi_o$ and at emission $\phi_s$. From Eq.~\eqref{Gphi} we have
\bea
\D\phi^{(m)} =\phi^{(m)}_o -\phi_s = I_{\phi}(r_s, \rho^{(m)}(r_s)) + m\pi,
\label{mapping2}
\eea
where we have used Eq.~\eqref{polecrossing} for the pole-crossing orbits to get the last term. For convenience, we introduce $[\D\phi]=\mod_{2\pi}(\D\phi)$. In Fig.~\ref{equatorial2}, we  show the results of $[\D\phi^{(0)}]$, $[\D\phi^{(1)}]$ and $[\D\phi^{(2)}]$ as functions of $\rho$. 
Note that a static disk cannot exist inside the ergosphere at $r_e$, therefore we only show the results for $r_s\in(r_e,\infty)$. We can see that for a black hole with given parameters $a$ and $Q$, the images with even $m$ have $[\D\phi]\in(0,\pi)$ while those with odd $m$ have $[\D\phi]\in(\pi,2\pi)$. For the primary image, the influence of black hole parameters on $[\D\phi]$ is quite small in general, while the influences become larger and larger as the image level goes higher. For the images at each level, $[\D\phi]$ increases with  the spin $a$ or the charge $Q$, and the influence of spin is overall greater than that of charge.

\subsection{Thin equatorial disk viewed by an off-axis observer}\label{disk off axis}
\begin{figure}[h]
\centering
\includegraphics[scale=0.35]{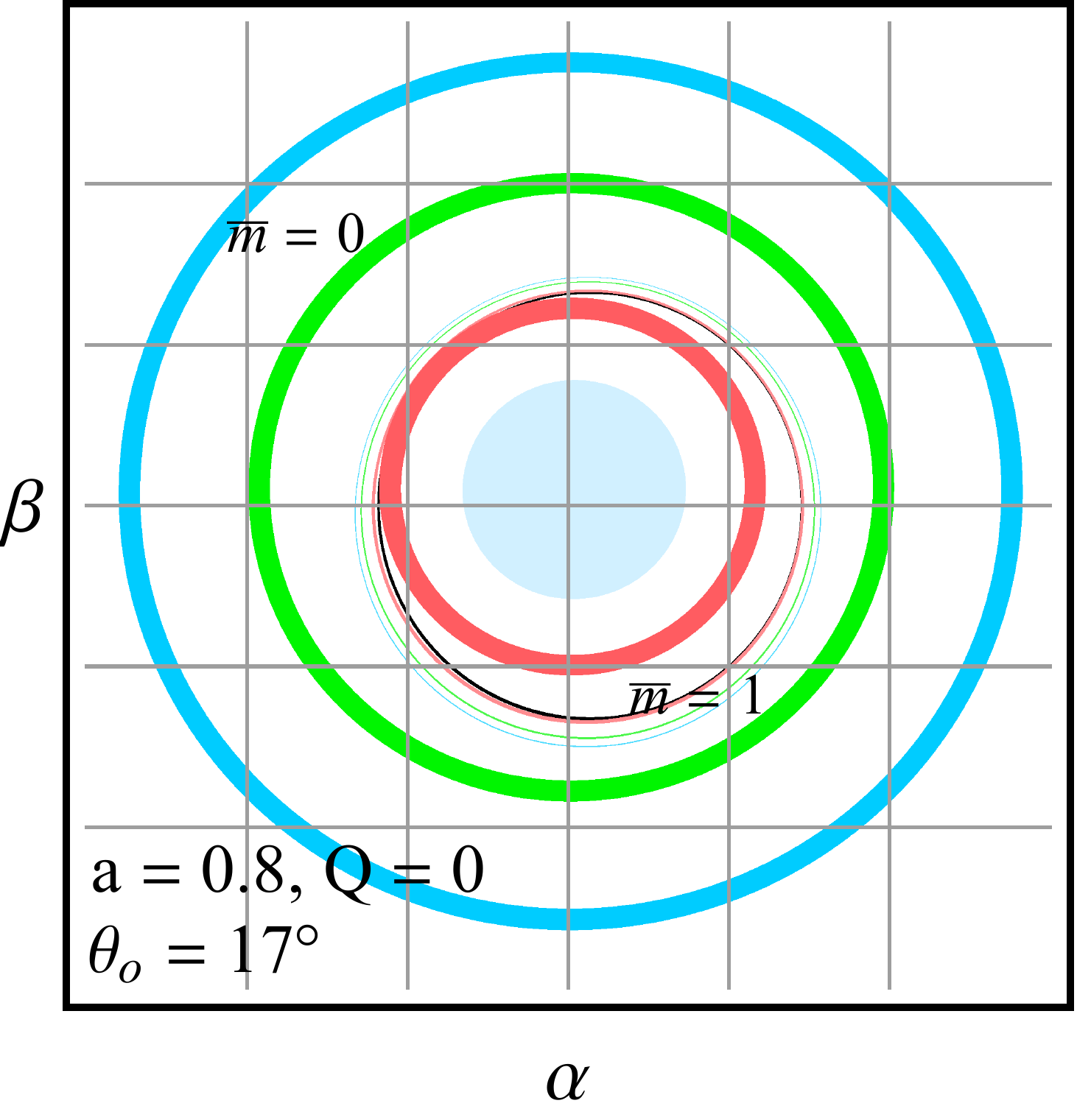} \
\includegraphics[scale=0.35]{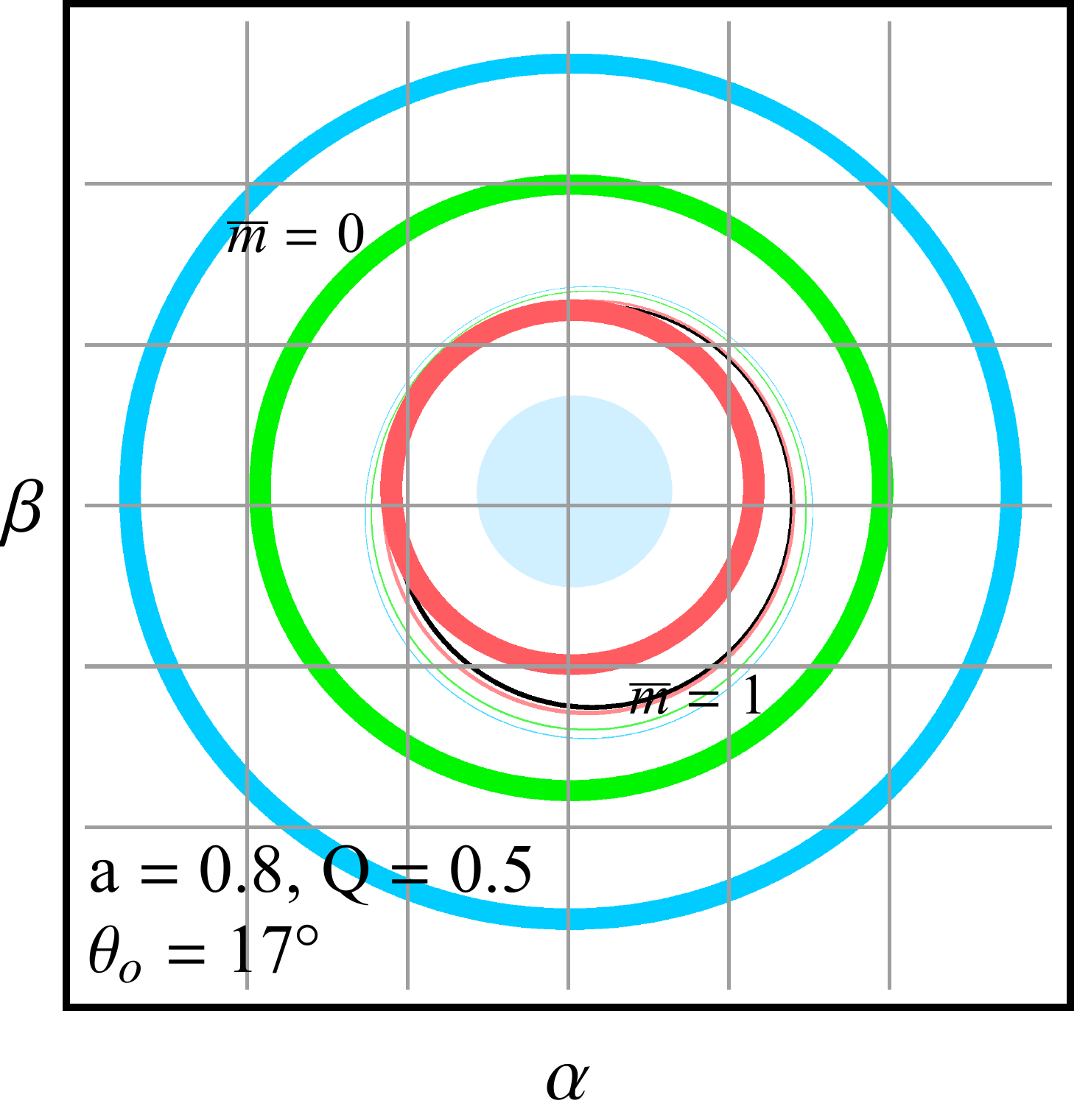} \
\includegraphics[scale=0.35]{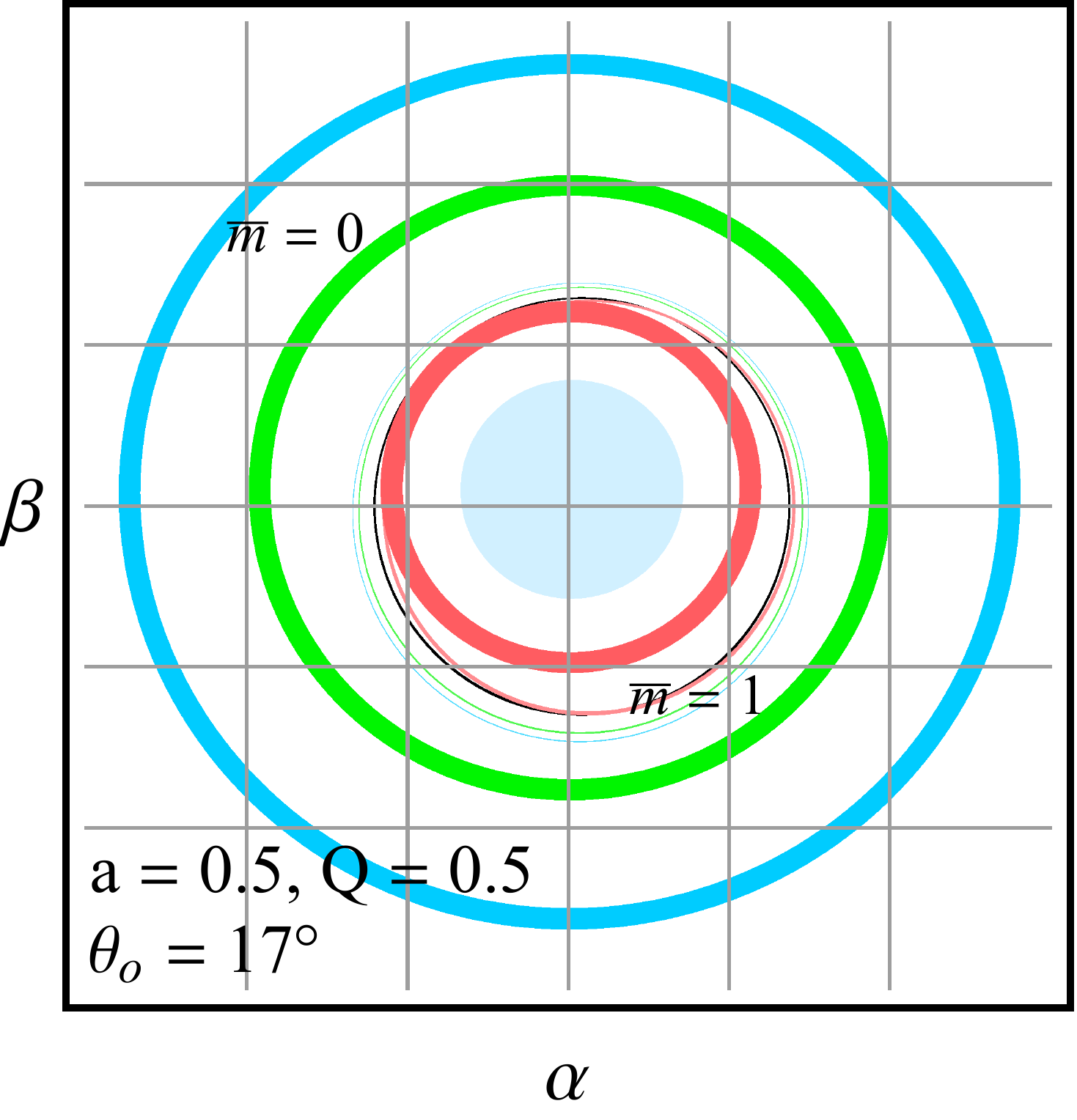}\\
~\\
\includegraphics[scale=0.35]{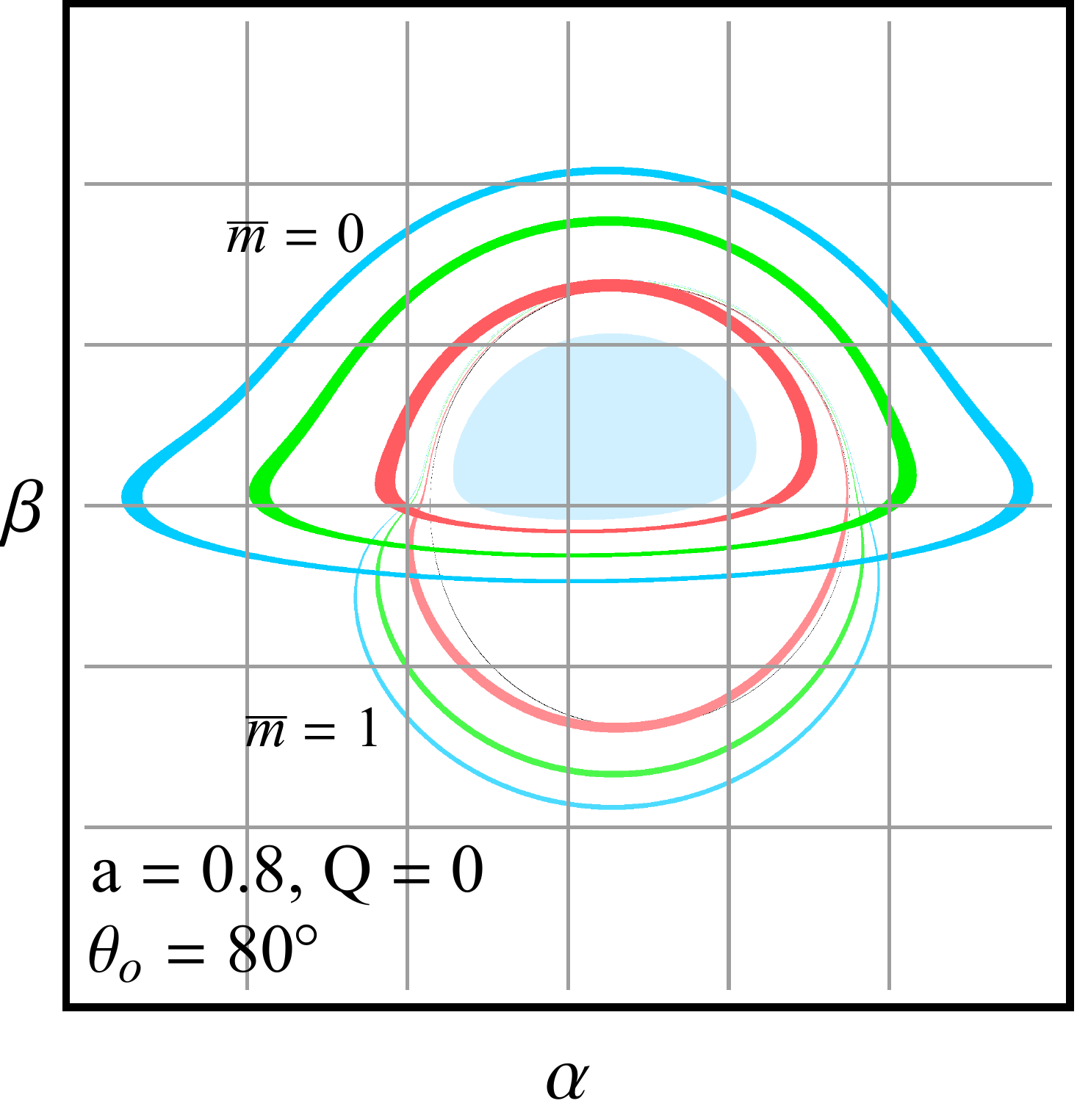} \
\includegraphics[scale=0.35]{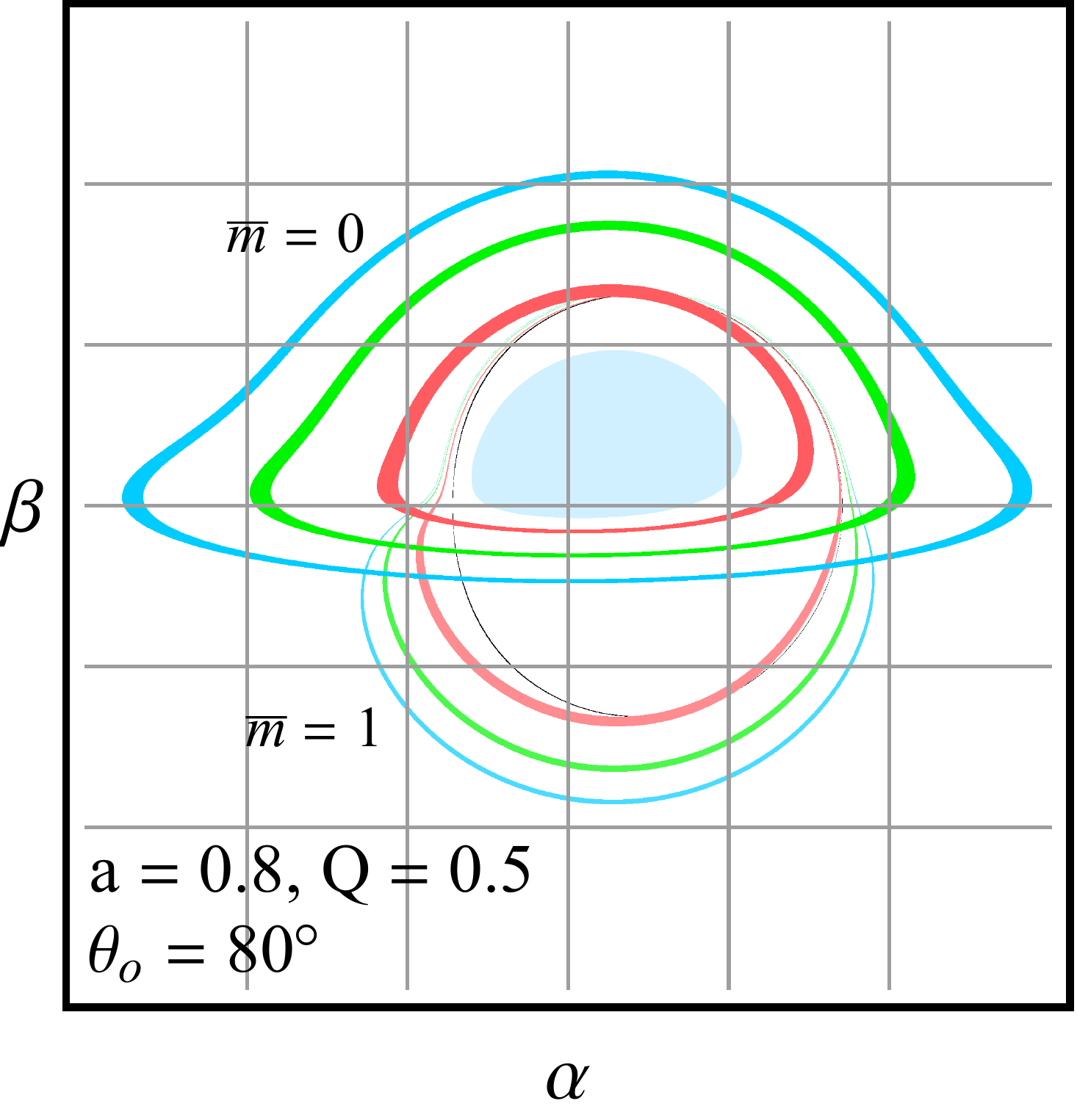} \
\includegraphics[scale=0.35]{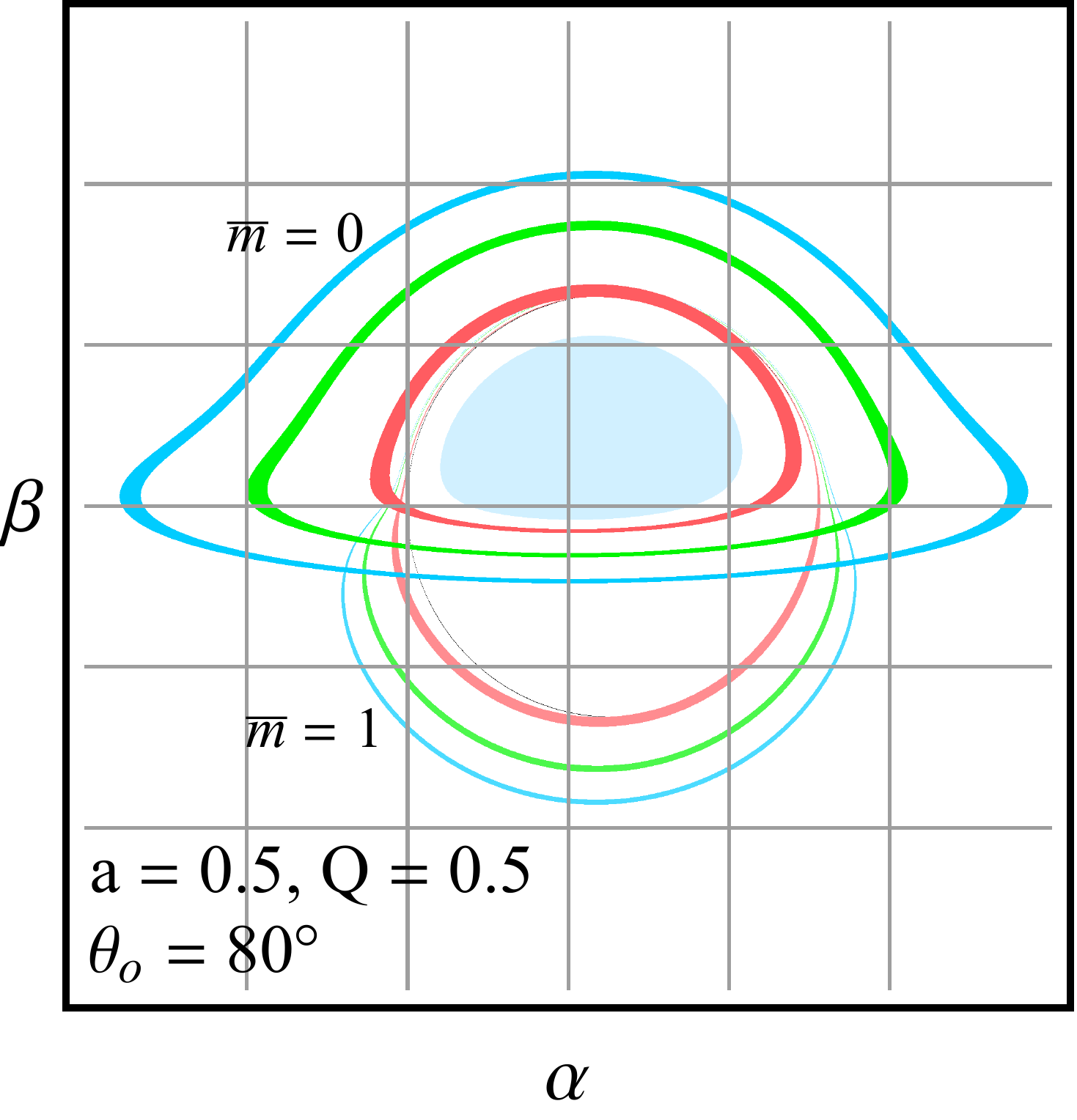}\\
\caption{Images positions of equatorial rings with $r_s\in[3,3.5]$ (pink), $r_s\in[6,6.5]$ (green) and $r_s\in[9,9.5]$ (blue) described in Cartesian coordinates \eqref{offcoordinates}. In each plot, the darker and lighter rings correspond to $\bar{m}=0, 1$, respectively, the center lightblue region is the inner shadow \cite{Chael:2021rjo} with the edge corresponds to a direct map of $r_+$, and the black ring corresponds to the critical curve.}
\label{disk}
\end{figure}

\begin{figure}[h]
\centering
\includegraphics[scale=0.27]{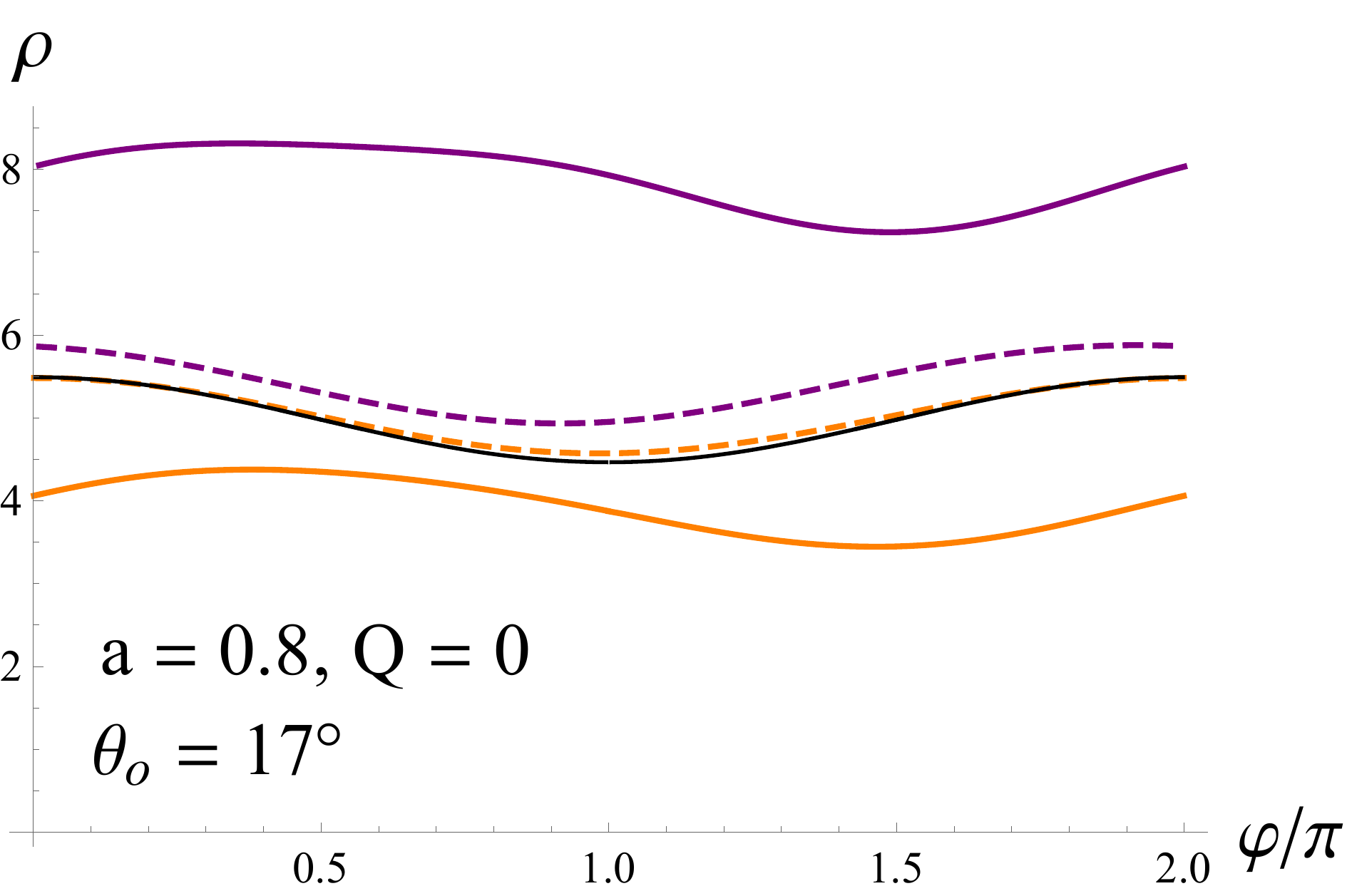} \
\includegraphics[scale=0.27]{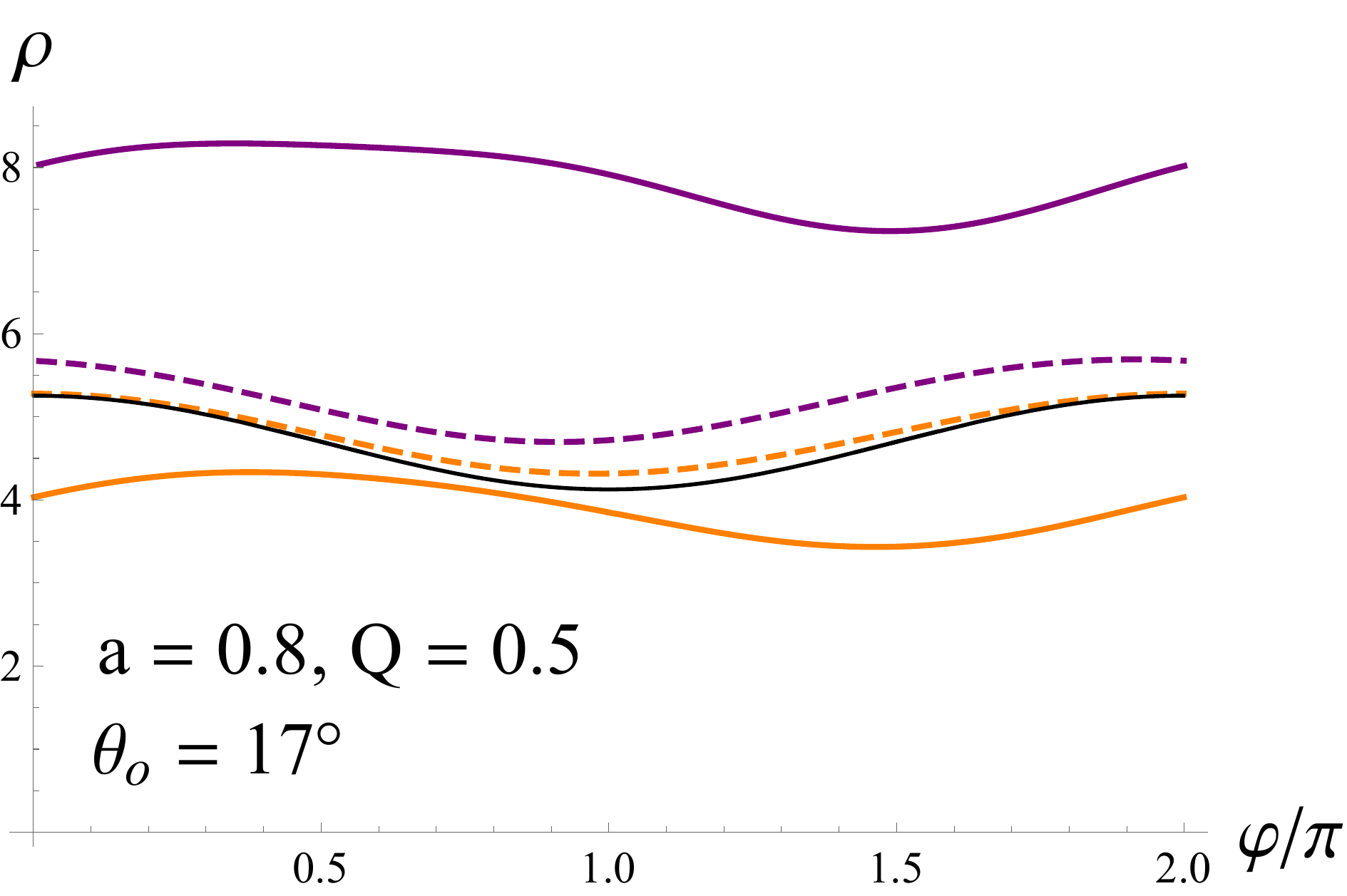} \
\includegraphics[scale=0.27]{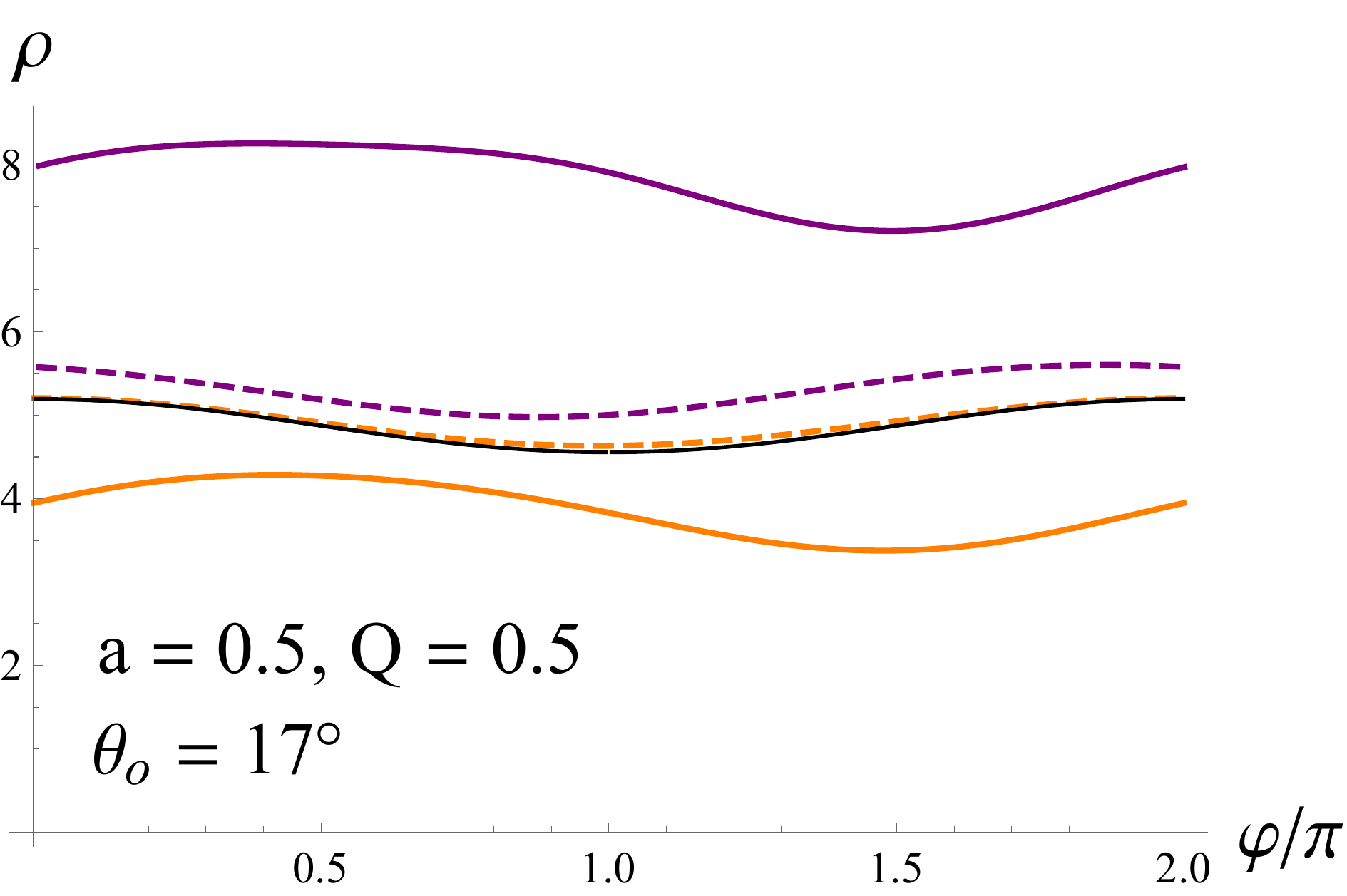} \\
~\\
\includegraphics[scale=0.27]{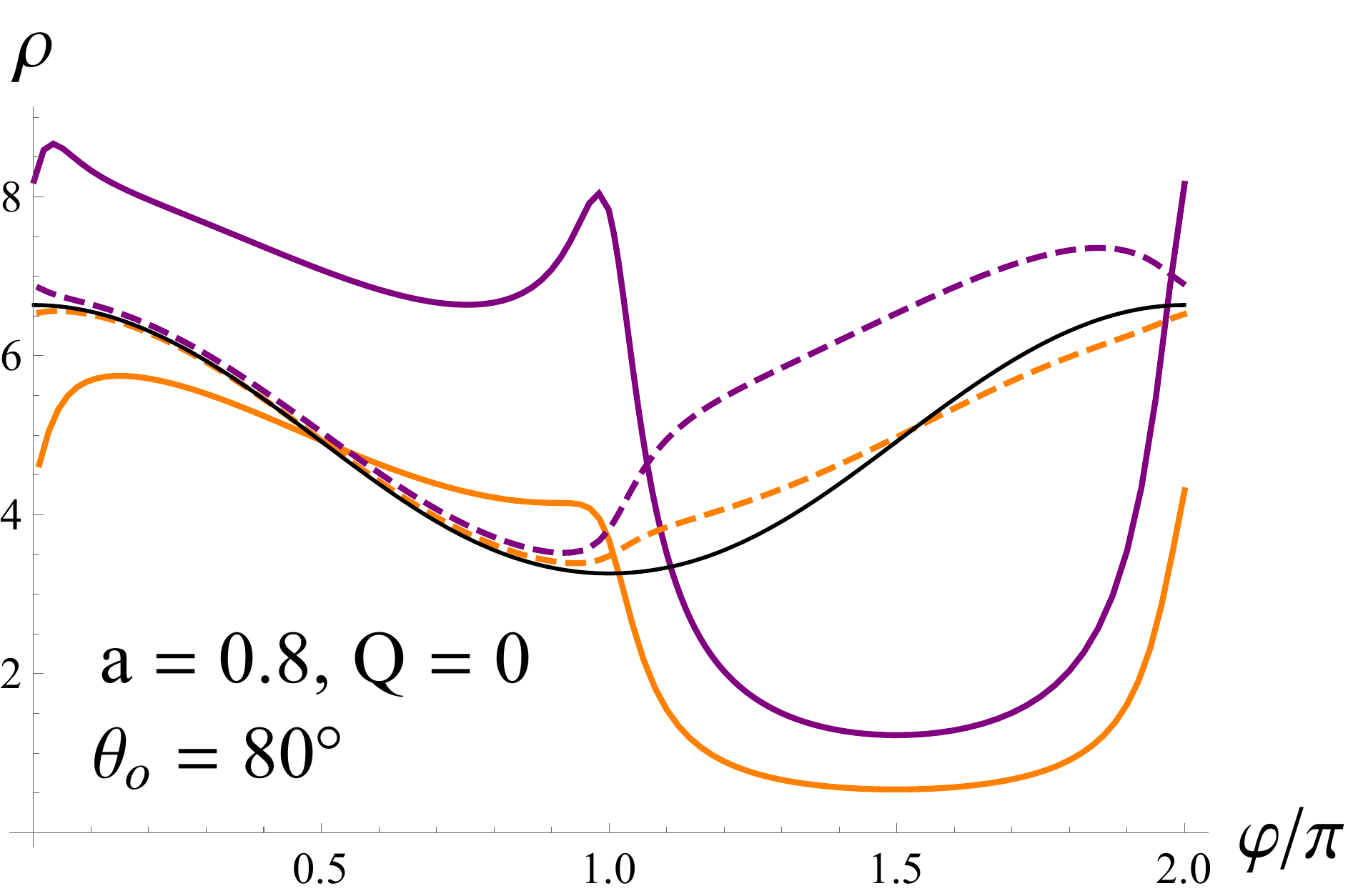}\
\includegraphics[scale=0.27]{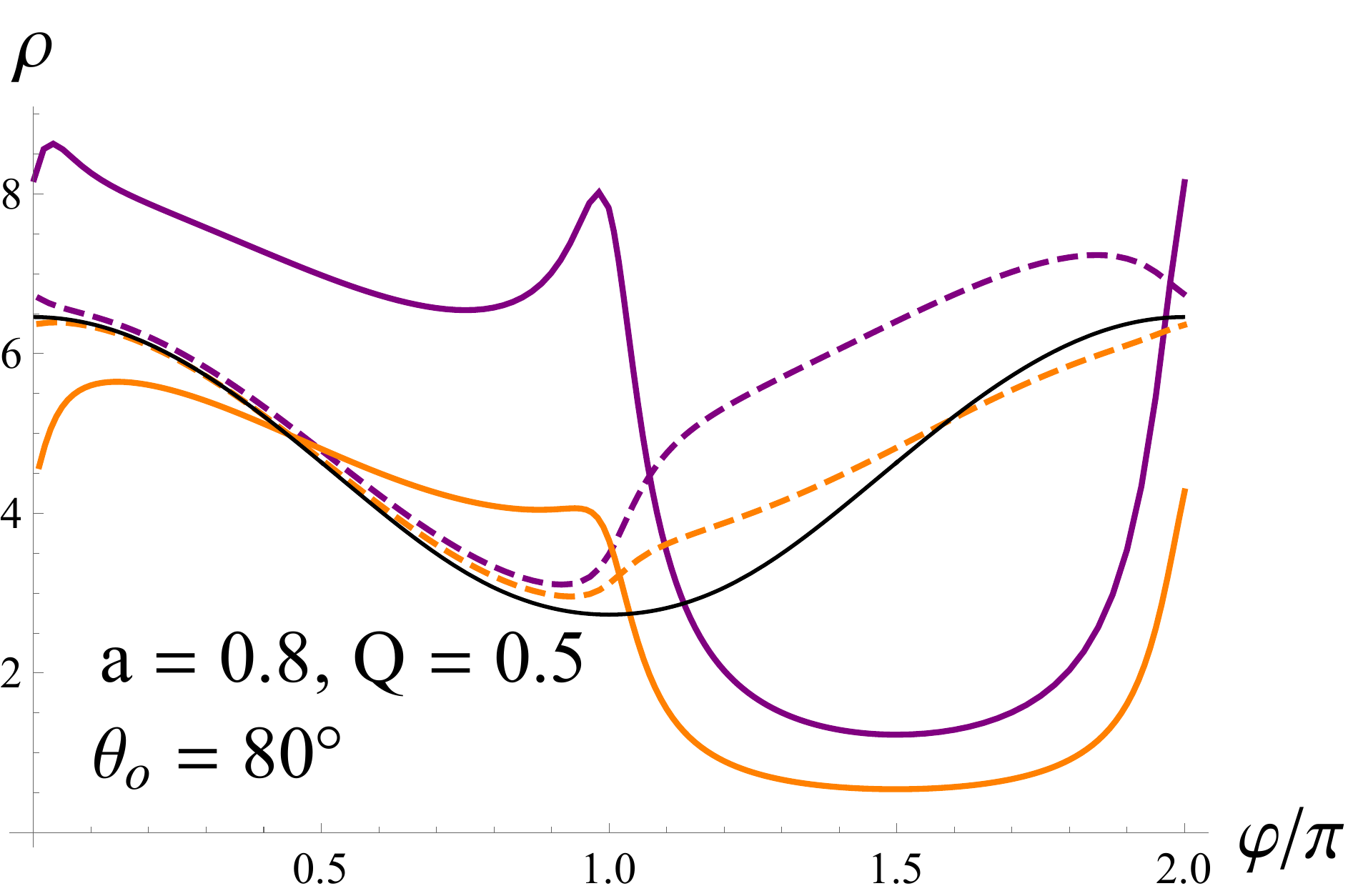}\
\includegraphics[scale=0.27]{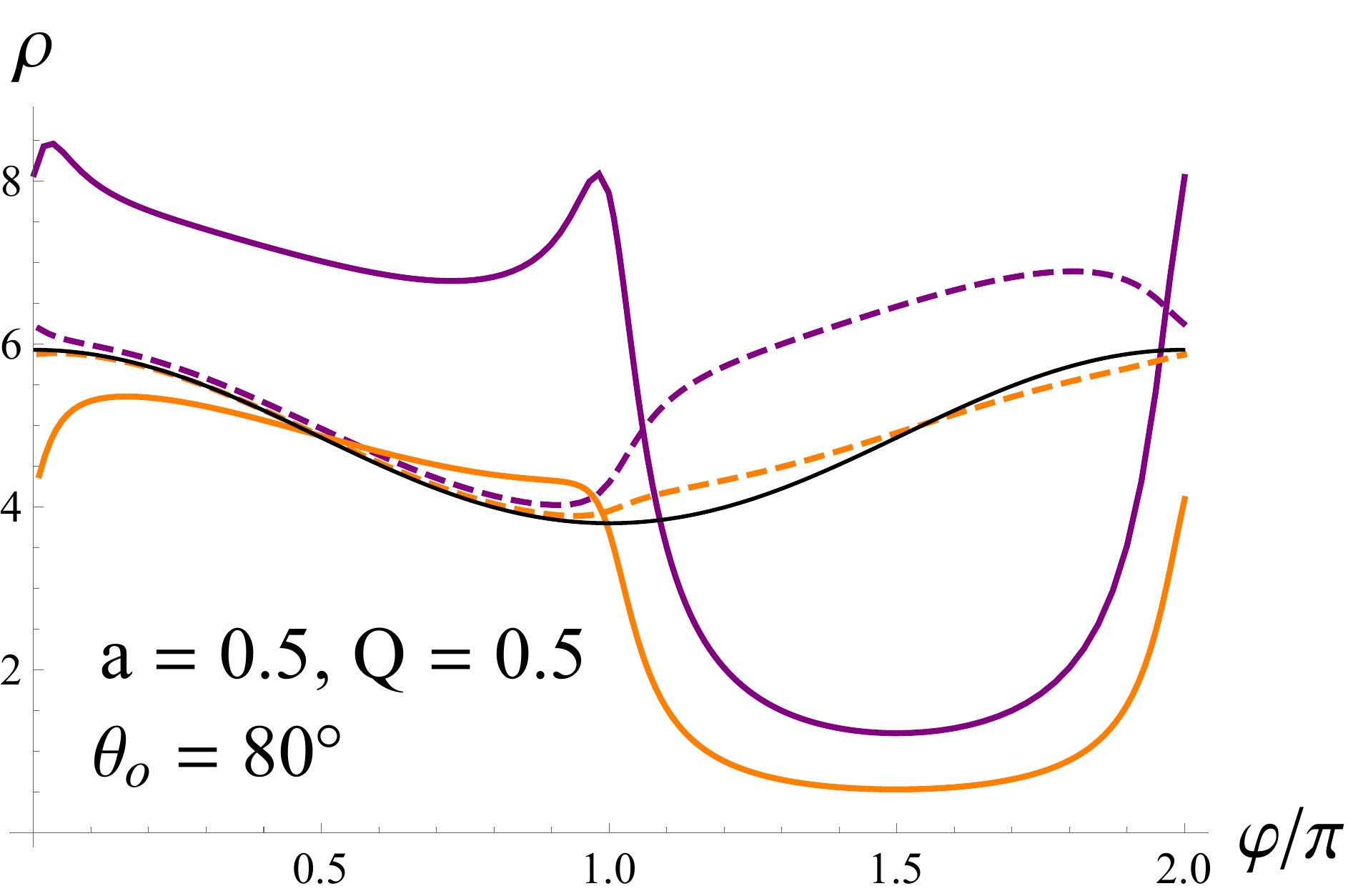} \\
\caption{Images positions of equatorial rings at $r_s=3$ (orange) and $r_s=7$ (purple) described in polar coordinates \eqref{offpolar}. In each plot, the solid and dashed curves correspond to $\bar{m}=0, 1$, respectively, and the black curve corresponds to the critical curve.}
\label{shapedisk}
\end{figure}

We now discuss the images of an equatorial thin disk consisting of emitting rings with $r_s>r_+$ and $\theta_s=\pi/2$, viewed by an off-axis observer [$\theta_o\neq0$ and $\pm_o =\si(\b)$].
We first consider the apparent positions of the equatorial emitting rings. The position of $m$-th image (denoted as $\mC_m$) of a equatorial ring at $r_s$ is determined by Eq.(\ref{G}) which now becomes
\bea
\mC_m: \ I_r^{(\omega)}(r_s, \a, \b) =\f{1}{a\sqrt{-u_-(\a, \b)}}[2mK(\a, \b)-\si(\b)F_o(\a, \b)].
\label{Cm}
\eea
However, due to the appearance of $\si(\b)$ in the last term, $m$ is no longer a good quantities to label continuous images for an inclined observer.
When crossing the $\a$-axis, Eq.~\eqref{Cm} yields
\bea
\mC_m: \ I_r^{(\omega)}(r_s, \a)=
\begin{cases}
\f{1}{a\sqrt{-u_-(\a)}}(2m-1)K(\a)& \text{if}\,\,\, \b \rightarrow 0^{+},\\
\f{1}{a\sqrt{-u_-(\a)}}(2m+1)K(\a)& \text{if}\,\,\, \b \rightarrow 0^{-}.
\end{cases}
\eea
Note that $\mC_{m}(\b\rightarrow 0^{-}) = \mC_{m+1}(\b\rightarrow 0^{+})$, which means that the lower part image of $\mC_m$ and the upper part image of $\mC_{m+1}$ can be smoothly connected. Thus we can instead introduce
\be
\label{mbar}
\bar{m} = m -H(\b)
\ee
as a label for continuous images, where $H$ is the Heaviside function. That is, we have $\{\mC_{\bar{m}},\bar m=0,1,2\dots\} = \{ \mC_{m}(\b<0)\cup\mC_{m+1}(\b>0), \, m=0,1,2,\dots\}$, this gives a set of smooth curves on the observer's screen. Moreover, one finds that $\bar{m}$ denotes how many times that a photon passes across the equatorial plane before it reaching the screen.
We will consider three source rings lying in the ranges $3<r_s<3.5$, $6<r_s<6.5$ and $9<r_s<9.5$, and consider three choices for the black hole parameters $a$ and $Q$.
In Fig.[\ref{disk}], we show the first two ($\bar m=0,\,1$) images of each selected equatorial source ring for each choice of the black hole parameters.
We also describe the first two image rings of source rings at $r_s=3$ and $7$ with the polar coordinates $(\rho, \varphi)$ [Eq.~\eqref{offpolar}] in Fig.[\ref{shapedisk}], to quantitatively show the variation of radius $\rho$ along polar angle $\vp$ for each image. 
We can see that, again, the primary images mainly reflect the feature of source profile (the inner bound $r_+$ of the disk) while the higher-order images become less dependent on the source detail and mainly reflect the property of the spacetime (the critical curve $\M C$).

\begin{figure}[h]
\centering
\includegraphics[scale=0.27]{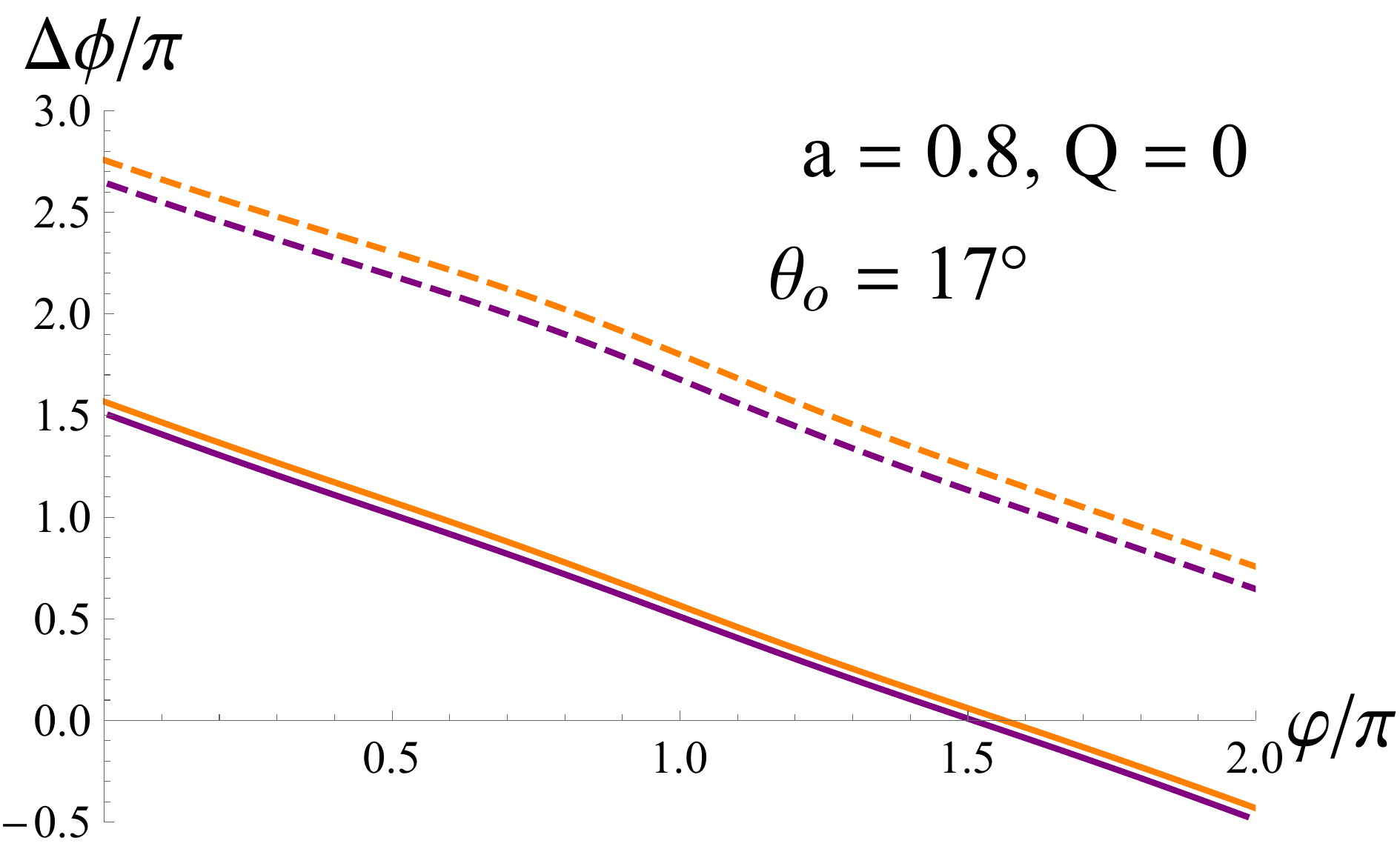} \
\includegraphics[scale=0.27]{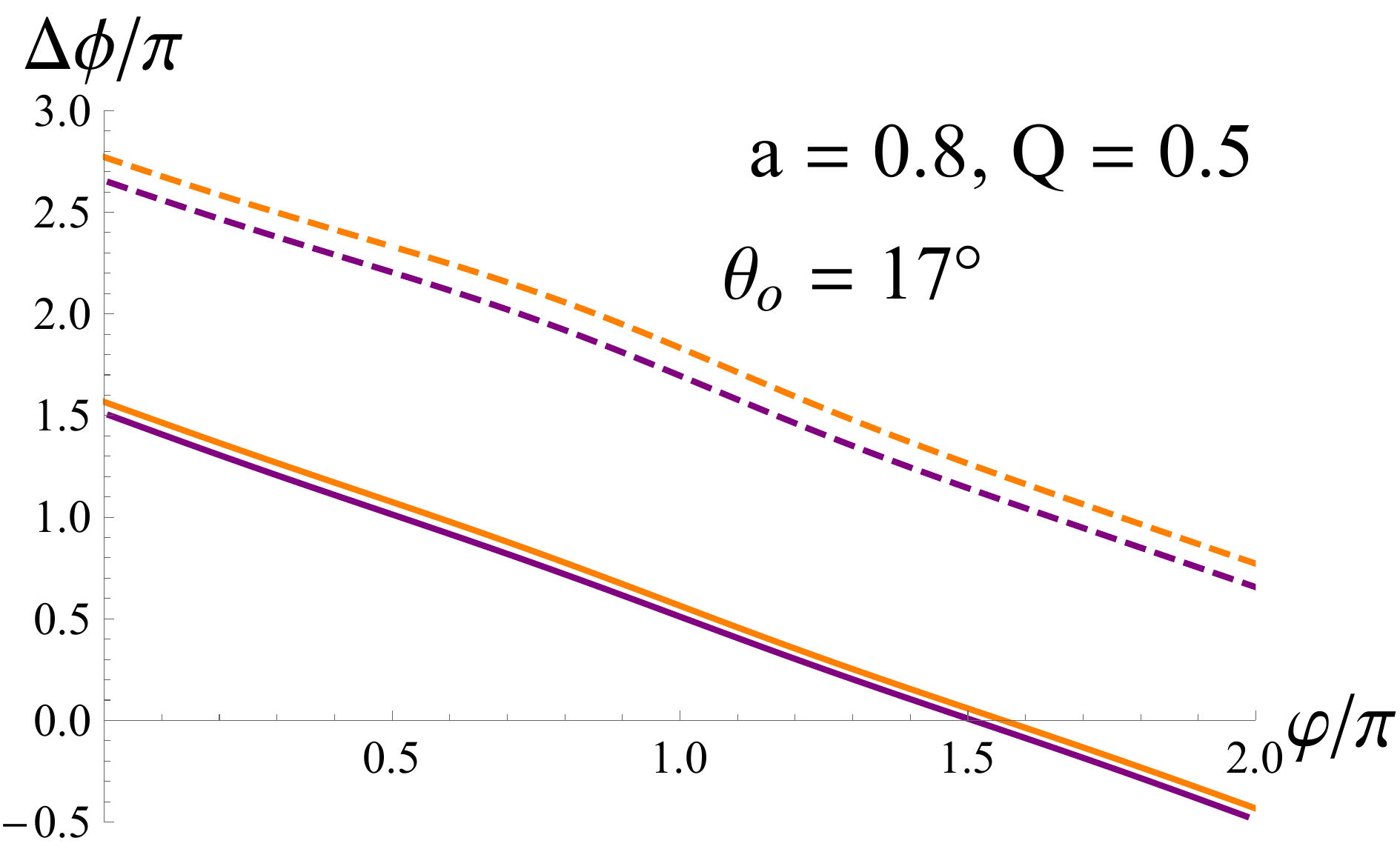} \
\includegraphics[scale=0.27]{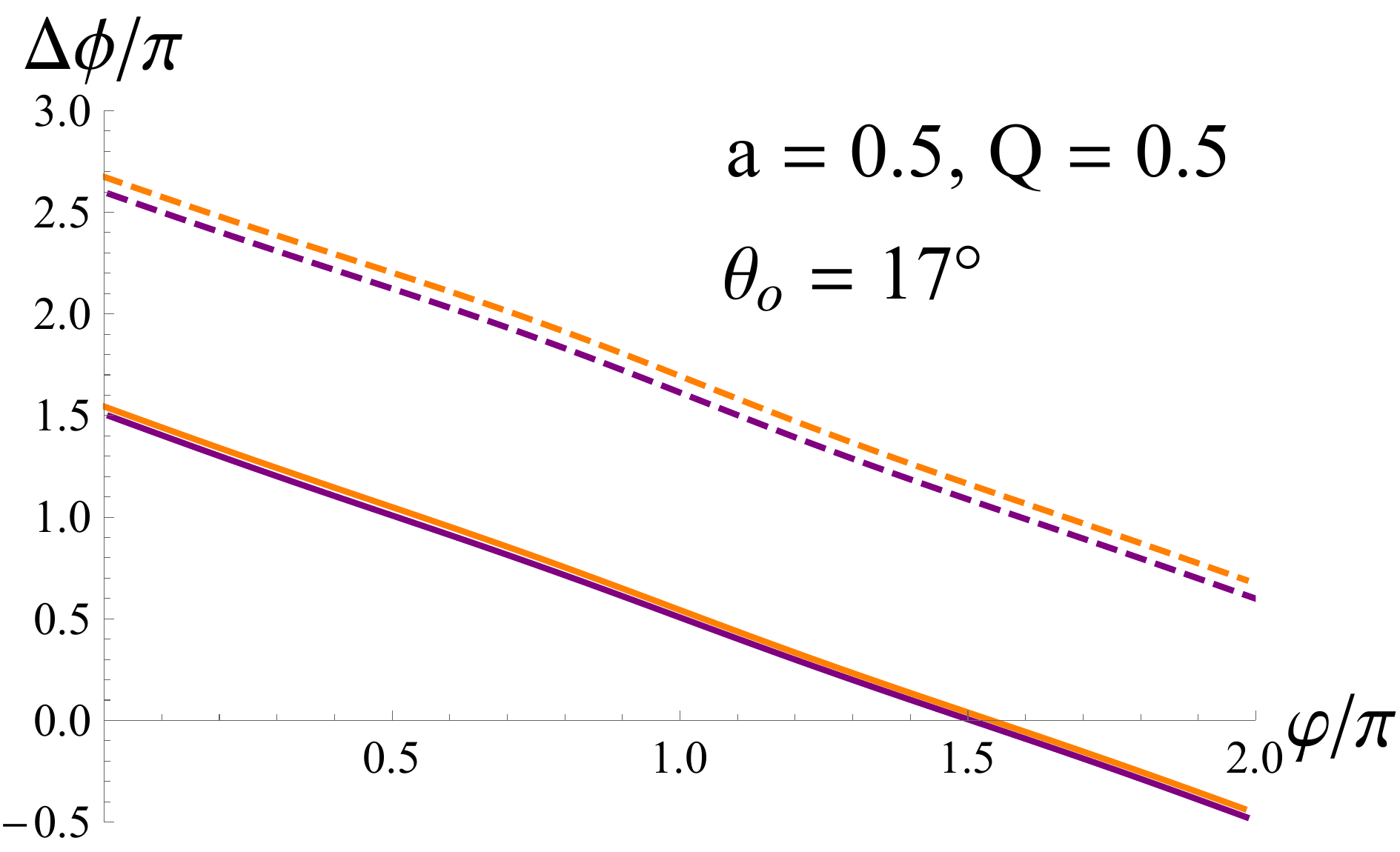} \\
~\\
\includegraphics[scale=0.27]{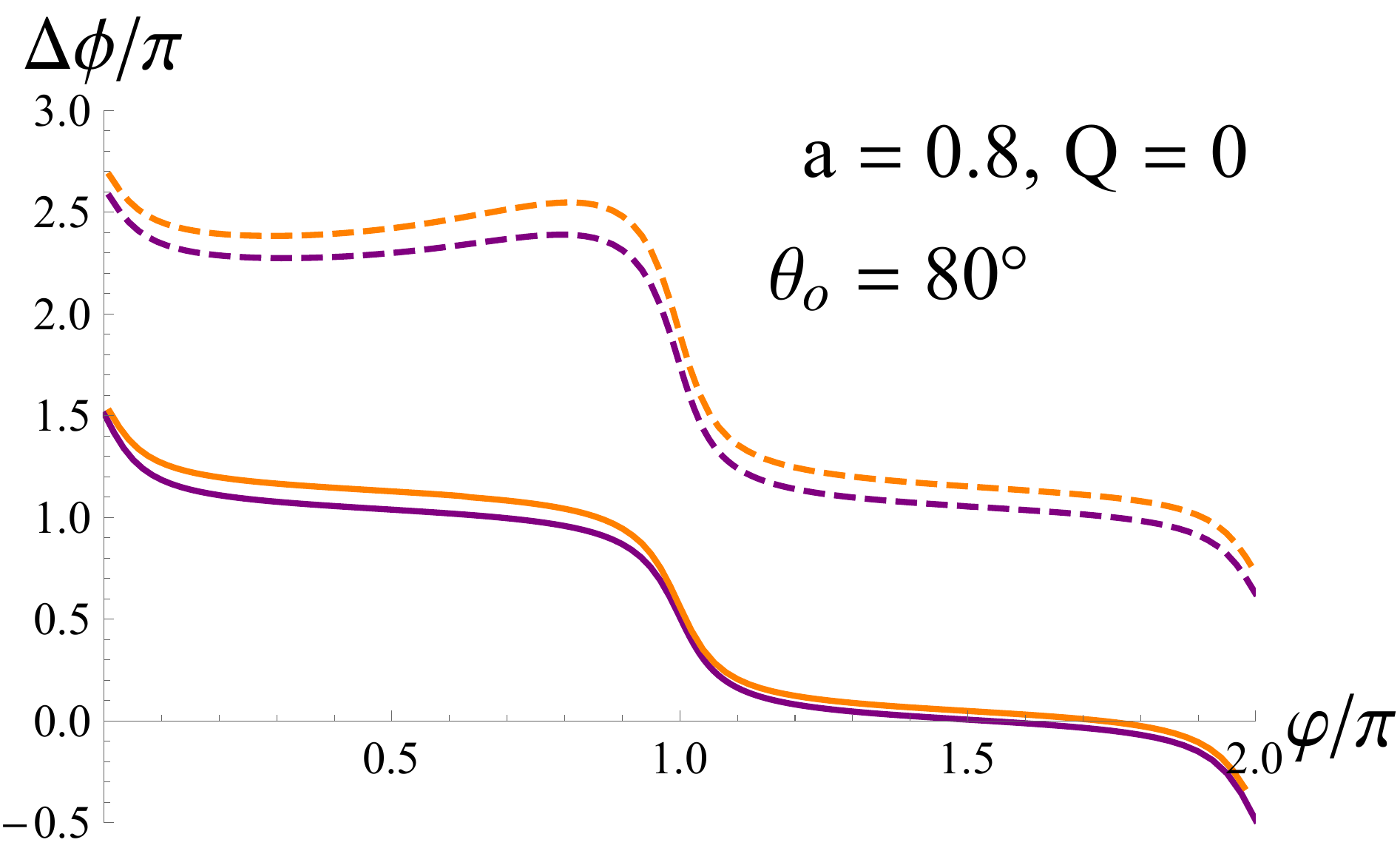}\
\includegraphics[scale=0.27]{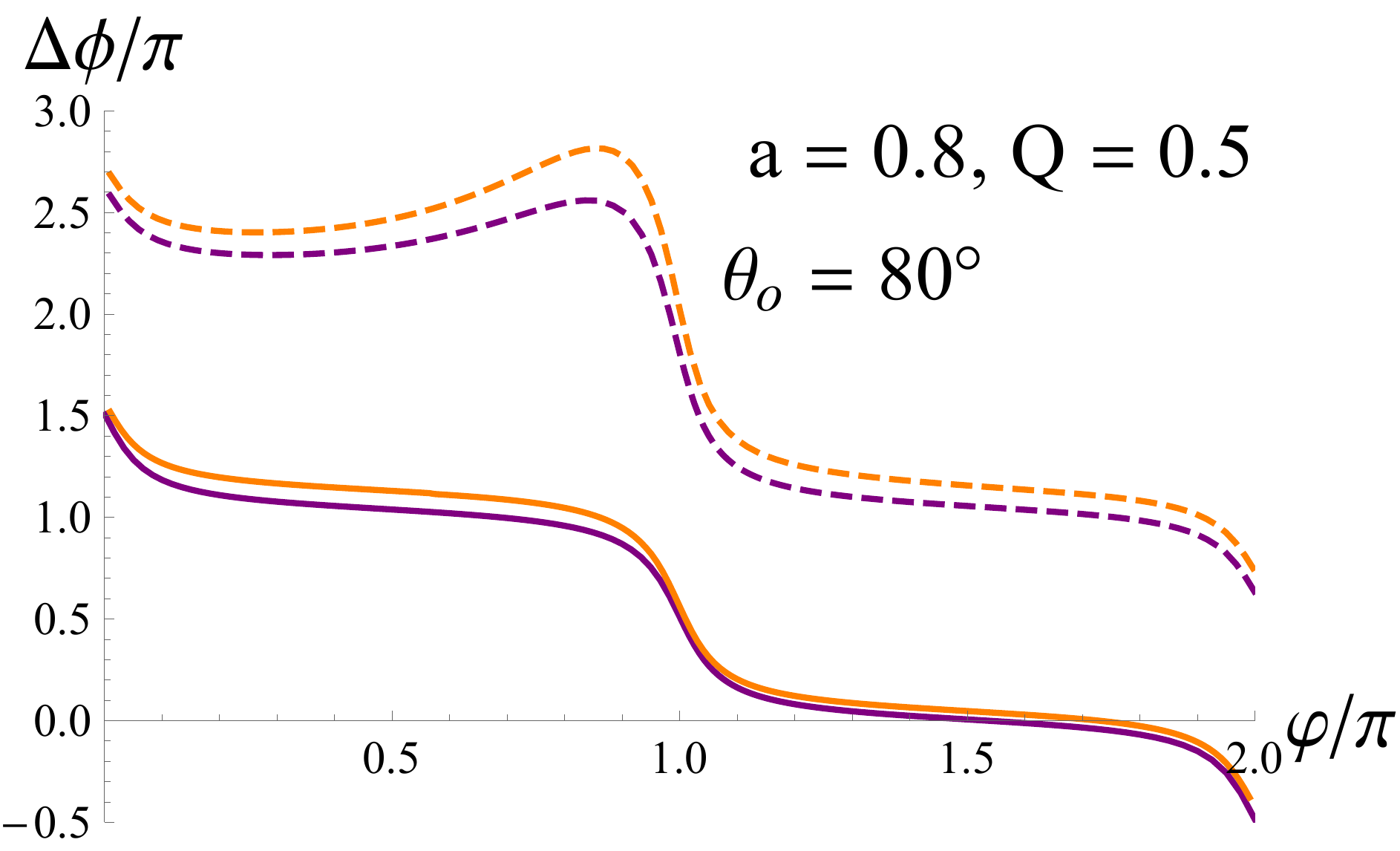}\
\includegraphics[scale=0.27]{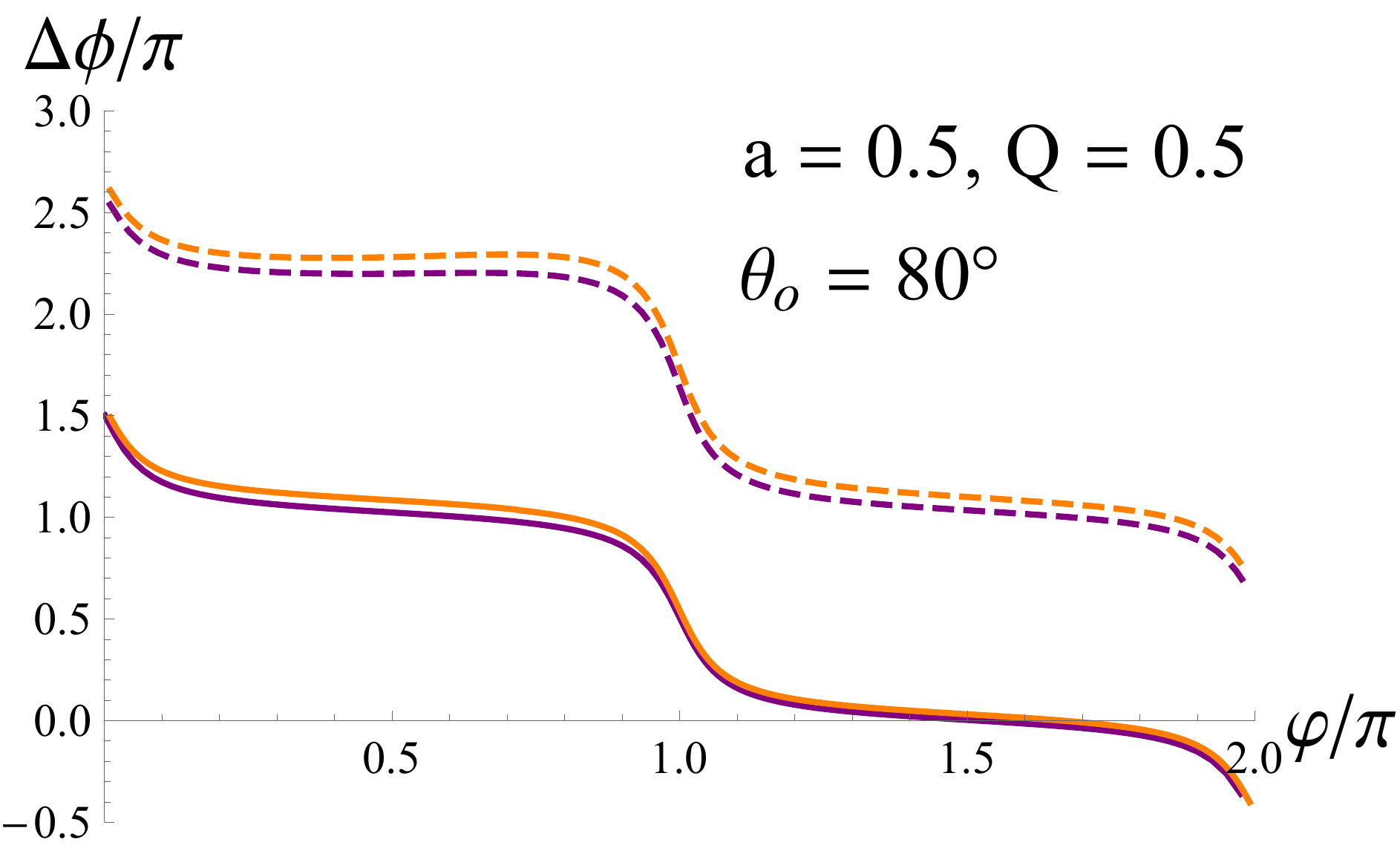}\\
\caption{Image rotations $\D\phi$ of equatorial rings at $r_s=3$ (orange) and $r_s=7$ (purple) described in polar coordinates \eqref{offpolar}. In each plot, the solid and dashed curves correspond to $\bar{m}=0, 1$, respectively.}
\label{Dphidisk}
\end{figure}

Next, we consider the change of azimuthal angles between the angles at arrival $\phi_o$ and at emission $\phi_s$.
Rewriting Eq.~\eqref{Gphi} in terms of the polar coordinates \eqref{offpolar} we get
\bea
\D\phi = \rho \cos{\varphi} \sin{\t_o} G_{\phi}^{(\bar{m})}(\rho,\varphi)-I^{(\omega)}_{\phi}(r_s,\rho,\varphi).
\label{phis}
\eea
Note that for an off-axis observer we can set $\phi_o=0$, then we have $\D\phi=-\phi_s$. Therefore, this equation actually gives the relation between the source position $\phi_s$ and its image position on the observer's screen.
We show this relation for selected black hole parameters and observer's inclinations in Fig.[\ref{Dphidisk}].
We find that in each plot the primary image is very close to the optical images of inclined source rings in flat spacetime\footnote{For example, the primary images viewed from $17^\circ$ show that the observed angles $\varphi=0, \ \pi/2, \ \pi$ and $3\pi/2$ correspond to the source angles $\phi_s \sim \pi/2, \ \pi, \ 3\pi/2$ and $ 0$, respectively.}, while the secondary image is rotated by roughly $\sim\pi$ relative to primary image, showing the lensing effects of black hole. We can also see that the image's rotation depends on the position $\vp$ along the image ring, the black hole spin $a$ and the charge $Q$. We will study the rotations in more detail in next section.

\section{Photon ring (higher-order images)}\label{sec:photonring}
In the previous section, we have discussed the overall appearances of various sources. In this section, we will move on to the details of higher-order images, which altogether form the so-called photon ring on the screen.
The photon ring is composed of the photons arriving near the critical curve $\M C$.
We have introduced the near-critical lens equations \eqref{ncequations} in Sec.~\ref{sec:nearcriticalrays}, which determine the behaviors of successive higher-order images. Here we will discuss the property of photon ring for various sources in terms of the key parameters $\g$, $\d$ and $\tau$, viewed on or off the pole.
In particular, we will study the dependence of these parameters on the black hole spin $a$ and charge $Q$.

\subsection{Sources viewed by an on-axis observer}\label{onimage}
\begin{figure}[hp]
\centering
\includegraphics[scale=0.32]{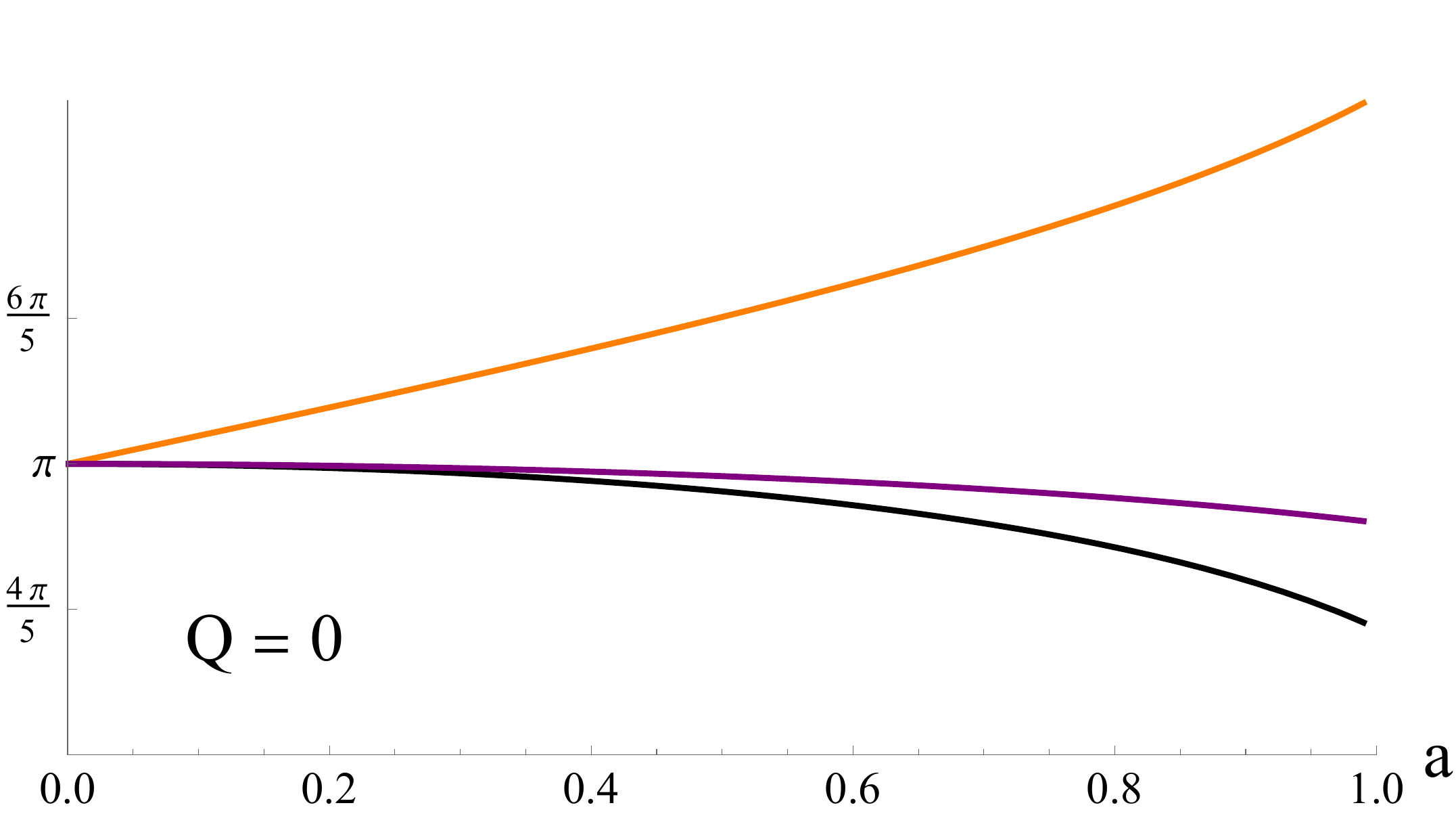} \qquad\,\,
\includegraphics[scale=0.32]{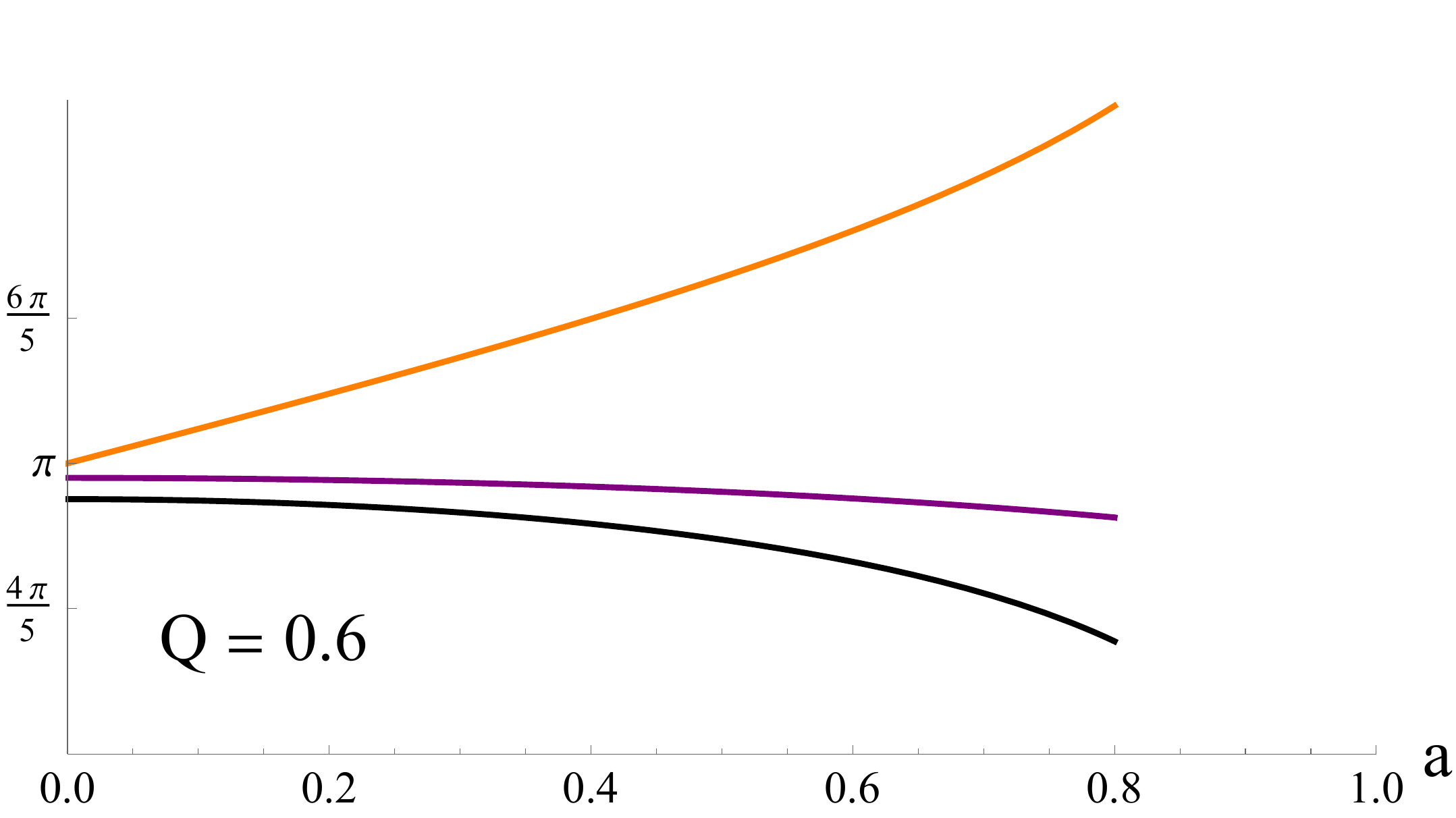}\\
\includegraphics[scale=0.32]{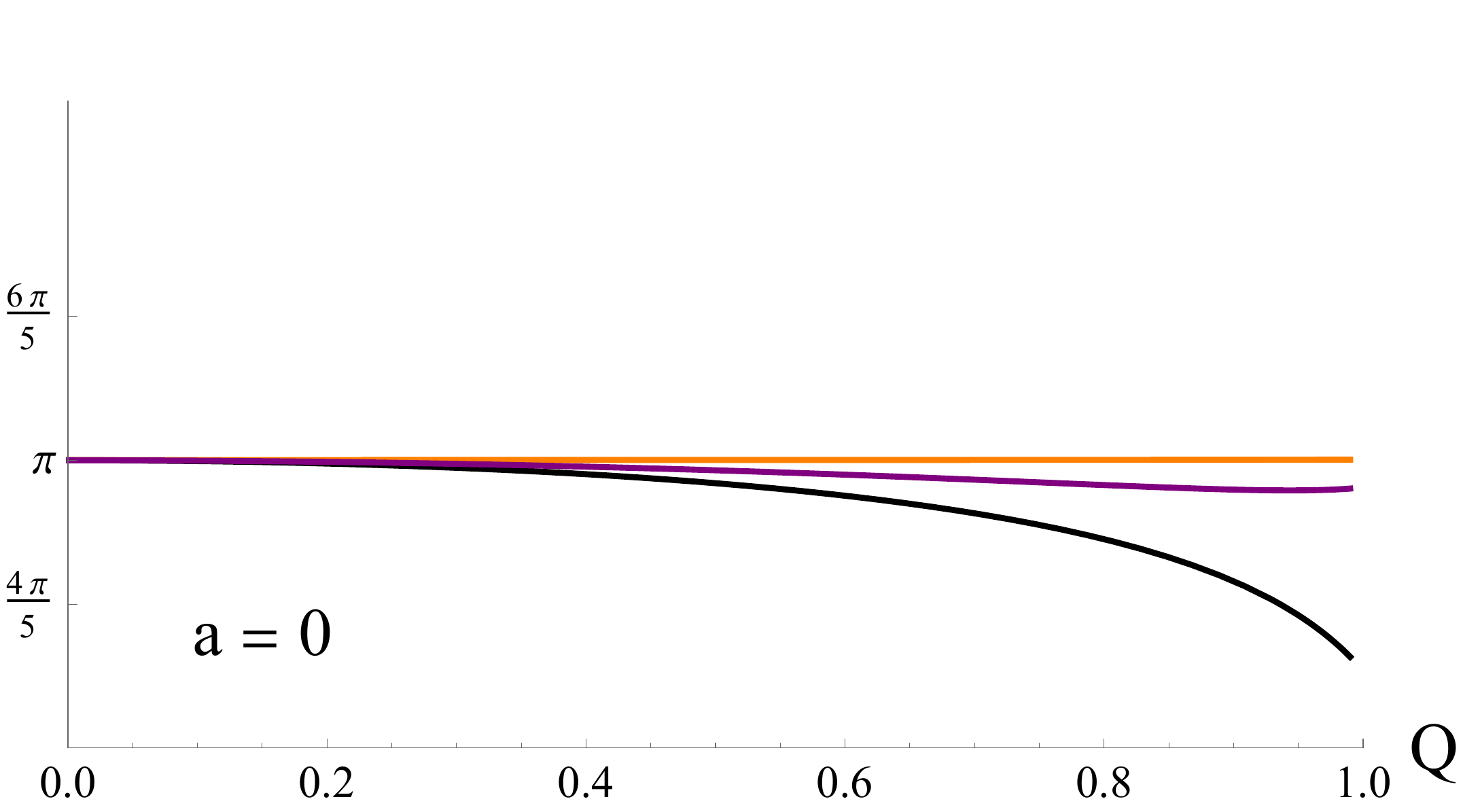} \qquad\,\,
\includegraphics[scale=0.32]{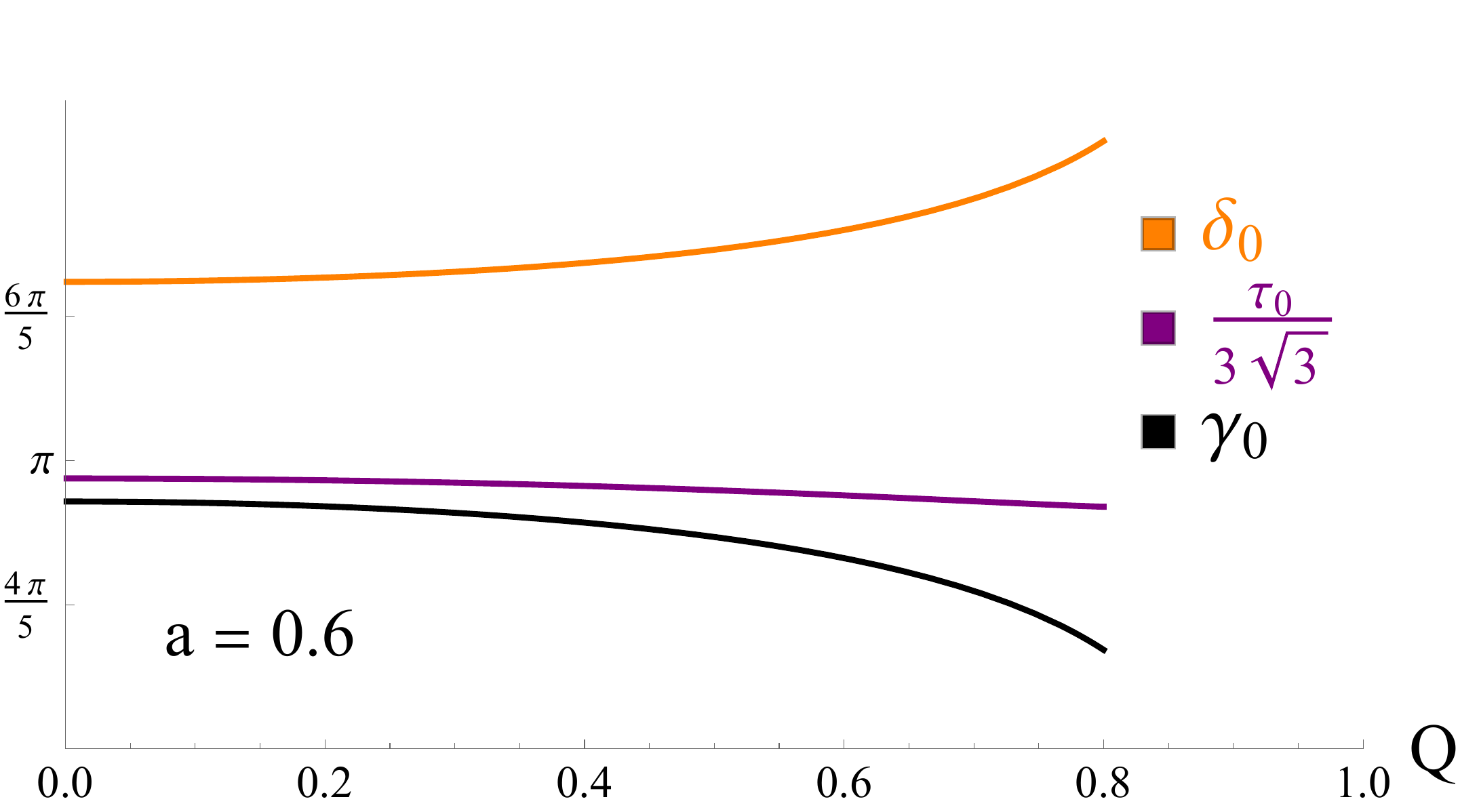}\\~\\
\includegraphics[scale=0.32]{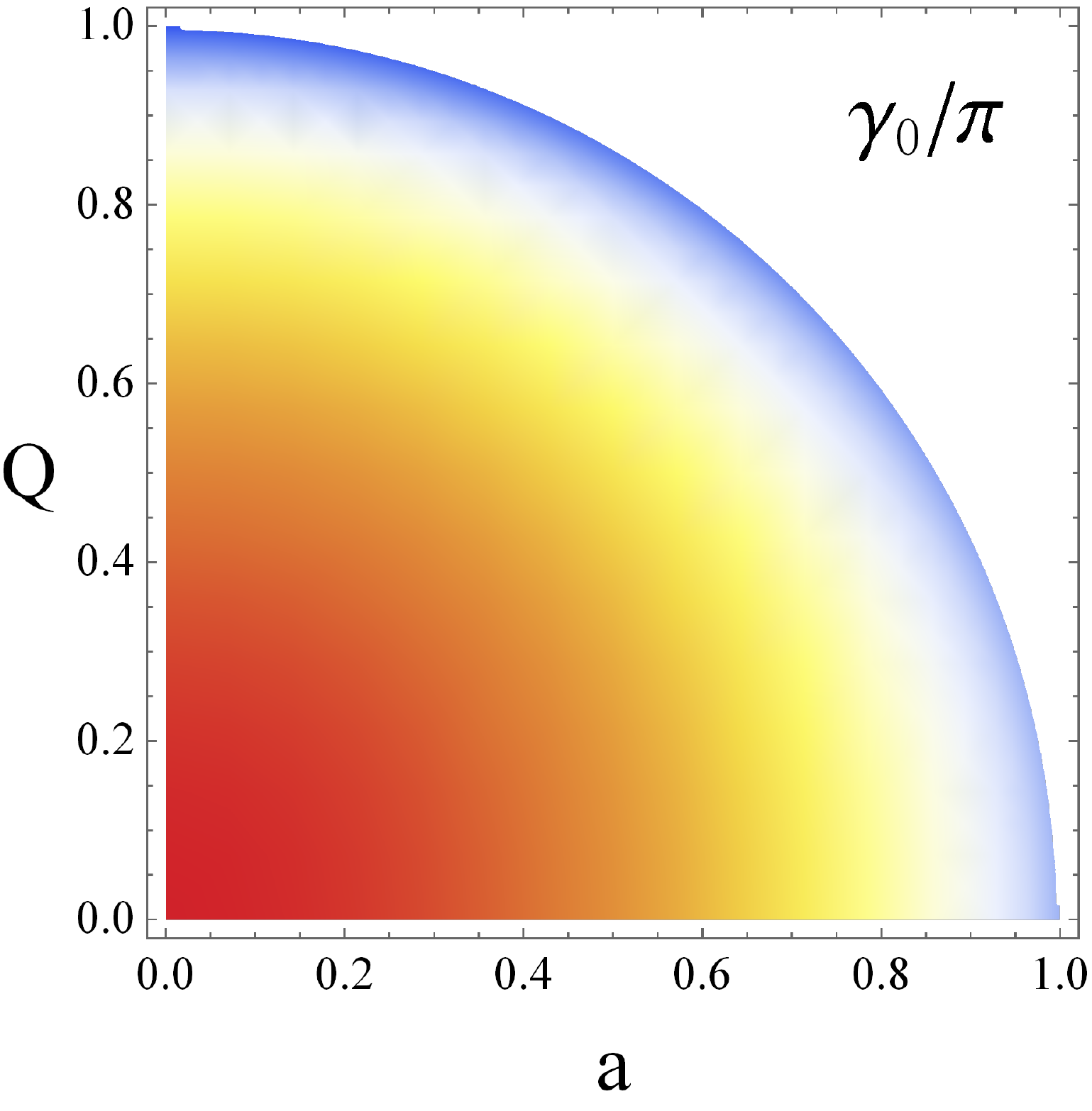} \
	 \includegraphics[scale=0.32]{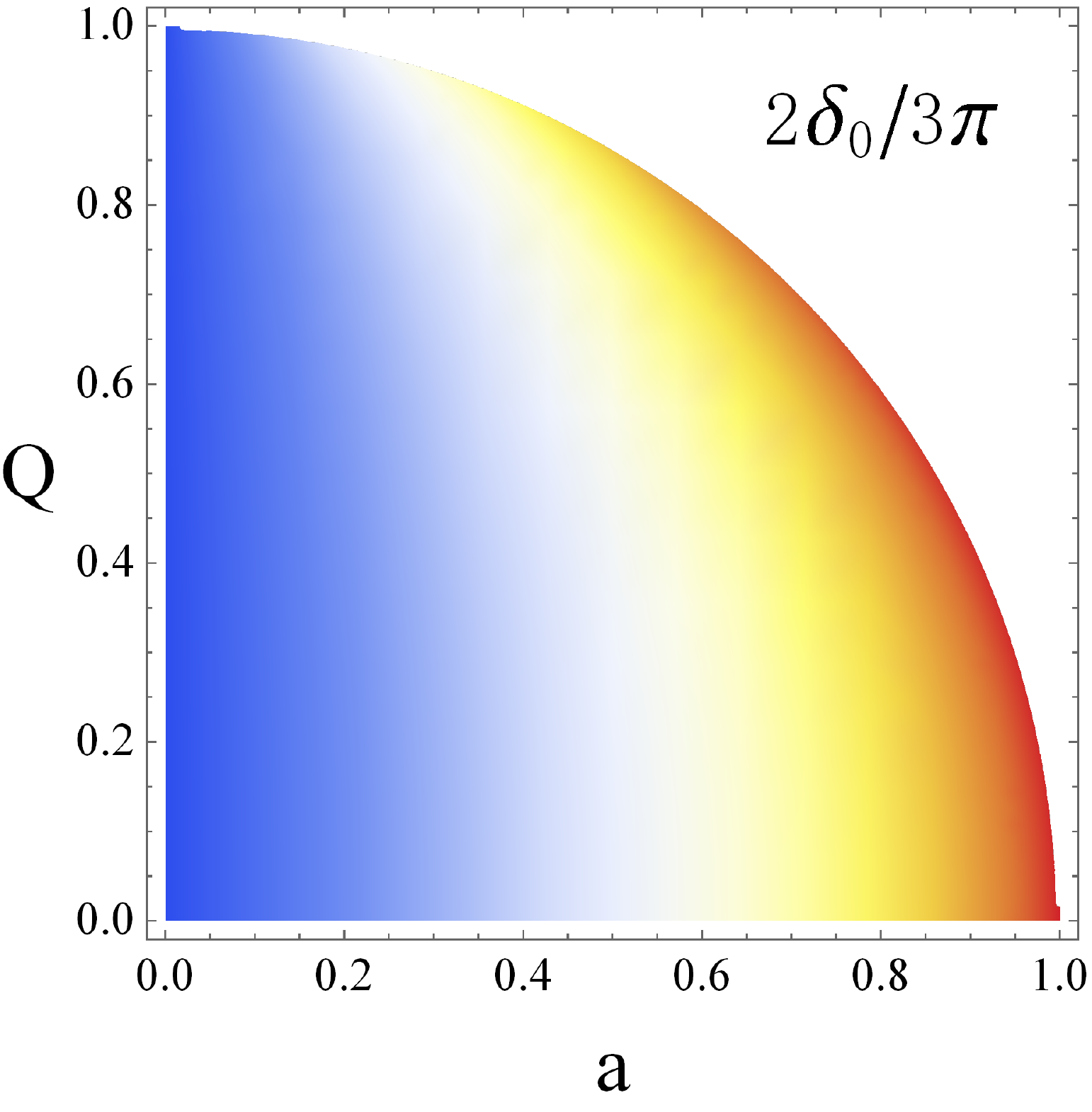} \
	\includegraphics[scale=0.32]{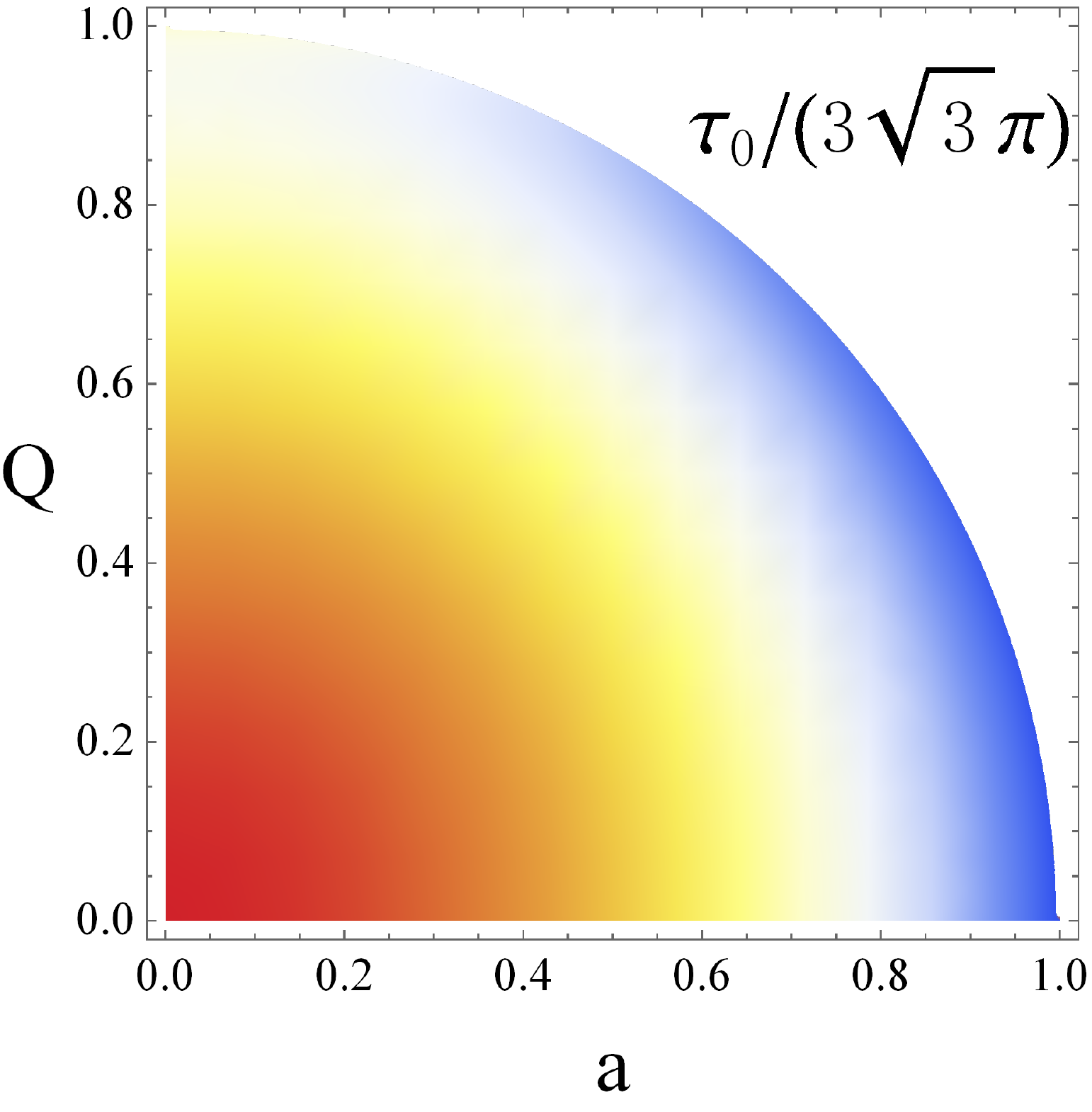} \
	 \includegraphics[scale=0.325]{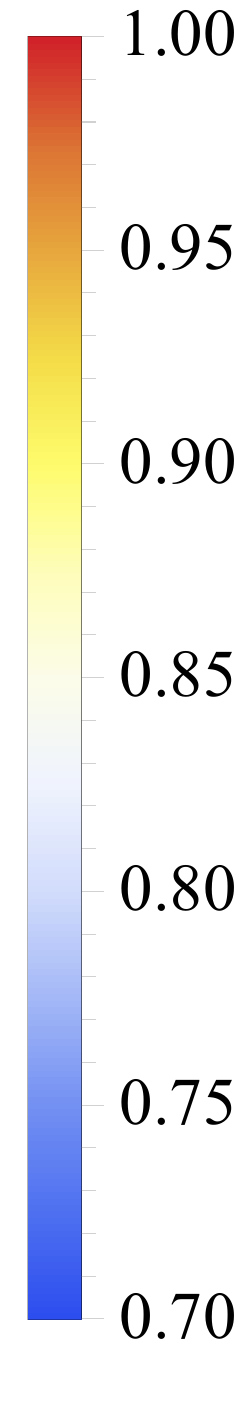}\\
\caption{The key parameters $\gamma_0$, $\d_0$ and $\tau_0$ for an on-axis observer. Top: we show the influences of spin $a$ on these parameters for black holes with charge $Q=0$ (left) and $Q=0.6$ (right). Middle: we show the influences of charge $Q$ on these parameters for black holes with spin $a=0$ (left) and $a=0.6$ (right). Bottom: we show the density plots of these parameters on the $(a,\,Q)$ plane within the whole allowed range for $a$ and $Q$.}
\label{fig:onrings}
\end{figure}

We first consider a general source viewed by an on-axis observer using the near-critical equations \eqref{ncequations}. In this case, we have $\theta_o=0$, thus $\pm_o=-1$, $F_o=K$, $\Pi_o=\Pi$ and $E'_o=E'$.
Moreover, the photon shell radii $\td r\in(\td r_-,\td r_+)$ reduces to a single photon sphere radius $\td r_0$, and the key parameters $\g$ [Eq.~\eqref{gamma}], $\d$ [Eq.~\eqref{deltaexp}] and $\tau$ [Eq.~\eqref{tauexp}] reduce to $\g_0$, $\d_0$ and $\tau_0$, respectively. The critical parameters $\g_0$, $\d_0$ and $\tau_0$ are independent of the position around a critical curve.
Now we consider successive images labeled with $m$ and $m+1$, then from Eqs.~\eqref{ncequations} we can obtain \cite{Gralla:2019drh}
\bea
\label{ondeq}
&&\frac{d_{m+1}}{d_m}\approx\exp\l[-[1+(-1)^m\frac{\td F_s}{\td K}]\g_0\r],\\
\label{onphieq}
&&(\D \phi)_{m+1}-(\D \phi)_m\approx [1+(-1)^m\frac{\td F_s}{\td K}]\d_0-(-1)^m\frac{\td F_s}{\td K}\pi,\\
\label{onteq}
&&(\D t)_{m+1}-(\D t)_m\approx [1+(-1)^m\frac{\td F_s}{\td K}]\tau_0+(-1)^m \f{4a\td{u}_+}{\sqrt{-\td{u}_-}}\td E'
[ \f{\td{F}_s}{\td K}-\f{\td{E}'_s}{\td E'}].
\eea
We have used Eq.~\eqref{polecrossing} for the pole-crossing orbits to obtain the last term in Eq.~\eqref{onphieq}.
Note that $d=\rho-\td \rho$ for the on-axis case and $C_\pm, D_\pm, H_\pm$ in Eqs.~\eqref{ncequations} were dropped out from the final results.
If we consider the images that execute $m$ and $m+2$ angular turnings, the above equations are simplified dramatically, and reduce to
\be
\label{onmplus2}
\frac{d_{m+2}}{d_m}\approx e^{-2\g_0},\qquad
(\D \phi)_{m+2}-(\D \phi)_m\approx 2\d_0,\qquad
(\D t)_{m+2}-(\D t)_m\approx 2\tau_0.
\ee

Next, we consider an equatorial source, for which we have $\theta_s=\pi/2$, $F_s=0$, $\Pi_s=0$ and $E'_s=0$. Then from Eqs.~\eqref{ondeq}, \eqref{onphieq} and \eqref{onteq} we get
\be
\label{oneq}
\frac{d_{\text{eq},m+1}}{d_{\text{eq},m}}\approx e^{-\g_0},\qquad
(\D \phi)_{\text{eq},m+1}-(\D \phi)_{\text{eq},m}\approx \d_0,\qquad
(\D t)_{\text{eq},m+1}-(\D t)_{\text{eq},m}\approx \tau_0.
\ee

Now we consider the $m$-th image ring for an equatorial source ring with $r_s\in(r_{s-},\,r_{s+})$. We use $\rho^{(m)}_\pm$ to denote the outer/inner edges for the image ring. Recalling that $d_\pm=\rho_\pm-\td\rho$, then the position $\rho^{(m)}_\pm$ can be obtained from Eq.~\eqref{ncequations} and the width of the $m$-th image ring is given by
\be
\label{defDrho}
\D \rho^{(m)}=\rho^{(m)}_{+}-\rho_{-}^{(m)}.
\ee
It follows from Eq.~\eqref{ncequations} that $\D\rho^{(m+1)}/\D\rho^{(m)}\approx e^{-\g_0}$, meaning that successive images are narrower by a factor of $e^{-\g_0}$.

The above expressions [from Eq.~\eqref{ondeq} to Eq.~\eqref{oneq}] have the same form as those for the Kerr case \cite{Gralla:2019drh}, however, the effects of charge $Q$ were included implicitly.
Like in the Kerr case, we can see that the three key parameters still control the behaviors of successive images in the KN case. The parameters $\g_0$, $\d_0$ and $\tau_0$ encode the effects of demagnification, angle lapse and time delay, respectively, between successive images.
For a general source, even though these effects are not precisely encoded by $\g_0$, $\d_0$ and $\tau_0$ when $m$ is shifted to $m+1$ [see Eqs.~\eqref{ondeq}, \eqref{onphieq} and \eqref{onteq}], the effects become evident when we consider images with $m$ and $m+2$. As we can see from Eq.~\eqref{onmplus2} that, relative to the images at $m$ level, those at $m+2$ level have demagnification $2\g_0$, rotation $2\d_0$ and time delay $2\tau_0$. For an equatorial source, we can further see from Eq.~\eqref{oneq} that each successive image has demagnification $\g_0$, rotation $\d_0$ and time delay $\tau_0$ relative to the reference image.
In order to see the influences of spin $a$ and charge $Q$, we show the three key parameters for black holes with different choices of $a$ and $Q$ in Fig.~\ref{fig:onrings}.

\subsection{Sources viewed by an off-axis observer}\label{offimage}
We now consider a general source viewed by an off-axis observer, for which we have $\theta_o\neq0$ and $\pm_o=\si(\beta)$.
For an off-axis observer, the key parameters $\g$, $\d$ and $\tau$ depend on $\td r$ nontrivially [see Eqs.~\eqref{gamma}, \eqref{deltaexp} and \eqref{tauexp}]. Near a fixed point $[\td r, \si(\td\b)]$ on the critical curve, the near-critical lens equations \eqref{ncequations} for successive images become \cite{Gralla:2019drh}
\bea
&&\frac{d_{m+1}}{d_m}\approx\exp\l[-[1-\si(\b)(-1)^m\frac{\td F_s}{\td K}]\g\r],\\
&&(\D \phi)_{m+1}-(\D \phi)_m\approx [1-\si(\b)(-1)^m\frac{\td F_s}{\td K}]\hat\d+\si (\b)(-1)^m \f{2\td \lambda\td \Pi}{a\sqrt{-\td{u}_-}}
[ \f{\td{F}_s}{\td K}-\f{\td{\Pi}_s}{\td \Pi}],\\
&&(\D t)_{m+1}-(\D t)_m\approx [1-\si(\b)(-1)^m\frac{\td F_s}{\td K}]\tau-\si(\b)(-1)^m \f{4a\td{u}_+}{\sqrt{-\td{u}_-}}\td E'
[\f{\td{F}_s}{\td K}-\f{\td{E}'_s}{\td E'}],
\eea
where $\hat\d=\d-2\pi H(\td r-\td r_0)$, and the functions $C_\pm$, $D_\pm$ and $H_\pm$ in Eqs.~\eqref{ncequations} were dropped out again from the final results.

For the images labeled with $m$ and $m+2$, we can still get similar expressions as Eqs.~\eqref{onmplus2}, in which the key parameters $\g_0,\,\d_0,\,\tau_0$ are replaced by those without the subscript.
Likewise, for an equatorial source, Eqs.~\eqref{oneq} still hold for the off-axis case once we made the same minor replacement for the key parameters.

\begin{figure}[H]
\centering
\includegraphics[scale=0.34]{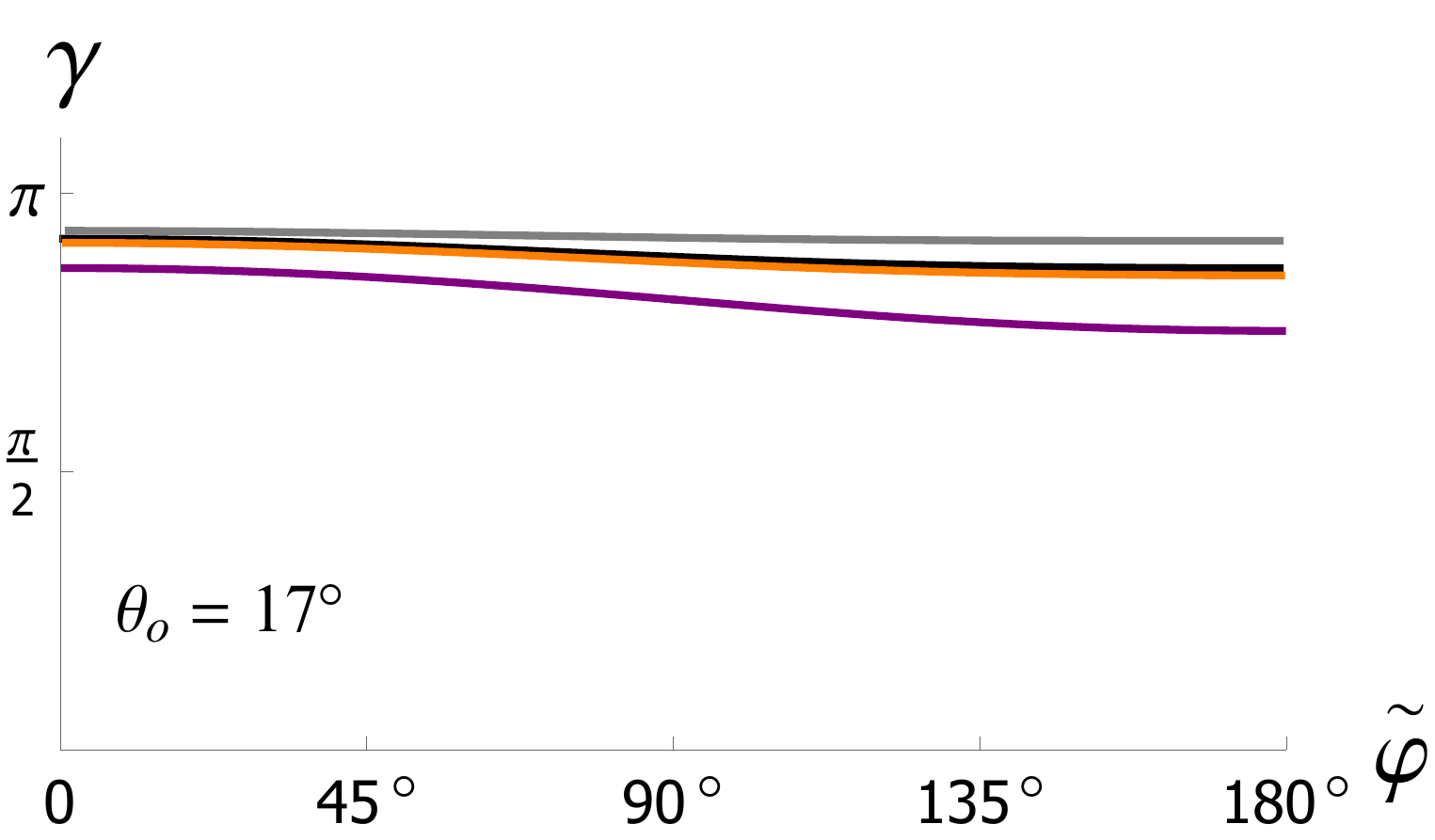}
\includegraphics[scale=0.34]{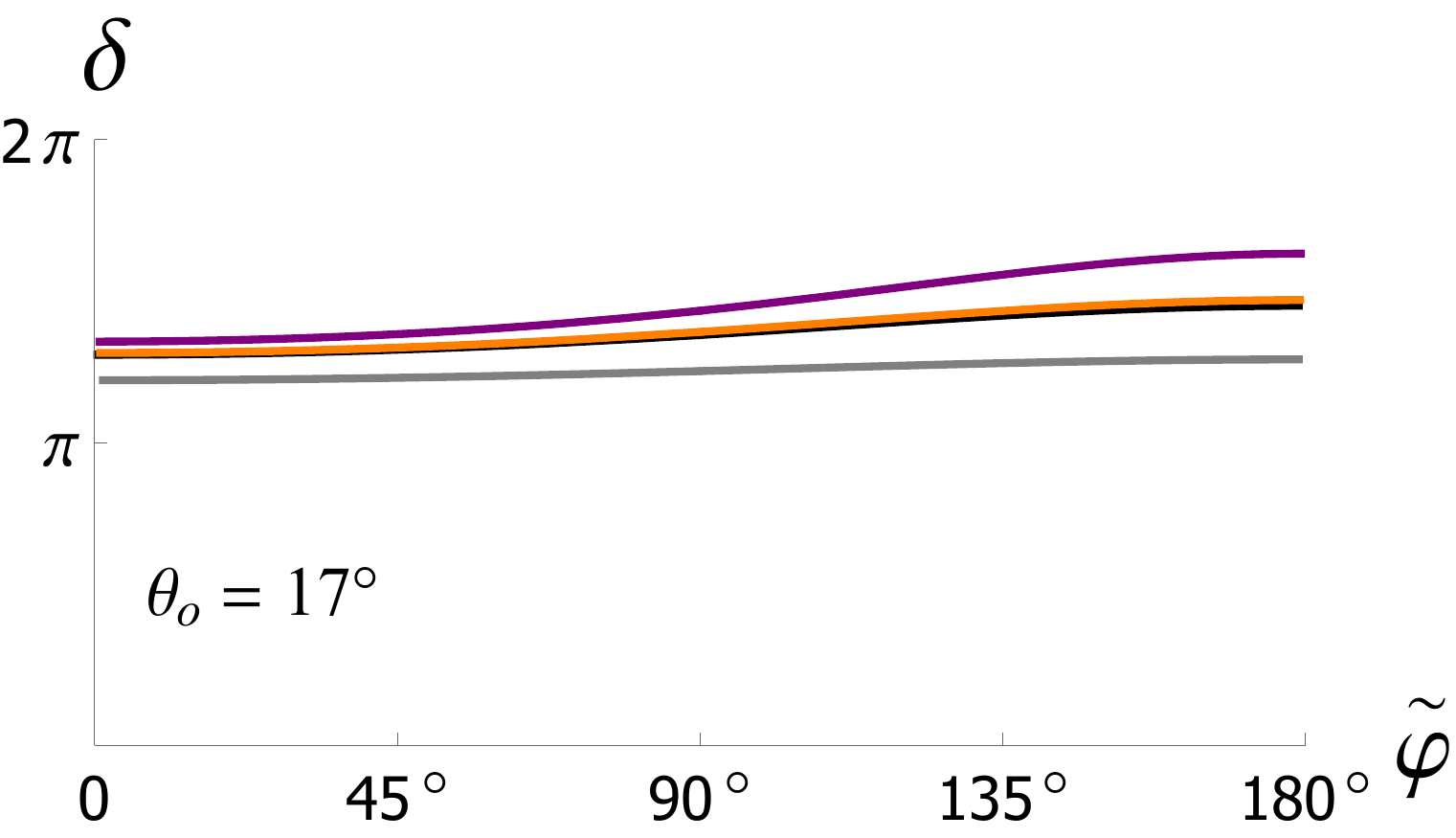} \
\includegraphics[scale=0.34]{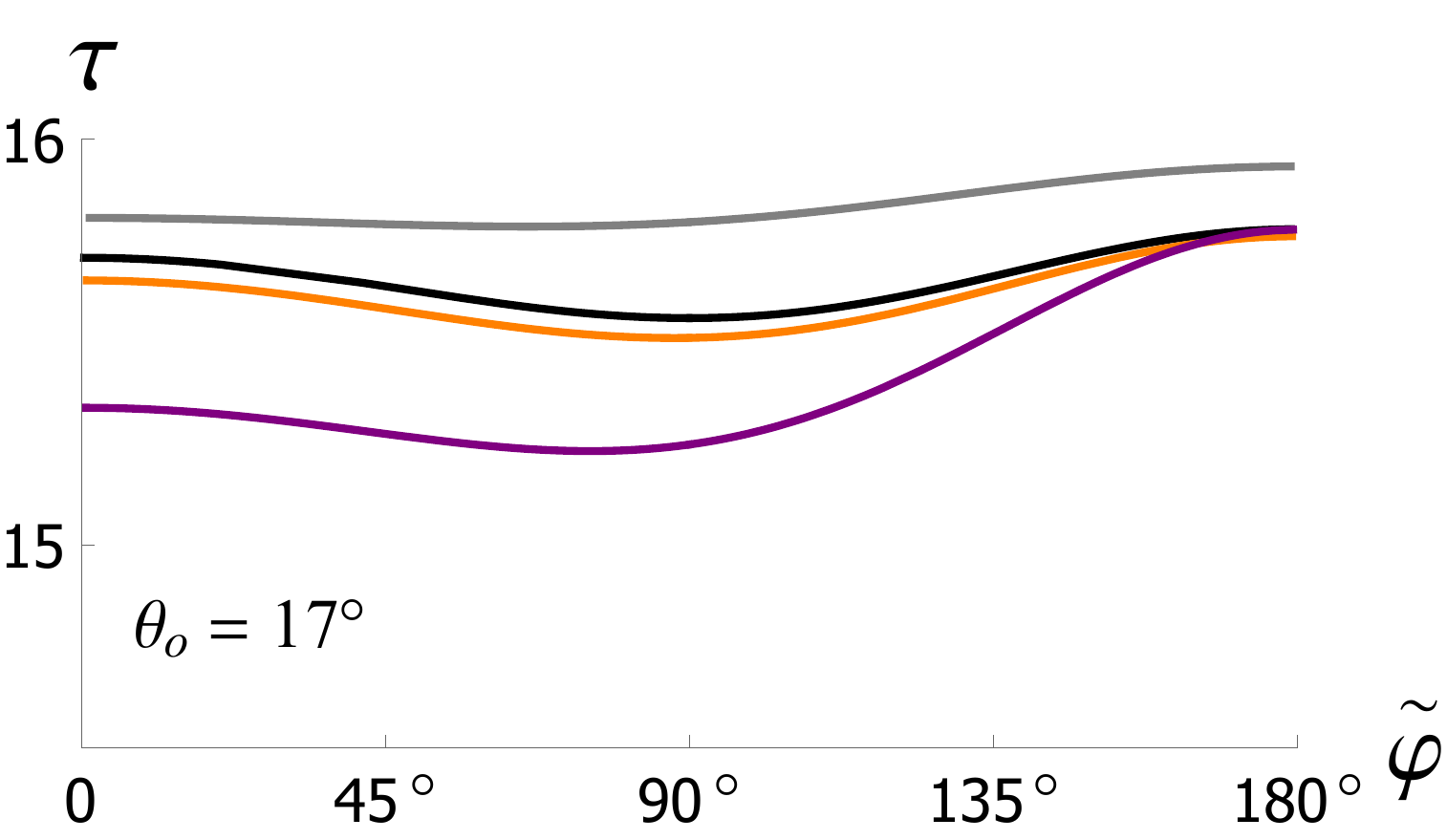}\\
~\\
\includegraphics[scale=0.34]{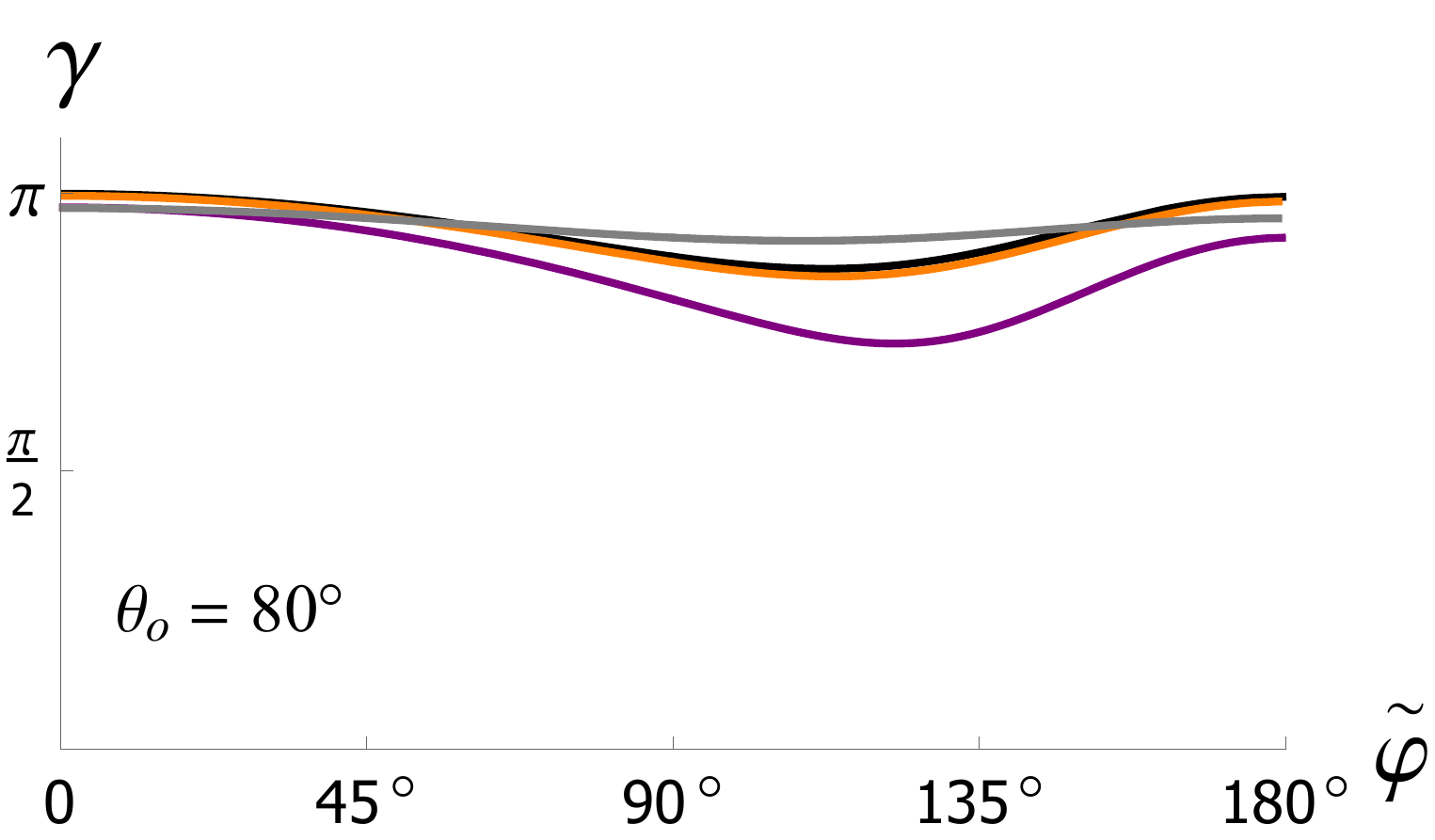}
\includegraphics[scale=0.34]{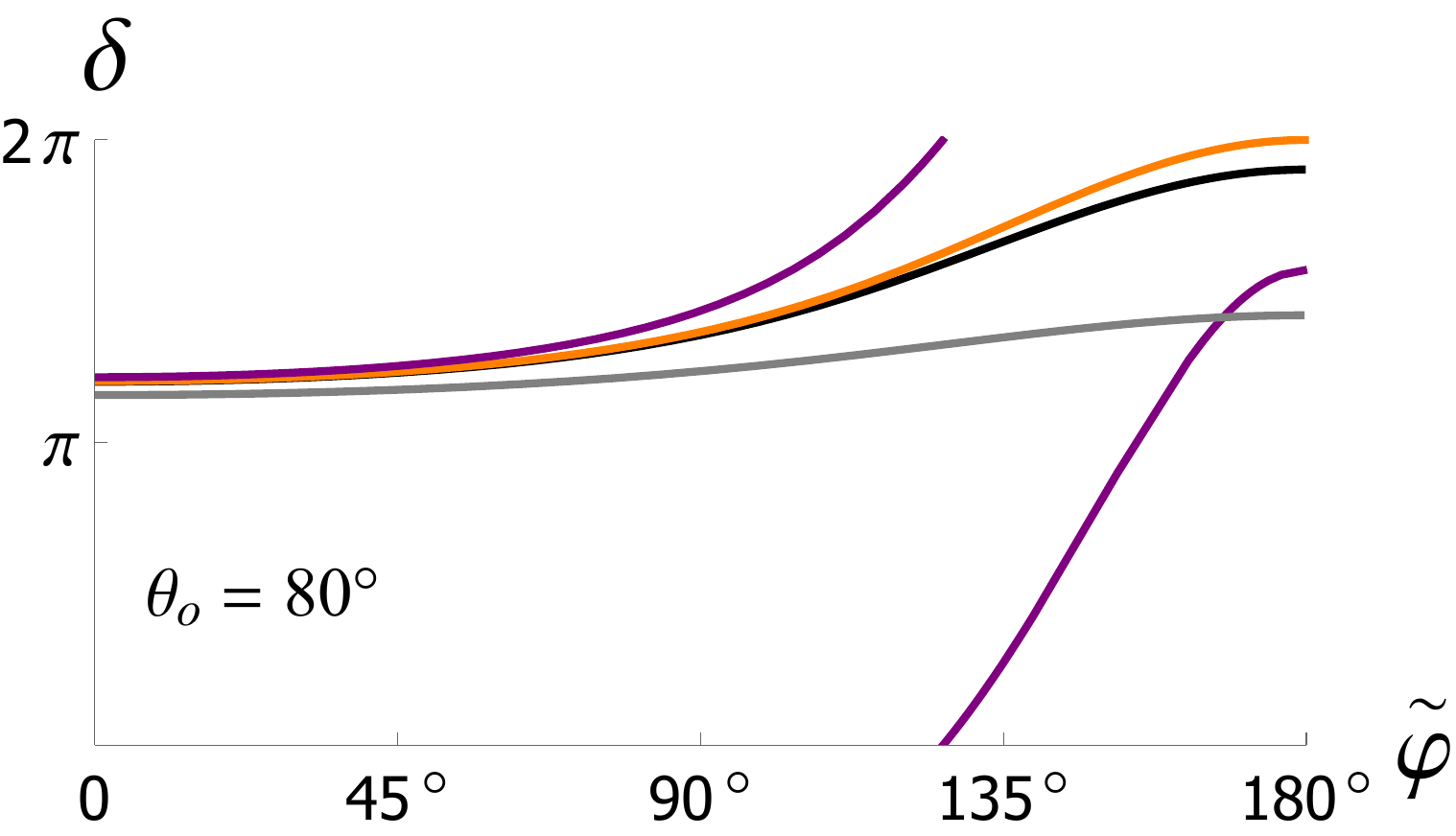} \
\includegraphics[scale=0.34]{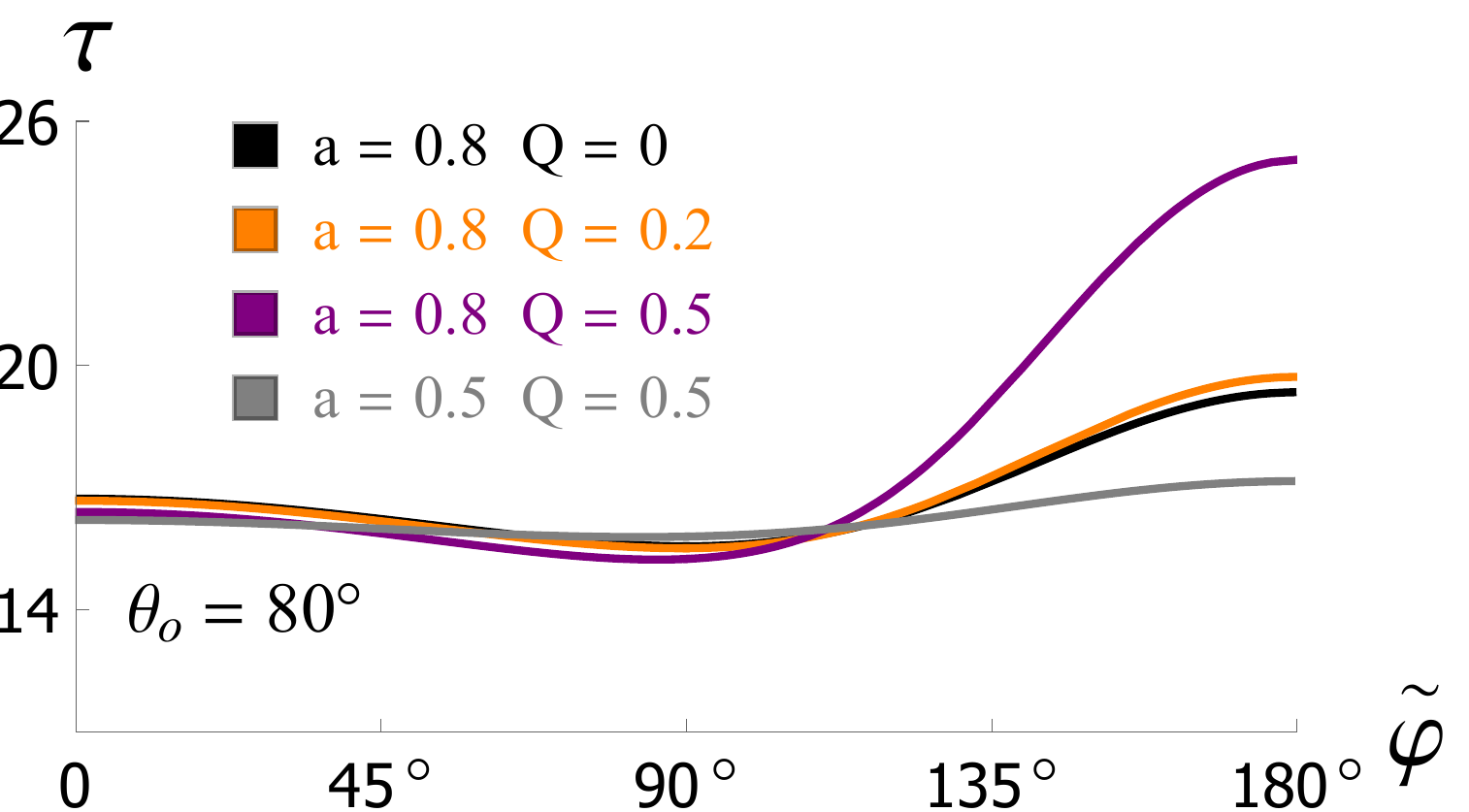}\\
\caption{The key parameters $\gamma$, $\d$ and $\tau$ for off-axis observers with inclinations $\t_o=17^\circ$ (above) and $80^\circ$ (below). For several selected choices of black hole parameters $a$ and $Q$, we show the values of the key parameters along the critical curve which is parameterized by the polar coordinate $\td\vp$ [see Eq.~\eqref{offpolar}].}
\label{fig:offrings}
\end{figure}

When a given source is observed off the pole, the key parameters $\g$, $\d$ and $\tau$ also control the behaviors of its successive images. However, in contrast with the on-axis case, the parameters for the off-axis case depend on the position $\td r$ along the critical curve $\M C$. Near each point on the critical curve, the properties of successive images are the similar as those for the on-axis case described above in Sec.~\ref{onimage} and we will not repeat them again.
In Fig.~\ref{fig:offrings} we show the value of the key parameters along the critical curve for selected inclination $\t_o$, spin $a$ and charge $Q$.

\section{Summary and discussion}\label{sec:summary}
In this paper, we studied the lensing effects of charged rotating KN black holes, closely following the work \cite{Gralla:2019drh} on the lensing by  Kerr black holes. Comparing with the results in Kerr case, we found the modifications in the integral form of the photon motion and three key parameters $\gamma$, $\delta$ and $\tau$ of critical photon emissions induced by the charge $Q$ of a KN black hole, see \eqref{criticalIr}, \eqref{criticalIphit}, \eqref{gamma}, \eqref{deltaexp} and \eqref{tauexp}. Then we investigated the optical appearances of luminous sources around KN black holes, and our main results can be seen in Fig.~\ref{fig:criticalcurves} to Fig.~\ref{fig:offrings}. In particular, we paid special attention to the higher-order images ($\bar{m}\ge2$) of luminous sources, which have not been discussed carefully in the literature.

From our studies, we noticed that for the primary and secondary images the effect of $Q$ are not evident, see Fig.~\ref{equatorial2} to Figs. \ref{Dphidisk}. However it is worth emphasizing that the charge $Q$ would have a significant effect for higher-order images, see Figs.~\ref{fig:onrings} and \ref{fig:offrings}. The reason why the charge $Q$ is able to have a noticeable effect on the higher-order images is that the photon emissions of higher order images are closer to the critical curve $\mathcal{C}$.  Since the lights travel around the black hole more times before reaching the observer, the higher-order images can encode more information of critical curve. On the other hand, from Fig.~\ref{fig:criticalcurves}, we can see that the charge $Q$ has a nonnegligible effect on the critical curve. As a result, it is quite reasonable that charge $Q$ would give a detectable effect on higher-order images. We believe that as the observational accuracy of EHT improves \cite{EventHorizonTelescope:2019uob,Falcke:2017ukt,Kardashev:2014sjq}, observations of higher-order images will become feasible and allows us to constrain the black hole parameters, including the charge   \cite{Johnson:2019ljv,Gralla:2020srx,EventHorizonTelescope:2021dqv,Broderick:2021ohx,
Ayzenberg:2022twz,Paugnat:2022qzy}.

\appendix

\section{Radial integrals for near-critical rays}\label{app:radialmotion}
In this appendix, we will compute the radial integrals for near-critical rays following \cite{Gralla:2019drh}.
We will outline the main steps and provide the final results for reference.

\subsection{Asymptotic expansions}\label{app:expansions}
\begin{figure}[H]
\centering
\includegraphics[scale=0.5]{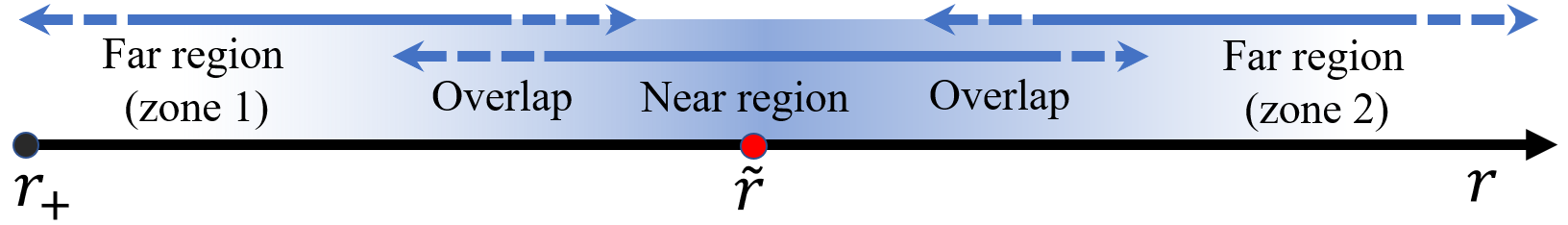}
\caption{Schematic diagram for near, far and overlapping regions with regards to $\td{r}$. Note that zone 1 ($\d r<0$) and zone 2 ($\d r>0$) are in two disjoint far regions.}
\label{mae}
\end{figure}

For the light rays crossing the equatorial plane we may introduce a nonnegative parameter $q=\sqrt{\eta}$ for convenience. For near-critical rays, the impact parameters can be written as
\bea
\lm=\td{\lm}(1+\delta\lm), \qquad
q=\td{q}(1+\delta q),
\eea
with $|{\delta\lm}|\sim|{\delta q}|\sim\e \ll 1$.
We introduce the $\d r$-coordinate around the critical radius $\td r$ as well
\be
r=\td{r}(1+\delta r).
\ee
Then we can define the near and far regions as follows,
\bea
&&\text{Near region}: \qquad |\d r|\ll 1, \\
&&\text{Far region}: \quad\quad\,\,\,\,  |\d r|\gg \e.
\eea
The far region is disjoint around $\td{r}$, as can be seen in Fig.\ref{mae}.
Apparently, there are overlapping regions here in which $\e \ll |\delta r|\ll 1$. This ensures the usability of the matched asymptotic expansion (MAE) method.

In the near region, the radial potential $\M R(r)$ may be expanded in the regime $|\d r|\sim\sqrt{\e}$, then
\be
\M R(r)
= \frac{1}{2}\M R^{\prime\prime}(\td r)\delta r^2+\M R_\lm \d \lm+\M R_q \d q+\M O(\e^{3/2}),
\ee
where
\bea
\M R^{\prime\prime}(\td r)=\td{r}^2(6\td{r}^2+a^2-\td{\lm}^2-\td{q}^2),\quad
\M R_{\lm}=-2\td{\lm}\td{r}\D(\td{r})\frac{2Q-3M\td{r}+\td{r}^2}{a(M-\td{r})},\quad
\M R_q=-2\td{q}^2\D(\td{r}).
\eea
It is convenient to define
\be
\M R_n(\d r):=\frac{1}{2}\M R^{\prime\prime}(\td r)\delta r^2+\M R_\lm \d \lm+\M R_q \d q
=4\td{r}^4\td\chi(\d r^2-\d r_0^2),
\ee
where
\be
\label{dr0}
\d r_0^2=\frac{\td\D}{2\td r^2\td \chi}\left[\frac{\td q^2}{\td r^2}\d q-\frac{\td \lambda}{a}\left(\frac{\td r-3M}{\td r-M}\right)\d\lambda\right],\qquad
\td\chi=1-\frac{M\D(\td{r})}{\td{r}(M-\td{r})^2}.
\ee

In the far region, we have $|\d r|\gg\e$ and the radial potential becomes
\be
\M R(r)
=\M R(\td r(1 +\d r))|_{\lm=\td \lm,q=\td q}+\M O(\e).
\ee
It is convenient to define
\be
\M R_f(\d r):=\M R(\td r(1 +\d r))|_{\lm=\td \lm,q=\td q}=4\td{r}^4\td\chi \d r^2\M Q(\d r)
\ee
where
\be
\M Q(\d r)=1+\frac{\d r}{\td \chi}+\frac{\d r^2}{4\td \chi}.
\ee

Similarly, we can also expand the functions $\M I_\phi$ and $\M I_t$ [Eq.~\eqref{defintegrals0}] in the near and far regions, respectively. Then the radial integrals $I_i$ ($i=r,\phi,t$) [Eq.~\eqref{defintegrals}] can be computed explicitly in each of these regions. Combining the near-zone and far-zone results for a physical photon trajectory by using the MAE method, one can obtain the final results for the radial integrals connecting a source to an observer.

Note that a physical light ray never encounter a turning point in the far region thus we should always have $\M R_f(\d r)>0$. On the other hand, a light ray has a turning point $r_t$ (corresponding to the larger root of $\M R_n(\d r)=0$) in the near region if $\d r_0^2>0$. In that case, $r_t=\td r(1+\d r_0)$. We have described in Sec.~\ref{sec:nearcriticalrays} that $\d r_0^2$ is related to the perpendicular distance $d$ from the critical curve and $\d r_0^2\gtrless0$ corresponds to the rays arriving outside/inside the critical curve, respectively.

\subsection{Matched asymptotic expansion computations}\label{app:mae}
In the near region, we use $I^n_i(\d r_a, \d r_b)$ to denote the leading order results in $\e$ of the radial integral from $\d r_a$ to $\d r_b$ with $\d r_a<\d r_b\ll1$. The radial integrals in \eqref{defintegrals} can be written as
\be
\label{nearintegrals}
I^n_r
=\frac{1}{2\td{r}\sqrt{\td\chi}}\int_{\d r_a}^{\d r_b}\frac{d(\d r)}{\sqrt{\d r^2-\d r_0^2}},\quad
I_{\phi}^n
=a\l(\f{\td{r}+M}{\td{r}-M}+\f{Q}{\td{\D}}\r)I_r^n, \quad
I_{t}^n
=\td{r}^2\l(\f{\td{r}+3M}{\td{r}-M}\r)I_r^n.
\ee
And the result for $I_r^n$ is
\be
I_r^n(\d r_a,\d r_b)=\frac{\si(\d r)}{2\td{r}\sqrt{\td\chi}}\log{\l(\frac{|\d r|+\sqrt{\d r ^2-\d
r^2_0}}{|\d r_0|}\r)}\Bigg|^{\d r_b}_{\d r_a}.
\ee

In one zone of the far region (either in zone 1 or in zone 2 in Fig.~\ref{mae}), we use $I^f_i(\d r_a, \d r_b)$ to denote the radial integral from $\d r_a$ to $\d r_b$ with $\e\ll\d r_a<\d r_b$.
The radial integrals in \eqref{defintegrals} can be written as
\be
I^f_i
=\frac{1}{2\td{r}\sqrt{\td\chi}}\int_{\d r_a}^{\d r_b}\frac{\M I^f_i(\d r) d(\d r)}{\sqrt{\d r^2 \M Q(\d r)}},\qquad
\M I^f_i(\d r)=\M I_i(\td r(1+\d r))|_{\lm=\td \lm,q=\td q},
\ee
where $\M I_i(r)$ are introduced in \eqref{defintegrals0}.
Following the procedure in \cite{Gralla:2019drh} we get
\bea
\label{If}
&&I_r^f(\d r_a,\d r_b)
 = \f{1}{2\td{r}\sqrt{\td\chi}} \ \si (\d r) \log{\l|\f{\sqrt{\M Q(\d r)}+\d r/2\td\chi+1}{\d r}\r|} \Bigg|^{\d r_b}_{\d r_a},\\
\label{iphif}
&&I_{\phi }^f(\d r_a,\d r_b)
=a\l(\f{\td{r}+M}{\td{r}-M}+\f{Q}{\td{\D}}\r)I_r^f(\d r_a,\d r_b) -\si(\d r)\f{M a}{\td{r}^2\sqrt{\td \chi}}\M Q_{\phi}(\d r)\bigg|^{\d r_b}_{\d r_a},\\
\label{itf}
&&I_{t}^f(\d r_a,\d r_b)=\td{r}^2\l( \f{\td{r}+3M}{\td{r}-M}\r)I_r^f(\d r_a,\d r_b)-\si(\d r)\f{\td{r}}{2\sqrt{\td\chi}}\M Q_t(\d r)\bigg|^{\d r_b}_{\d r_a},
\eea
where
\bea
\M Q_{\phi}(\d r)&=&-\frac{c_0+\d r}{(\d r_+-\d r_-)\d r_-\sqrt{\M Q(\d r)}} \arctanh{\M Q_2(\d r, \d r_-)}+(\d r_-\leftrightarrow \d r_+),\\
\M Q_t(\d r)&=&-4\td \chi\sqrt{\M Q(\d r)}-\f{4M\sqrt{\td \chi}}{\td{r}}\arctanh{\M Q_2(\d r, \infty)}\nn \\
&&+\l[ \f{c_1(1+\d r_-)+c_2(1+\d r_-)^2+(1+\d r_-)^4}{\d r_-(\d r_{-}-\d r_+)\sqrt{\M Q(\d r_-)}}\arctanh{\M Q_2(\d r,\d r_-)}+(\d r_-\leftrightarrow \d r_+) \r] ,\qquad\,\, \,\,\,\,
\eea
with $\d r_\pm$ being the radii of the event horizon in the $\d r-$coordinates, and
\bea
&&c_0=1-\frac{a\td{\lm}}{2M\td{r}},\quad
c_1=\f{2M a}{\td{r}^3}(a-\td{\lm}), \quad c_2=\f{a^2}{\td{r}^2}, \\
&&\M Q_2(\d r_a,\d r_b)=\f{2\sqrt{\M Q(\d r_a)\M Q(\d r_b)}}{\M Q(\d r_a)+\M Q(\d r_b)-(\d r_a-\d r_b)^2/4\td \chi}.
\eea

Next we use the MAE method to calculate radial integrals
for the photons connecting $|\d r_a|\ll1$ in the near region to $|\d r_b|\gg\e$ in the far region through their overlapping region $\e\ll|\d R|\ll1$.
There are several distinct scenarios. For the photons arriving outside the critical curve, we have $\d r_a>\d r_0$ such that the point is outside the turning point and we denote the integrals for this case as $I_{i,\text{out}}^{n\rightarrow 2}(\d r_a,\d r_b)$. For the photons arriving inside the critical curve, we may either have $0<\d r_a<\d r_b$ or $\d r_b<\d r_a<0$ and we denote the corresponding integrals as $I_{i,\text{in}}^{n\rightarrow 2}(\d r_a,\d r_b)$ and $I_{i,\text{in}}^{1\rightarrow n}(\d r_b,\d r_a)$, respectively.
We have used the superscripts ``$1\rightarrow n$'' to denote the path integrals from zone 1 of the far region to the near region, and  ``$n\rightarrow 2$'' to denote the path integrals from the near region to zone 2 of the far region (see Fig.~\ref{mae}).
We have also used the subscript ''in/out" to denote the photons arriving inside or outside the critical curve, respectively.

We will take $I_{r,\text{out}}^{n\rightarrow 2}(\d r_a, \d r_b)$ as an example to perform the calculation, and list the results for $I_{\phi,\text{out}}^{n\rightarrow 2}(\d r_a,\d r_b)$ and $I_{t,\text{out}}^{n\rightarrow 2}(\d r_a,\d r_b)$ without calculation.
We compute the integral $I_{r,\text{out}}^{n\rightarrow 2}(\d r_a, \d r_b)$ by splitting it into two pieces, and read
\bea
I_{r,\text{out}}^{n\rightarrow 2}(\d r_a, \d r_b)&=&I^n_r(\d r_a,\d R)+I^f_r(\d R,\d r_b)\nn\\
&=&\frac{1}{2\td{r}\sqrt{\td\chi}}\l[\log{\l(\frac{|\d r|+\sqrt{\d r ^2-\d
r^2_0}}{|\d r_0|}\r)}\Bigg|^{\d R}_{\d r_a}
+\log{\l|\f{\sqrt{\M Q(\d r)}+\d r/2\td\chi+1}{\d r}\r|} \Bigg|^{\d r_b}_{\d R}\r]\nn\\
&=&\f1{2\td{r}\sqrt{\td\chi}} \l[ \log\l( \f{8\td\chi}{|\d r_0|\sqrt{1-\td\chi}} \r)- \arctanh\l[\f{\sqrt{\d r_a^2-\d r_0^2}}{\d r_a}  \r]-\arctanh \f{\sqrt{\M Q(\d r_b)}}{1+\d r_b/2\td\chi} \r],\qquad\,\,\,\,
\eea
where in the last step $\d R$ was canceled out due to the scaling nature of an overlap region.
The results for $I_{\phi,\text{out}}^{n\rightarrow 2}$ and  $I_{t,\text{out}}^{n\rightarrow 2}$ are
\bea
&&I_{\phi,\text{out}}^{n\rightarrow 2}(\d r_a, \d r_b)=a\l( \f{\td{r}+M}{\td{r}-M}+\f{Q}{\td{\D}} \r)I_{r,\text{out}}^{n\rightarrow 2}(\d r_a, \d r_b)-\f{M a}{\td{r}^2\sqrt{\td\chi}}[\M Q_{\phi}(\d r_b)-\M Q_{\phi}(0)], \label{phinf}\\
&&I_{t,\text{out}}^{n\rightarrow 2}(\d r_a, \d r_b)=\td{r}^2\l( \f{\td{r}+3M}{\td{r}-M}\r)I_{r,\text{out}}^{n\rightarrow 2}(\d r_a, \d r_b)-\f{\td{r}}{2\sqrt{\td\chi}}[\M Q_{t}(\d r_b)-\M Q_{t}(0)].\label{tnf}
\eea
The results for other cases can be obtained in a similar way.

\subsection{Radial integrals connecting light sources to observers at infinity}\label{app:results}
What we really need are the integrals of near-critical photons emitted from $r_s=\td{r}(1+\d r_s)$ and reaching a distant observer $r_o\rightarrow \infty$.
We have the following four cases for the integrals $I_{i,\text{out/in}}(\d r_s,\infty)$ with $i\in\{r,\,\phi,\,t\}$ [to get Eqs.~\eqref{Ir}, \eqref{Iphi} and \eqref{It}].

(a). For the source in zone 2 and the photon arriving outside $\M C$, we have
\bea
&&I_{i,\text{out}}(\d r_s,\infty)\approx I_i^f(\d r_s,\infty)\qquad
\text{(direct)},\\
\label{Irout}
&&I_{i,\text{out}}(\d r_s,\infty)\approx I_{i,\Out}^{n\rightarrow2}(\d r_0, \d r_s) +I_{i,\Out}^{n\rightarrow2}(\d r_0, \infty)\qquad
\text{(reflected)}.
\eea

(b). For the source in the near region and the photon arriving outside $\M C$, we have
\bea
&&I_{i,\text{out}}(\d r_s,\infty)\approx I_{i,\Out}^{n\rightarrow2}(\d r_s,\infty)\qquad
\text{(direct)},\\
&&I_{i,\text{out}}(\d r_s,\infty)\approx I_{i}^{n}(\d r_0, \d r_s) +I_{i,\Out}^{n\rightarrow2}(\d r_0, \infty)\qquad
\text{(reflected)}.
\eea

(c). For the source in the near region and the photon arriving inside $\M C$, we have
\be
I_{i,\text{in}}(\d r_s,\infty)\approx I_{i,\In}^{n\rightarrow2}(\d r_s,\infty).
\ee

(d). For the source in zone 1 and the photon arriving inside $\M C$, we have
\be
\label{Irin}
I_{i,\text{in}}(\d r_s,\infty)\approx I_{i,\In}^{1\rightarrow n}(\d r_s, \d r_n) +I_{i,\In}^{n\rightarrow2}(\d r_n, \infty),
\ee
where $\d r_n$ represents a point in the near region and it would be cancelled out in the final result.

The explicit expressions for the above integrals can be obtained by substituting the relevant results from the previous subsection.
For example, let us compute the results for the complete integrals that either end up at infinity or at the horizon, that is $\d r_s=\infty$ for $I_{i,\Out}$ and $\d r_s=\d r_+$ for $I_{i,\In}$. Then from Eqs.~\eqref{Irout} and \eqref{Irin} we obtain [Eq.~\eqref{Ir}]
\bea
&&I_{r,\text{out}}(\infty,\infty)\approx -\frac{1}{2\td r\sqrt{\td \chi}}\log\l[\l(\frac{1+\sqrt{\td \chi}}{8\td \chi}\r)^2 \d r_0^2 \r]
=-\frac{1}{2\td r\sqrt{\td \chi}}\log[C_+(\infty,\td r) d],\\
&&I_{r,\text{in}}(\d r_+,\infty)\approx -\frac{1}{2\td r\sqrt{\td \chi}}
\log\l[ \frac{(1+\sqrt{\td \chi})\sqrt{1-\td\chi}}{(8\td \chi)^2} \sqrt{\f{1+\M Q_2(\d r_+,0)}{1-\M Q_2(\d r_+,0)}} |\d r_0^2| \r]
\nn\\
&&\qquad\qquad\quad\,\,\,\,\,
=-\frac{1}{2\td r\sqrt{\td \chi}}\log[C_-(\d r_+,\td r) d],
\eea
where [see Eq.~\eqref{distance}]
\bea
&&C_+(\infty,\td r)=\l(\frac{1+\sqrt{\td \chi}}{8\td \chi}\r)^2 \frac{\td{\D}}{2\td{r}^4\td\chi}\sqrt{\td{\b}^2+\l[ \td{\a}-\l( \f{\td{r}+M}{\td{r}-M}\r)a\sin{\t_o} \r]^2},\\
&&C_-(\d r_+,\td r)=-\frac{\sqrt{1-\td \chi}}{1+\sqrt{\td \chi}}\sqrt{\f{1+\M Q_2(\d r_+,0)}{1-\M Q_2(\d r_+,0)}}C_+(\infty,\td r).
\eea
Similarly, we can obtain the results for $I_\phi$ and $I_t$, which are written in the form as Eqs.~\eqref{Iphi} and \eqref{It}, respectively, in which $D_\pm$ and $H_\pm$ for the complete integrals are obtained as
\bea
&&D_+(\infty,\td r)=-\f{2Ma}{\td r^2\sqrt{\td\chi}}[\M Q_\phi(\infty)-\M Q_\phi(0)],\\
&&D_-(\d r_+,\td r)=-\f{Ma}{\td r^2\sqrt{\td\chi}}[\M Q_\phi(\d r_+)+\M Q_\phi(\infty)-2\M Q_\phi(0)],\\
&&H_+(\infty,\td r)=-\f{\td r}{\sqrt{\td\chi}}[\M Q_t(\infty)-\M Q_t(0)],\\
&&H_-(\d r_+,\td r)=-\f{\td r}{2\sqrt{\td\chi}}[\M Q_t(\d r_+)+\M Q_t(\infty)-2\M Q_t(0)].
\eea
Note that the radial integrals for the KN case are almost in the same form as those for the Kerr case \cite{Gralla:2019drh}.

\section*{Acknowledgments}
We are grateful to office mates in South 530 for helpful discussions. Special thanks to Zhenyu Zhang and Qiaomu Peng for their suggestions on drawing figures of this paper. The work is in part supported by NSFC Grant  No. 11735001, 11775022 and 11873044. MG is also supported by ``the Fundamental Research Funds for the Central Universities'' with Grant No. 2021NTST13.

\bibliographystyle{utphys}
\bibliography{kn}

\end{document}